\documentclass[amsmath,amssymb,nofootinbib,prd,showpacs,twocolumn,letterpaper]{revtex4-1}

\usepackage[english]{babel}
\usepackage[dvips]{graphicx}
\usepackage{amsthm}
\usepackage{bm}
\usepackage{color}
\usepackage{latexsym}
\usepackage{times}
\usepackage{stmaryrd}

\theoremstyle{definition}

\theoremstyle{plain}

\newcommand{\dd}{\mathrm{d}}

\newcommand{\cf}{\textit{cf.}~}
\newcommand{\ie}{\textit{i.e.}~}
\newcommand{\eg}{\textit{e.g.}~}
\newcommand{\st}{\textit{s.t.}~}

\begin{document}
\title[Discontinuous Galerkin methods for general-relativistic
  hydrodynamics]{Discontinuous Galerkin methods for
  general-relativistic hydrodynamics: formulation and application to
  spherically symmetric spacetimes}

\author{David Radice}
\affiliation{Max-Planck-Institut f\"ur Gravitationsphysik, Albert Einstein
Institut, Potsdam, Germany}

\author{Luciano Rezzolla}
\affiliation{Max-Planck-Institut f\"ur Gravitationsphysik, Albert Einstein
Institut, Potsdam, Germany}
\affiliation{Department of Physics and Astronomy, Louisiana State
University, Baton Rouge, USA}

\begin{abstract}
  We have developed the formalism necessary to employ the
  discontinuous-Galerkin approach in general-relativistic
  hydrodynamics. The formalism is firstly presented in a general
  $4$-dimensional setting and then specialized to the case of
  spherical symmetry within a $3+1$ splitting of spacetime. As a
  direct application, we have constructed a one-dimensional code,
  \texttt{EDGES}, which has been used to asses the viability of these
  methods via a series of tests involving highly relativistic flows in
  strong gravity. Our results show that discontinuous Galerkin methods
  are able not only to handle strong relativistic shock waves but, at
  the same time, to attain very high orders of accuracy and
  exponential convergence rates in smooth regions of the flow. Given
  these promising prospects and their affinity with a pseudospectral
  solution of the Einstein equations, discontinuous Galerkin methods
  could represent a new paradigm for the accurate numerical modelling
  in relativistic astrophysics.
\end{abstract}

\pacs{
04.25.Dm, 
02.70.Dh, 
47.11.Kb, 
04.40.Dg, 
}

\maketitle

\section{Introduction}

Special and general-relativistic hydrodynamics play a fundamental role
in a number of astrophysical scenarios characterized by strong
gravitational fields and flows with high Lorentz factors. These
scenarios are of central importance for many phenomena in high-energy
astrophysics or gravitational-wave astronomy, such as, core-collapse
supernovae, coalescing neutron stars, accretion flows and relativistic
jets.

It is not surprising therefore that the study of numerical methods for
relativistic hydrodynamics in astrophysical contexts was an active branch
of research already in the late '60s, starting with the seminal works by
May and White~\cite{May66} and by Wilson~\cite{wilson_1972_nsf} and with
the availability of the first massive computational facilities.  In these
investigations, the approach was to cast the relativistic hydrodynamics
equations as non-linear advection equations in a form that resembles the
Newtonian Euler equations. These were then solved using finite-difference
(FD) schemes, stabilized using a combination of upwinding and
artificial-viscosity methods to avoid excessive oscillations at shocks
(see~\cite{Font08} for a comprehensive list of references). Although
these methods allowed to perform the first numerical studies in
general-relativistic hydrodynamics, they also had several limitations,
such as the difficulty of tuning the artificial viscosity to avoid
excessive smearing of the shock fronts or, most importantly, the
limitation to mildly relativistic flows, \ie with Lorentz factor $W
\lesssim 2$~\cite{Font08}.

A major leap forward in numerical relativistic hydrodynamics took place
when it was realized that the major problem behind Wilson's approach was
the use of a formulation which breaks the conservative nature of the
equations~\cite{Marti91}. This realization lead to a formulation of the
equations of relativistic hydrodynamics in a conservative form, the
so-called ``Valencia formulation''~\cite{Banyuls97}, and to the use of
finite-volume (FV) and FD high-resolution shock capturing (HRSC) methods
for their numerical solution (see, \eg~\cite{Leveque92, Toro99}). These
methods were shown to be able to handle ultra-relativistic flows and to
sharply resolve shocks without spurious oscillations or need for
artificial viscosity. For these reasons they have been the key ingredient
in a number of recent achievements of numerical relativistic
hydrodynamics and magnetohydrodynamics (MHD; see,
\eg~\cite{Giacomazzo:2010,Rezzolla:2011} and references therein).

Even tough FV and FD schemes have been particularly successful and are
indeed the standard choice for modern numerical codes in relativistic
hydrodynamics and MHD, they also suffer from
some limitations, such as the difficulty of handling complicated grid
structures and boundary conditions, or those associated with achieving
high orders of accuracy. These are mainly due to the fact that
high-order accuracy is generally attained with the use of large
reconstruction stencils and expensive non-linear limiting operators,
which become quickly cumbersome to handle when the grid is not
structured and/or quadratures are required in the computation of the
fluxes,~\ie for higher than third order FV schemes. Large
reconstruction stencils also come with large ghost regions when doing
parallel calculations, leading to poor scalability results. Finally
these schemes are also often overly dissipative in situations in which
shock waves are not the dominant part of the dynamics and may fail to
properly resolve fine structures of the
flow~\cite{johnsen_2010_ahr}. This has important consequences for the
accuracy of general-relativistic hydrodynamics
codes~\cite{Baiotti:2009gk}.

For these reasons, alternative approaches to general relativistic
hydrodynamics such as finite-element methods~\cite{meier_1999_mas}, or
spectral methods \cite{gourgoulhon_1991_seg, grandclement_2009_smn} are
worth consideration. This latter approach is particularly interesting
because spectral methods are able to attain very high accuracy, but it is
also limited by the well known fact that these methods fail spectacularly
when the solution develops large gradients or discontinuities. For this
reason, spectral methods for relativistic hydrodynamics have been limited
to the generation of initial data~\cite{Bonazzola97, Ansorg01b} or to
situations in which strong gradients could be treated with shock-tracking
techniques within a multi-domain framework~\cite{gourgoulhon_1991_seg}.

More recently, however, a novel method was suggested by Dumbser and
Zanotti~\cite{Dumbser2009}, who presented a hybrid FV/discontinuous
Galerkin (DG) approach for special relativistic resistive
magnetohydrodynamics (MHD). In this approach a local spacetime DG
method was used as an implicit predictor step in the context of a
high-order FV scheme, in order to treat the stiff source term of
resistive MHD.

Indeed, Spectral Discontinuous Galerkin Methods (SDGM) and their
variation employing Gaussian numerical integration (SDGM-NI) were
developed to overcome some of the above limitations of FV and spectral
or pseudospectral methods respectively~\cite{canuto_2008_sme}. These
methods work essentially by combining the classical Runge-Kutta
discontinuous Galerkin approach by Cockburn~\cite{cockburn_2001_rkd}
with the spectral element method (SEM) of
Patera~\cite{patera_1984_sem}. For this reason they are also often
referred to as DG-SEM or DG-SEM-NI. These methods are particularly
well suited for the solution of conservation laws and have been
successfully applied to a number of classical hyperbolic, parabolic
and elliptic problems (see, \eg~\cite{cockburn_2000_dg,
  canuto_2008_sme}). Finally they have been also successfully applied to
the solution of the Einstein equation in vacuum by
Zumbush~\cite{zumbusch_2009_fed} and Field et al.~\cite{field10}.

We develop here the necessary formalism for the application of fully
explicit DG methods to relativistic hydrodynamics on curved spacetimes.
As an application we present a prototype code employing SDGM-NI for
general relativistic hydrodynamics in spherical symmetry. We show that
the proposed scheme is able to properly resolve strong shocks and achieve
high-order, spectral accuracy for smooth solutions. While we will not
discuss explicitly the coupling of the solution of the hydrodynamics
equations with that of the Einstein equations, it is clear that a natural
choice would be to use discontinuous Galerkin methods, such as the ones
recently proposed by~\cite{zumbusch_2009_fed} or~\cite{field10}, or
finite-element-methods such as the ones introduced
by~\cite{Sopuerta:2005rd, Sopuerta:2005gz}, also for the metric evolution
equations. This approach would have the advantage, with respect to the
solution proposed with the ``Mariage des
Maillages''~\cite{Dimmelmeier05a,Duez:2008rb}, that the fluid and the
spacetime variables would share the same grid and no expensive
interpolations would therefore be needed.

The paper is organised as follows. In Sect.~\ref{sec:theory} we derive
the general theory for the application of discontinuous Galerkin
methods to relativistic hydrodynamics in curved spacetimes and we
specialize it to the spherically-symmetric case. In
Sect.~\ref{sec:edges} we present our prototype numerical code,
\texttt{EDGES} (Extensible Discontinuous GalErkin Spectral library),
which was used to test DG methods for general relativistic
hydrodynamics in one-dimension (1D) and spherical symmetry. The
results obtained on a representative number of test cases are then
presented in Sect.~\ref{sec:results}. Finally
Sect.~\ref{sec:conclusions} is dedicated to the summary and
conclusions.

We use a spacetime signature $(-,+,+,+)$, with Greek indices running
from 0 to 3 and the Latin indices from 1 to 3. We also employ the
standard convention for the summation over repeated indices. Finally,
all the quantities are expressed in a system of units in which
$c=G=M_\odot=1$, unless otherwise stated.

\section{Discontinuous Galerkin methods for general-relativistic
hydrodynamics}
\label{sec:theory} 

Broadly speaking, Galerkin methods are projection methods for the weak
formulation of the equations. In the case of the general-relativistic
hydrodynamics equations, such formulation could be obtained in two
different ways. The first one consists in starting from the
relativistic hydrodynamics equations written in a conservative form in
a chosen coordinate system, \eg~the Valencia
formulation~\cite{Banyuls97}, and then integrating them against a test
function. The numerical scheme obtained with the Galerkin projection
would then be a direct generalization of the standard HRSC schemes
used in general-relativistic hydrodynamics. Indeed, when considered at
first-order only, DG schemes reduce to FV ones and it is for this
reason that they are often interpreted as an alternative way to attain
high-order FV methods

While this approach is certainly possible and would seem to be quite
natural, it has the limitation that since we start from the equations in
their coordinate form, we also have to choose a metric with respect to
which the volume integrals are performed. The choice of such a metric is
effectively arbitrary, but any choice different from that of a flat
metric corresponds to the absorption of a multiplicative factor into the
definition of the test function and thus it is equivalent to a
modification of the scheme in the higher-than-first-order case. As a
result, the choice of the metric is unimportant only in the FV limit of
Galerkin methods, but it plays a central role in higher-order Galerkin
schemes.

A second way to obtain the weak formulation of the equations and the
one actually outlined in this paper, is to follow an approach which is
instead manifestly covariant and thus does not require any assumed
background. After the formulation is obtained, it can then be
decomposed in the standard $3+1$ split of general relativity. This
choice has the advantage of producing the most natural extension of
the commonly used HRSC frameworks to the DG case. The resulting
schemes will be naturally covariant, suited for the use with standard
spacelike or null foliations or even independently of any foliation or
coordinate system. The reason why this is possible lays, as pointed
out by Meier~\cite{meier_1999_mas}, in the covariant nature of
finite-element methods and, by extension, of DG methods. In these
methods, in fact, the equations are formulated on reference elements
mapped into the physical space via diffeomorphisms, thus removing any
need for a (preferred) coordinate system. The important difference
between our approach and the one by Meier~\cite{meier_1999_mas} is in
the use of non-conforming, discontinuous, Galerkin methods. This gives
us the possibility of reducing the coupling of the numerical solution
across the elements to flux terms, thus enabling the construction of
globally explicit, local schemes, in contrast to the need of solving
implicit, global, nonlinear problems.

\subsection{Weak formulation of the equations of relativistic
hydrodynamics}

Let $(\mathcal{M}, g_{\alpha \beta})$ be a strongly hyperbolic, $C^2$,
spacetime with metric $g_{\alpha \beta}$ and let $\nabla$ be the
covariant derivative associated with $g_{\alpha \beta}$. We consider a
perfect fluid described by a rest-mass-density 4-vector $J^{\alpha}$
and a stress energy tensor $T^{\alpha \beta}$ defined by
\begin{equation}\label{eq:perfect.fluid}
  J^{\alpha} \equiv \rho u^{\alpha}\,, \qquad T^{\alpha \beta} \equiv
  \rho h u^{\alpha} u^{\beta} + p g^{\alpha \beta}\,,
\end{equation}
where $\rho$ is the rest-mass density, $u^{\alpha}$ is the fluid
4-velocity, $p$ is the pressure, $\varepsilon$ is the specific
internal energy and $h \equiv 1 + \varepsilon + p / \rho$ is the
specific enthalpy.

If we assume baryon-number conservation and a generic equation of
state (EOS) of the form $p = p(\rho,\varepsilon)$, then the equations
of motion for the fluid on $\mathcal{M}$ read
\begin{equation}\label{eq:euler}
\nabla_{\alpha} J^{\alpha} = 0\,, \qquad \nabla_{\beta} 
T^{\alpha \beta} = 0\,, \qquad p =
p(\rho,\varepsilon)\,.
\end{equation}
In general these equations are to be intended in the sense of
distributions\footnote{Here and hereafter we will denominate as
  distributions generalized functions, such as Dirac delta.}, since we
expect the solution to develop singularities in the form of shock
waves.

In general we are interested in solving (\ref{eq:euler}) on an open,
regular\footnote{See~\cite{chen_2009_ggt} for a detailed discussion of
the regularity requirements. Broadly speaking this amounts to having a
domain which has a normal defined everywhere except for at most a
discrete set of points (vertexes); \ie a cubic box is a regular domain.},
finite domain $\Omega \subset \mathcal{M}$, with suitable
initial/boundary data.  A precise mathematical formulation of this
problem can be done within the context of bounded divergence-measure
vector fields using the theory developed in~\cite{chen_2003_edm,
chen_2009_ggt}. In particular, we will look for solutions in the
functional space, $\mathcal{V}$, of all the $L^\infty$, \ie ``bounded'',
vector fields over $\Omega$, whose divergence, in the sense of the
distributions, are Radon measures\footnote{For a precise definition
see~\cite{chen_2003_edm}. Broadly speaking this condition means that we
restrict ourselves to cases in which the solution presents at most mild
singularities, such as jump-discontinuities.}

As a first step we introduce a triangulation of $N$ \emph{``elements''}
of $\Omega$, $\{\Omega_j\}_{j=1}^N$, by selecting a family of
diffeomorphisms $\varphi_j\colon K \subset \mathbb{R}^4 \rightarrow
\Omega$, $\Omega_j = \varphi_j(K)$ such that
\begin{equation}
  \bigcup_{j=1}^N \Omega_j = \Omega\,, \qquad \mathring{\Omega}_i \cap
  \mathring{\Omega}_j = \emptyset\,, \ \ \forall\ \  i \neq j\,,
\end{equation}
where $K$ is the, so-called, \emph{``reference element''}, usually an
hypercube or a 4D simplex and $\mathring{\Omega}_j$ denotes the interior
of $\Omega_j$. We also arrange the local coordinate system,
$\{x^\mu\}_j$, induced by $\varphi_j$, so that $\partial_0$ is timelike
or null.

If we now look for solutions $J^{\alpha} \in \mathcal{V}$, the first
of the equations (\ref{eq:euler}) is equivalent, in the sense of
distributions, to
\begin{equation}\label{eq:continuity}
\sum_{j=1}^N \left[ \int_{\Omega_j} J^{\alpha}\, \nabla_{\alpha} \phi\,
\pmb{\epsilon} -
\int_{\partial\Omega_j} \phi\, \mathcal{J}^{\alpha}\,
\epsilon_{\alpha\beta\gamma\delta} \dd x^\beta \dd x^\gamma \dd
x^\delta \right] = 0\,,
\end{equation}
for all $\phi \in C^1_0(\mathcal{M})$\footnote{We recall that a
  function of class $C^n_0(\Omega)$ is a function of class
  $C^n(\Omega)$ and, in addition, with compact support in
  $\Omega$.}. Note that in expression~\eqref{eq:continuity} the symbol
$\pmb{\epsilon}$ refers to the proper volume form of the spacetime,
\ie in any local chart, $\{x^\mu\}$, \hbox{$\pmb{\epsilon} =
  \sqrt{-g}\, \pmb{e}^{0} \wedge \pmb{e}^{1} \wedge \pmb{e}^{2} \wedge
  \pmb{e}^{3}=\epsilon_{\alpha\beta\gamma\delta} \dd x^{\alpha} \dd
  x^\beta \dd x^\gamma \dd x^\delta$}, and $\mathcal{J}^{\alpha}$ is
the internal normal trace of $J^{\alpha}$.  This object reduces simply
to $J^{\alpha}$, when $J^{\alpha}$ and $\Omega_k$ are regular, but in
the general case the second integral has to be intended as the action
of a measure, $\mathcal{J}^{\alpha}
\epsilon_{\alpha\beta\gamma\delta}\dd x^\beta \dd x^\gamma \dd
x^\delta$, on $\phi$~\cite{chen_2009_ggt}.

In the same way, if we look for solutions $T^{\alpha \beta} \in
\mathcal{V}\otimes\mathcal{V}$, the second of the equations
(\ref{eq:euler}) is equivalent, in the sense of distributions, to
\begin{equation}\label{eq:momentum}
\begin{split}
  \sum_{j=1}^N \int_{\Omega_j} T^{\alpha \beta}\,& \nabla_{\beta}
  \phi_{\alpha}\, \pmb{\epsilon} = \\
  &\sum_{j=1}^N \int_{\partial\Omega_j} \phi_\alpha\, \mathcal{T}^{\alpha
  \beta}\, \epsilon_{\beta\gamma\delta\mu} \dd x^\gamma \dd x^\delta \dd
  x^\mu
\end{split}
\end{equation}
for all the one-forms $\phi_{\alpha} \in C^1_0\big(\mathcal{M};
T^\ast\!\mathcal{M}\big)$, $T^\ast\!\mathcal{M}$ being the co-tangent
bundle of $\mathcal{M}$. Again, $\mathcal{T}^{\alpha \beta}$ is a
generalization of $T^{\alpha \beta}$ and the integral has to be
interpreted as the action of $\mathcal{T}^{\alpha \beta}
\epsilon_{\beta\gamma\delta\mu}\dd x^\gamma \dd x^\delta \dd x^\mu$ on
$\phi_{\alpha}$ in the non-smooth case.

The solution of the relativistic hydrodynamics equations consists then in
finding 
\begin{subequations}
\label{eq:weak.euler}
\begin{align}
&J^\alpha \in \mathcal{V} & \st &
(\ref{eq:continuity}) \textrm{~holds~} & & \forall \phi \in
C^1_0(\mathcal{M})\,, \\
&T^{\alpha\beta} \in \mathcal{V}\! \otimes\! \mathcal{V}
& \st &
(\ref{eq:momentum}) \textrm{~holds~} & & \forall \phi_\alpha \in
C^1_0\big(\mathcal{M}; T^\ast\!\mathcal{M}\big),
\end{align}
\end{subequations}
together with an EOS and proper boundary-initial data, to be specified
through $\mathcal{J}^{\alpha}$ and $\mathcal{T}^{\alpha \beta}$ on
$\partial\Omega$. We remark, again, that (\ref{eq:weak.euler}) is
perfectly equivalent, in the sense of distributions, to
(\ref{eq:euler}) and that the triangulation $\{\Omega_j\}_{j=1}^N$
has been introduced mainly for later convenience.

\subsection{Spacetime discontinuous Galerkin formulation}

As mentioned above, within the Galerkin approach a numerical scheme is
obtained by projecting (\ref{eq:weak.euler}) on a finite dimensional
subspaces $V\subset\mathcal{V}$. In general this space is constructed
starting from the space of piecewise polynomials, in particular we
define
\begin{equation}
  X = \big\{ u \in L^\infty(\Omega) \colon u \circ \varphi_j \in
  \mathbb{P}_D(K), \ j = 1,\ldots, N \big\}\,,
\end{equation}
where $\mathbb{P}_D(K)$ is the space of polynomials with at most
degree $D$ on $K$. Notice that the functions in $X$ are allowed to be
discontinuous at the edges of the elements, hence the name
``discontinuous Galerkin'' for the resulting numerical scheme. The
space $V$ is taken as the space of all the vector fields ``whose
components are elements of $X$'', more precisely
\begin{equation}
  V = \big\{ u^{\alpha} \in \mathcal{V} \colon [\varphi_j]_\ast u^{\alpha} \in
  \big[\mathbb{P}_D(K)\big]^4, \ j = 1,\ldots, N \big\}\,,
\end{equation}
where $\left[\mathbb{P}_D(K)\right]^4$ is the space of 4-tuples of
polynomials with at most degree $D$ on $K$ and $[\varphi_j]_\ast$ is
the pull-back associated with $\varphi_j$. The Galerkin method is then
simply the restriction of (\ref{eq:weak.euler}) to $V$ so that it
consists in finding $J^\alpha \in V$ and $T^{\alpha\beta}\in V\otimes
V$ such that (\ref{eq:continuity}) and (\ref{eq:momentum}) hold for
suitable choices of the test functions, $\phi$ and $\phi_\alpha$, that
we are going to discuss in the following. The resulting equations can
be solved numerically, because a finite number of conditions suffice
to fully determine $J^{\alpha}$ and $T^{\alpha \beta}$ as long as we
have a way to evaluate the fluxes $\mathcal{J}^{\alpha}$ and
$\mathcal{T}^{\alpha \beta}$ on $\partial\Omega_j$.

As discussed above, in the continuous case these fluxes are simply the
restriction of $J^{\alpha}$ and $T^{\alpha \beta}$ to $\partial\Omega_j$.
More explicitly if we write symbolically $\mathcal{P} = \{ \rho, u^1,
u^2, u^3, \epsilon \}$ for the primitives variables and consider
$J^{\alpha}$ and $T^{\alpha \beta}$ as functions of $\mathcal{P}$, \ie
$J^{\alpha} = J^{\alpha}(\mathcal{P})$ and $T^{\alpha \beta} = T^{\alpha
\beta}(\mathcal{P})$, then $\mathcal{J}^{\alpha} =
J^{\alpha}\big(\mathcal{P}^\ast\big)$ and $\mathcal{T}^{\alpha \beta} =
T^{\alpha \beta}\big(\mathcal{P}^\ast\big)$, $\mathcal{P}^\ast$ being the
restriction of $\mathcal{P}$ on $\partial \Omega_j$, as the fluxes can
only depend on the location in the spacetime through $\mathcal{P}$.
Stated differently, $\mathcal{J}^{\alpha}$ and $\mathcal{T}^{\alpha
\beta}$ are simply the Godunov fluxes for the conservation law. In the
general case $\mathcal{J}^{\alpha}$ and $\mathcal{T}^{\alpha \beta}$ can
be determined with causality considerations on spacelike
boundaries\footnote{Where the solution has different limits,
$\mathcal{P}_1$ and $\mathcal{P}_2$ from different sides of
$\Gamma\subset\partial\Omega_j$ and $\Gamma$ is spacelike, we proceed as
in the Godunov method and we set $\mathcal{J}^{\alpha} = J^{\alpha}
\big[\mathcal{P}^\ast (\mathcal{P}_1, \mathcal{P}_2)\big]$.  Causality
requires that $\mathcal{P}^\ast$ must depend only on the past limit of
$\mathcal{P}$ at $\Gamma$, say $\mathcal{P}_1$, this implies
$\mathcal{P}^\ast = \mathcal{P}_1$, thus $\mathcal{J}^{\alpha} =
[J^{\alpha}]_1$. The same argument can be applied to evaluate
$\mathcal{T}^{\alpha \beta}$.}, or as solutions of generalized Riemann
problems, on timelike and null-like boundaries, as they are known as soon
as $\mathcal{P}^\ast$ is known on those boundaries. This is basically
analogous to the Newtonian case when using spacetime DG methods to
discretize balance laws. For this reason we refer
to~\cite{palaniappan_2004_sdg} for a more in-depth discussion. We limit
ourselves to note that, in the context of a numerical scheme, the
computation of the fluxes can be greatly simplified with the use of
approximate Riemann solvers, such as the HLLE scheme. In that case
$\mathcal{J}^\alpha$ and $\mathcal{T}^{\alpha\beta}$ are approximated
directly without the need to explicitly compute the solution of the
Riemann problem, $\mathcal{P}^\ast$, at the interface. This is discussed
in some more detail, for the 3+1 case, in the Appendix \ref{sec:example}.

Once we have a way to compute the fluxes, the fully discrete equations
are readily obtained by testing (\ref{eq:continuity}) and
(\ref{eq:momentum}) on a set of linearly-independent test functions,
$\phi$ and $\phi_{\alpha}$. The resulting finite set of equations can
be cast in a set of nonlinear equations for the spectral coefficients
of the numerical solutions, when these are expanded over a linear
basis of $V$, or $V\otimes V$, for $J^{\alpha}$ and $T^{\alpha \beta}$
respectively, and by following a standard finite-element method
procedure, see \eg~\cite{quarteroni_1997_nap}. In particular, it is
sufficient to consider test functions $\phi\in X$ and $\phi^{\alpha}
\in V$, so that the final numerical scheme consists in finding
\begin{subequations}
\label{eq:spacetime.dg}
\begin{align}
&J^\alpha \in V & \st &
(\ref{eq:continuity}) \textrm{~holds~} && \forall \phi \in X\,,\\
&T^{\alpha\beta} \in V\otimes V & \st &
(\ref{eq:momentum}) \textrm{~holds~} && \forall \phi^\alpha \in V\,,
\end{align}
\end{subequations}
where the test functions on $\partial\Omega_j$ in the boundary
integrals appearing in~\eqref{eq:spacetime.dg} has to be interpreted
as being a $C^1_0$ extension to $\mathcal{M}$ of the original test
function, created in such a way as to smoothly match the one-sided
limit, from the interior of $\Omega_j$, of the original test function.

To explicitly write down the method, in every finite element, $\Omega_j$,
we select a set of conserved quantities, $\mathcal{C} = \{J^0, T^{0\mu}
\}$\footnote{Other choices are possible, for example in the context of a
$3+1$ split we could use the same conserved quantities as the ones used
in the Valencia formulation.}, for which there exists a one-to-one
relation, involving the EOS, with the set of primitive
variables~\cite{Papadopoulos-Font-1999}, $\mathcal{P}$, so that we can
formally write $J^i = J^i(\mathcal{C})$ and $T^{i\mu} =
T^{i\mu}(\mathcal{C})$.  We can then obtain a set of nonlinear equations
for $\mathcal{C} \in X^5$ simply expanding the Galerkin conditions
$(\ref{eq:continuity})$ and $(\ref{eq:momentum})$
\begin{equation}\label{eq:spacetime.dg.continuity}
\begin{split}
  \sum_{j=1}^N \Bigg[ \int_{\Omega_j} & J^0\, \partial_0 \phi\,
  \pmb{\epsilon} + \int_{\Omega_j} J^j(\mathcal{C})\, \partial_j \phi
  \, \pmb{\epsilon} \Bigg] = \\
  &\sum_{j=1}^N\int_{\partial\Omega_j} \mathcal{J}^\mu\, \phi\,
  \epsilon_{\mu\alpha\beta\gamma}\dd x^\alpha \dd x^\beta \dd x^\gamma
\end{split}
\end{equation}
and, setting $\phi_\alpha = \phi \delta^{\mu}_{\phantom{\mu}\alpha}$,
\begin{equation}\label{eq:spacetime.dg.momentum}
\begin{split}
\sum_{j=1}^N\Bigg[ \int_{\Omega_j}& T^{0\mu}\, \partial_0 \phi\,
  \pmb{\epsilon} + \int_{\Omega_j} T^{i\mu}(\mathcal{C})\,
  \partial_i \phi\, \pmb{\epsilon} \Bigg]=\\
  & \sum_{j=1}^N\Bigg[\int_{\partial\Omega_j} \mathcal{T}^{\nu\mu}\,
  \phi\,\epsilon_{\nu\alpha\beta\gamma}\dd x^\alpha \dd x^\beta \dd
  x^\gamma + \\
  &\ \ \qquad\qquad\qquad \int_{\Omega_j} T^{\nu\lambda}(\mathcal{C})\,
    \Gamma^{\mu}_{\phantom{\mu}\lambda\nu}\, \phi\, \pmb{\epsilon}\Bigg]\,,
\end{split}
\end{equation}
where $\Gamma^{\alpha}_{\phantom{\alpha}\beta\gamma}$ are the Christoffel
symbols and $\phi\in X$.

The key point here is that, as the functions are discontinuous across
the $\partial\Omega_j$'s, these equations are local equations for the
spectral coefficients within the $\Omega_j$'s coupled only through the
fluxes. In particular this implies that, if the computational grid
``follows the causal structure of the spacetime'', in the sense that
it can be traversed with a succession of ``causal slices'' satisfying
some sort of generalized Courant-Friedrichs-Lewy (CFL) condition
ensuring the causal disconnection between the timelike boundaries of
the elements, then the discontinuous Galerkin method becomes globally
explicit.  Under these conditions, in fact, the fluxes between the
elements of the grid slice depends only on the data on the previous
slice. Once these are computed, we are left with a set of formally
decoupled equations involving the spectral coefficients of the
numerical solution in the different elements. The precise mathematical
definitions of ``causal slices'' and of ``grid that follows the causal
structure of the spacetime'' are given in
Appendix~\ref{sec:caus_triang}.

\subsection{Discontinuous Galerkin formulation in the $3+1$ split}

Clearly the strategy outlined above could be useful in situations in
which stiff sources are present, such as in the resistive MHD case, where
implicit DG methods have already been shown to be suitable as predictors
to treat stiff sources~\cite{Dumbser2009}, but in the unmagnetized case
an implicit time stepping is unnecessarily expensive in most situations.
The generation of a triangulation which follows the causal structure of
the spacetime could also be highly non-trivial, especially if the
spacetime is evolved dynamically. For these reasons, instead of directly
solving (\ref{eq:spacetime.dg.continuity}) and
(\ref{eq:spacetime.dg.momentum}), one can use them in order to derive a
fully explicit scheme. This can be accomplished by performing a $3+1$
split directly at the level of the discrete problem
(\ref{eq:spacetime.dg}).

As customary, we foliate the spacetime along $t = \textrm{const.}$
hypersurfaces, $\Sigma_t$, and consider a vector $t^{\alpha}$ such
that $t^{\alpha}\, \nabla_{\alpha} t = 1$.  Using this vector we
define the three-volume form $\eta_{\alpha\beta\gamma} =
\epsilon_{\delta\alpha\beta\gamma} t^\delta$.  Also, as usually done
in this context, we can use the integral lines of $t^{\alpha}$ to
identify points on $\Sigma_t$ with points on $\Sigma \equiv \Sigma_0$
and interpret the variation of the fields across the $\Sigma_t$'s as
being the result of a dynamics on a three-manifold, $\Sigma$. We are
then interested in studying (\ref{eq:euler}) in a world-tube $S \times
(0,t)$, $S \subset \Sigma$ being an open, bounded, regular domain in
$\Sigma$, together with proper boundary-initial conditions. Clearly
this is a particular case of the general problem studied above.

In order to apply the discontinuous Galerkin formulation, we consider a
triangulation $\{S_j\}_{j=1}^N$ of $S$, by selecting a family
of diffeomorphisms $\Phi_j\colon T\subset
\mathbb{R}^3 \rightarrow \Sigma$, $S_j = \Phi_j(T)$ such that
\begin{equation}
  \bigcup_{j=1}^N S_j = S, \qquad \mathring{S}_i \cap \mathring{S}_j =
  \emptyset, \quad \forall\ \  i \neq j\,,
\end{equation}
where $T$ is now a three-dimensional reference element, a cube or a
tetrahedron. This induces a triangulation of $\Omega$
\begin{equation}
\{\Omega_{j,n}\}_{j=1, n=1}^{N,Q} = \Big\{S_j \times \big(n \Delta t, (n+1)
\Delta t\big)\Big\}_{j=1, n=1}^{N,Q}\,,
\end{equation}
that follows the causal structure of the spacetime on the $n$-th
thin-sandwich $\Omega_n = S \times (t_n,t_n+\Delta t)$, at least for
small $\Delta t$.

As we intend to use the method of lines in order to integrate the
equations in time, we can factor out the time dependence from the test
functions by considering the functional space
\begin{equation}
Y = \big\{ u \in L^\infty(S) \colon u \circ \Phi_j \in
  \mathbb{P}_D(T), \ j = 1,\ldots, N \big\}\,,
\end{equation}
and vector functions ``whose components are elements of $Y$'':
\begin{equation}
W = \big\{ u^{\alpha} \in \mathcal{V} \colon [\Phi_j]_\ast u^{\alpha} \in
  \big[\mathbb{P}_D(T)\big]^4, \ j = 1,\ldots, N \big\}\,.
\end{equation}
Given a function $u \in Y$, we can consider it as a function,
$\tilde{u}$, over $S\times (0,t)$ with the identification
$\tilde{u}(x^i,s) \equiv u(x^i)$ so that $Y \hookrightarrow \tilde{Y}
\subset X$ and, with a slight abuse of the notation, we can consider
$Y$ as being a subspace of $X$. In a similar way we can consider $W$
to be a subspace of $V$. For these reasons we can choose $\phi \in Y$
in (\ref{eq:spacetime.dg.continuity}) and
(\ref{eq:spacetime.dg.momentum}). We can thus obtain a fully explicit
method by re-projecting (\ref{eq:spacetime.dg}) onto a new couple of
subspaces or, equivalently, by finding
\begin{subequations}
\label{eq:3p1.base.implicit.dg}
\begin{align}
&J^\alpha \in C^1\big((0,t); W\big) &\st (\ref{eq:continuity})
\textrm{~holds~} &\forall \phi \in Y\,,\\ 
&T^{\alpha\beta} \in C^1\big((0,t); W\!\otimes\! W\big) &\hspace{-5pt}  \st
(\ref{eq:momentum}) \textrm{~holds~} &\forall \phi^\alpha \in W.
\end{align}
\end{subequations}
%
In the expressions above, and as customary when dealing with evolution
problems, we have used the notation $u(\cdot) \in
C^k\big((a,b);X\big)$, where $a,b\in\mathbb{R}$ and $X$ is a Banach
space, to indicate that, when $u$ is a regarded as a function only of
time, it describes a regular, $C^k$, curve in $X$. In other words, when
$u(t,x_1^i)$ is interpreted as a function of time only, it is of class
$C^k$, while when $u(t_1,x^i)$ is interpreted as a function of $x^i$
only, it is an element of the function space $X$. Note also that in
this new formulation we do not allow the numerical solution to be
discontinuous in time, so that the time integration can be performed
with a standard solver for ordinary differential equations (ODEs).

We can derive a more explicit form for the
(\ref{eq:3p1.base.implicit.dg}) by projecting
(\ref{eq:spacetime.dg.continuity}) with $\phi \in Y$, dividing both
terms by $\Delta t$ and by letting $\Delta t \to 0$, to obtain
\begin{equation}\label{eq:dg.continuity}
\begin{split}
  \sum_{j=1}^N \partial_t \int_{S_j} J^0\, \phi\, \pmb{\eta} =
  \sum_{j=1}^N \Bigg[ &\int_{S_j} J^i(\mathcal{C})\, \partial_i \phi\,
  \pmb{\eta} - \\ & \int_{\partial S_j} \mathcal{J}^i\, \phi\,
  \eta_{i\alpha\beta}\dd x^\alpha \dd x^\beta \Bigg]\,.
\end{split}
\end{equation}

Reasoning along the same lines, we can derive an explicit discretization
of the second of the (\ref{eq:euler}), starting from the
(\ref{eq:spacetime.dg.momentum}), to obtain
\begin{equation}\label{eq:dg.momentum}
\begin{split}
  &\sum_{j=1}^N \partial_t \int_{S_j} T^{0\mu}\, \phi\, \pmb{\eta} =
  \sum_{j=1}^N \Bigg[ \int_{S_j} T^{i\mu}(\mathcal{C})\, \partial_i
  \phi\, \pmb{\eta} - \\ &\quad \int_{\partial S_j} \mathcal{T}^{i\mu}\,
  \phi\, \eta_{i\alpha\beta}\dd x^\alpha \dd x^\beta - \int_{S_j}
  T^{\alpha\beta}(\mathcal{C})\, \Gamma^{\mu}_{\phantom{\mu}\beta\alpha}
  \, \phi\, \pmb{\eta} \Bigg]\,.
\end{split}
\end{equation}
Finally the $3+1$ discontinuous Galerkin formulation can be summarized
as in finding
\begin{subequations}
\label{eq:3p1.base.dg}
\begin{align}
&J^\alpha \in C^1\big((0,t); W\big) & \st (\ref{eq:dg.continuity})
\textrm{~holds~} &\forall \phi \in Y\,,\\
&T^{\alpha\beta} \in C^1\big((0,t); W\!\otimes\! W\big)\!\! & \st
(\ref{eq:dg.momentum}) \textrm{~holds~} &\forall \phi^\alpha \in W.
\end{align}
\end{subequations}

This scheme can be interpreted as a higher-order generalization of a
FV discretization of the manifestly covariant formulation of
relativistic hydrodynamics proposed
by~\cite{Papadopoulos-Font-1999}. As a consequence, our scheme
inherits properties such as hyperbolicity and the flexibility to work
with spacelike or null-like foliations directly from
\cite{Papadopoulos-Font-1999}. This can be seen considering the case
in which $D=0$, that is, when looking for solutions that are constant
over each element, $S_j$.  Then a sufficient number of Galerkin
conditions can be obtained by simply choosing $\phi = \chi_{S_j}$ for
$j = 1,2,\ldots, N$, where $\chi_E$ is the indicator function of the
set $E$, \ie a function which is equal to one in $E$ and identically
zero elsewhere. With this choice we obtain the set of equations
\begin{equation}
  \partial_t \int_{S_j} J^0\, \pmb{\eta} + \int_{\partial S_j}
  J^i(\mathcal{C})\, \eta_{i\alpha\beta}\dd x^\alpha \dd x^\beta = 0 \,,
\end{equation}
and
\begin{equation}
\begin{split}
  \partial_t \int_{S_j} T^{0\mu}\, \pmb{\eta} + \int_{\partial S_j}
  \mathcal{T}^{i\mu}\,& \eta_{i\alpha\beta}\dd x^\alpha \dd x^\beta = \\
  &- \int_{S_j} T^{\alpha\beta}(\mathcal{C})\,
  \Gamma^{\mu}_{\phantom{\mu}\beta\alpha} \, \pmb{\eta}\,,
\end{split}
\end{equation}
for all $j = 1,2,\ldots, N$. These can easily be recognised as being
the FV discretization of the formulation by~\cite{Papadopoulos-Font-1999}.

\subsection{Discontinuous Galerkin formulation in spherical symmetry}
\label{sec:spherical.symmetry}

As a particular case of the formalism outlined above we consider the case
in which the spacetime is spherically symmetric. This case is
particularly interesting because the equations become $1+1$ dimensional
and are therefore well-suited for rapid prototyping and testing of new
methods and techniques\footnote{Of course in $1+1$ dimensions a
Lagrangian approach such as the one presented in~\cite{Rezzolla1994,
musco05} is by far superior as it allows for natural spatial adaptivity
and conservation properties. However, both of these advantages disappear
in more than one spatial dimension. On the other hand Lagrangian
approaches to multidimensional relativistic hydrodynamics have been
recently proposed within the context of smoothed-particle-hydrodynamics
schemes in special~\cite{rosswog_2010_csr} and general
relativity~\cite{Siegler00}.}.

In particular we consider a spherically symmetric spacetime in a
radial-polar gauge
\begin{equation}
\label{eq:line.element}
  \dd s^2 = - \alpha^2 \dd t^2 + A^2 \dd r^2 + r^2 (\dd \theta^2 + \sin^2
  \theta \dd \phi^2)\,,
\end{equation}
where $\alpha$ and $A$ are functions of $t$ and $r$ only.  We next
introduce the Bondi mass function, $m$, and the metric potential,
$\nu$, by
\begin{equation}
  A(t,r) = \bigg(1-\frac{2m(t,r)}{r}\bigg)^{-1/2}\,,\quad \alpha(t,r) =
  e^{\nu(t,r)}\,.
\end{equation}
Following~\cite{Romero96}, we define the physical velocity, $v$, by $v
\equiv A u^r / \alpha u^t$, where $W = \alpha u^t = (1-v^2)^{-1/2}$ is
the Lorentz factor. Furthermore we introduce the ``conserved''
quantities
\begin{subequations}
\label{eq:conserved.variables}
\begin{align}
   \mathcal{D} &\equiv \alpha A J^t = \rho A W\,, \\
   S &\equiv \alpha T^{tr} = \rho h W^2 v\,, \\
   E &\equiv \alpha^2 T^{tt} = \rho h W^2 - p\,, \\
\tau &\equiv E - {\cal D}\,.
\end{align}
\end{subequations}

With these definitions, the Einstein equations reduce to the
Hamiltonian constraint
\begin{equation}\label{eq:hamiltonian.constraint}
  \partial_r m = 4 \pi r^2 E\,,
\end{equation}
and the slicing condition $\partial_t K_{\theta\theta} =
K_{\theta\theta} = 0$
\begin{equation}\label{eq:slicing.condition}
  \partial_r \nu = A^2 \bigg[ \frac{m}{r^2} + 4 \pi r ( p + S v) \bigg]\,,
\end{equation}
where $K_{ij}$ is the extrinsic curvature.

These equations have to be integrated with the boundary conditions
given by $m(0) = 0$ and by the requirement that $\nu$ matches the
Schwarzschild solution at the outer boundary of the computational
domain (see \eg~\cite{noble_2003_nsr} for a detailed derivation and
discussion of this equation). 

The equations for the hydrodynamics are simply
(\ref{eq:dg.continuity}) and (\ref{eq:dg.momentum}), where the
elements, $S_j$, are taken to be the spherical shells $r_j < r <
r_{j+1}$. These can be written in terms of the conserved quantities
(\ref{eq:conserved.variables}) and specialized for the metric
(\ref{eq:line.element}) by substituting the explicit expression for
the Christoffel symbols and the determinant of the metric, to obtain
\begin{widetext}
\begin{equation}\label{eq:dg.spherical.symmetry}
  \sum_{j=1}^N \int_{r_j}^{r_{j+1}} \partial_t \pmb{F}^t(\mathcal{C})\,
  \phi\, r^2\, \dd r = \sum_{j=1}^N \Bigg\{ \int_{r_j}^{r_{j+1}} X\,
  \pmb{F}^r(\mathcal{C})\, \partial_r \phi\, r^2 \dd r - \Big[ r^2 \, X\,
  \pmb{\mathcal{F}}^r\, \phi \Big]_{r_j}^{r_{j+1}} + \int_{r_j}^{r_{j+1}}
  \pmb{s}(\mathcal{C})\, \phi\, r^2\, \dd r\, \Bigg\}\,,
\end{equation}
\end{widetext}
where we define $X \equiv \alpha / A$. Here, the fluxes are given by
\begin{equation}
\pmb{F}^t = \left\{{\cal D},\ S,\ \tau\right\}\,,\quad 
\pmb{F}^r = \left\{{\cal D}v,\ Sv+p,\ S-{\cal D}v\right\}\,,
\end{equation}
while the source term is
\begin{equation}
\begin{split}
\pmb{s} = \biggl\{0,\ (Sv-\tau-{\cal D})\Big(8 \alpha A &\pi r p + \alpha A
\frac{m}{r^2}\Big)\\ &+ \alpha A p \frac{m}{r^2} + 2 \frac{\alpha p}{A
  r},\ 0 \biggr\}\,.
\end{split}
\end{equation}
In the derivation of (\ref{eq:dg.spherical.symmetry}) the momentum
constraint was used to substitute the derivatives of the metric in the
source term and a factor $X$ was absorbed into the test function in the
derivation of the equation for $\tau$.

A close examination of the (\ref{eq:dg.spherical.symmetry}) reveals
that, again, this formulation of the equations can be interpreted as
an higher-order generalization of a classical FV discretization of the
equations of relativistic hydrodynamics. In particular it can be seen
that, in the case in which ${\cal D}\,,S$ and $\tau$ are constant over
each element, (\ref{eq:dg.spherical.symmetry}) reduces to the FV
method discussed in~\cite{Romero96}.

\section{The \texttt{EDGES} code}
\label{sec:edges}

In order to test discontinuous Galerkin methods for relativistic
hydrodynamics and their reduction to spherical symmetry, we have
developed a new 1D code, \texttt{EDGES}. This consists of a general
discontinuous Galerkin library which is then used by a code solving
the general-relativistic hydrodynamics equations in spherical
symmetry.  \texttt{EDGES} makes extensive use of advanced generic
programming techniques such as static polymorphism via recursive
templates, and expression templates see \eg~\cite{yang_2000_oon}. The
code employs the \texttt{Blitz++} high-performance array
library~\cite{Veldhuizen98, blitz} and makes use of the
\texttt{UMFPACK} multi-frontal sparse factorization method
\cite{Davis-2002a-UMFPACK-report, Davis-2002b-UMFPACK-report,
  Davis-Duff-1997-UMFPACK, Davis-Duff-1999-UMFPACK, umfpackweb} for
linear systems inversion. In what follows we describe in more detail
the implementation of the discontinuous Galerkin approach within
\texttt{EDGES}.

\subsection{The DG equations in a fully discrete form}

\begin{figure}
\begin{center}
  \includegraphics[width=8.0cm]{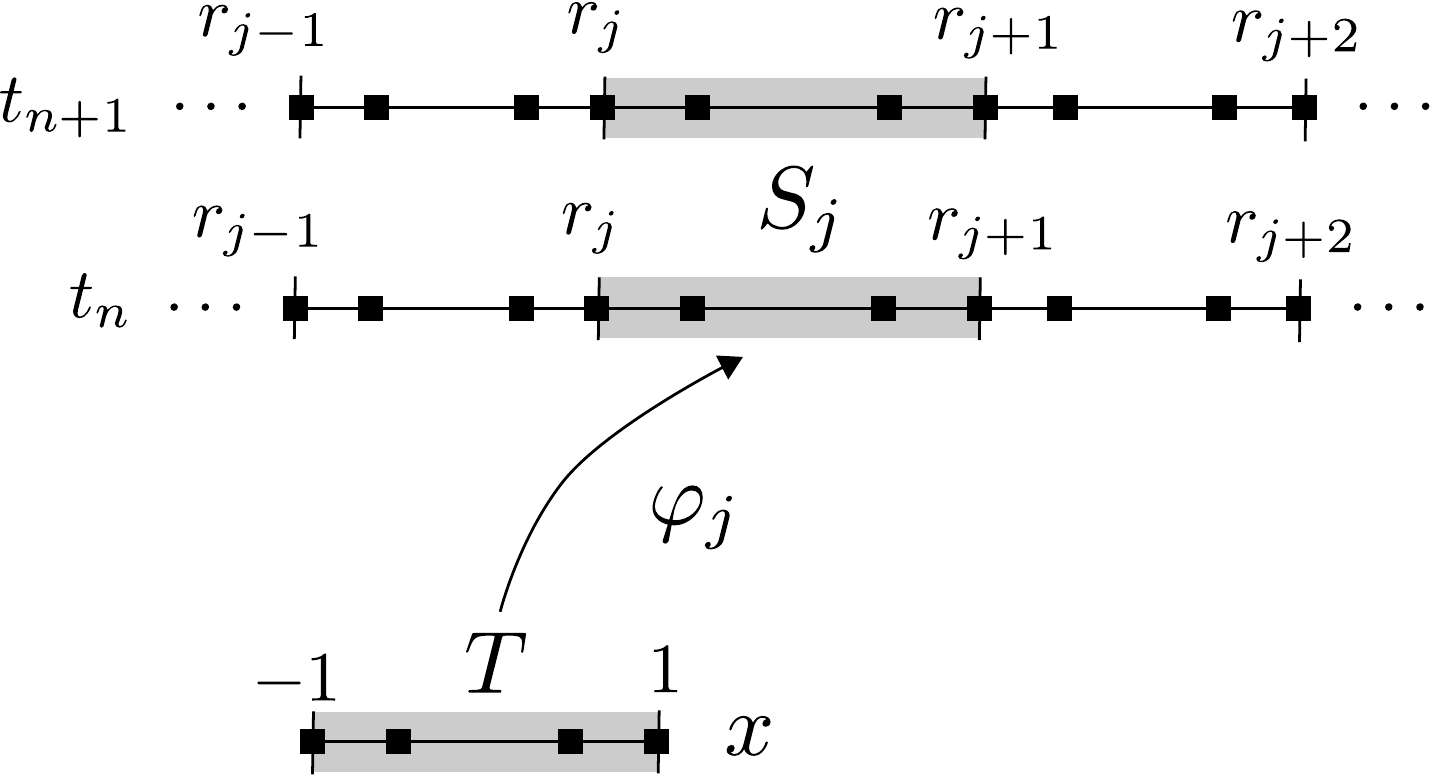}
  \caption{\label{fig:edges.grid} Scheme of the spacetime grid
    structure in \texttt{EDGES}. The collocation nodes (filled
    squares) are generated and stored on the reference element, $T$,
    and then mapped onto the finite element $S_j$ with an affine
    transformation, $\varphi_j$. Note that each point on the boundary
    of an element is associated with two distinct degrees of freedom,
    thus allowing the functions to have two different one-sided
    limits. Because the number of collocation points needed is $D+1$,
    where $D$ is the order of the polynomial representation, the
    figure refers to a polynomial of order three.}
\end{center}
\end{figure}

We consider a spherical shell $S = [0,R]$ in the spacetime
(\ref{eq:line.element}), containing a fluid described by
(\ref{eq:perfect.fluid}) and (\ref{eq:euler}). Furthermore we consider
a ``triangulation'' of $S \equiv \bigcup_{j=1}^N S_j$ where (see
Fig.~\ref{fig:edges.grid} for a scheme of the triangulation)
\begin{equation}\label{eq:edges.grid}
  S_i \cap S_j = \emptyset\,,\quad \forall\ \  i \neq j\,,\qquad S_j =
  \varphi_j\big([-1,1]\big)\,.
\end{equation}
The functional space that we consider in \texttt{EDGES} is,
\begin{equation}
 Z = \big\{ u \in L^\infty(S)\colon u \circ \varphi_j \in
 \mathbb{P}_D[-1,1]\big\}\,,
\end{equation}
the set of all the functions that are polynomials of degree $D$ over each
$S_j$.

If we denote by $L_D(x)$ the $D$-th \textit{Legendre} polynomial on
$[-1,1]$, a Gaussian quadrature of order $2D-1$ can be obtained with
the formula
\begin{equation}
\label{eq:gaussian.quadrature}
  \int_{-1}^1 f(x) \, \dd x \approx \sum_{i=0}^D w_i \, f(x_i)\,,
\end{equation}
where $\{x_i\}_{i=0}^D$ are the zeros of $(1-x^2) {\dd L_D(x)}/{\dd
  x}$, $w_i$ are a set of weights given by
\begin{equation}
  w_i = \int_{-1}^1 l_i(x) \, \dd x\,, \quad i = 0,1,\ldots, D
\end{equation}
and $\{l_i(x)\}_{i=0}^D$ are the \textit{Lagrange} polynomials associated
with the nodes $\{x_i\}_{i=0}^D$, \ie a set of polynomials of degree $D$
such that $l_i(x_k) = \delta_{ik}$ for $i,k = 0,1, \ldots, D$.  Given two
regular functions $f$ and $g$ in $r\in (0,R)$ we now define their
\textit{continuous} scalar product as
\begin{equation}
\label{eq:scalar.product.continuous}
  (f,g) \equiv \int_{0}^{R} f(r)\, g(r)\, r^2 \dd r\,,
\end{equation}
and use the quadrature formula~\eqref{eq:gaussian.quadrature} to
introduce the \textit{discrete} scalar product
\begin{equation}
\label{eq:scalar.product}
(f,g)_D \equiv \sum_{j=1}^N (f,g)_{j,D}\,,
\end{equation}
where
\begin{equation}
  (f,g)_{j,D} \equiv \sum_{i=0}^D w_i\, | \varphi_j' |\,
  f\big[\varphi_j(x_i)\big]\, g\big[\varphi_j(x_i)\big]
\end{equation}
and $|\varphi_j'|$ is the Jacobian of the affine transformation
$\varphi_j\colon[-1,1] \rightarrow S_j$. 

With these definitions in place we can construct a fully discrete systems
by looking for solutions $\mathcal{D},S,\tau \in Z$ and computing the
integrals in (\ref{eq:dg.spherical.symmetry}) using the Gaussian
quadrature (\ref{eq:gaussian.quadrature}) over each element. In
particular, using the notation~\eqref{eq:scalar.product}, we obtain
\begin{equation}
\begin{split}
  \big(r^2 \partial_t \pmb{F}^t, \phi&\big)_D = \big( r^2 X \pmb{F}^r,
  \partial_r \phi\big)_D \\ &- \sum_{j=1}^N \Big[ r^2\, X\,
  \pmb{\mathcal{F}}^r \phi \Big]_{r_j}^{r_{j+1}} + \big( r^2 \pmb{s},
  \phi \big)_D\,.
\end{split}
\end{equation}
where $r_j$ is the left vertex of the $j$-th element in the discretization
of the radial coordinate in the line element~\eqref{eq:line.element}.

Considering now test functions with support contained within a given
element, $S_j$, say
\begin{equation}
  \phi(r) = l_i\big(\varphi_j^{-1}(r)\big) \, \chi_{[r_j, r_{j+1}]}(r)
\end{equation}
where $l_i\big(\varphi_j^{-1}(r)\big)$ is the Lagrange polynomial or
order $i$ evaluated at the position $x = \varphi_j^{-1}(r)$ of the
reference element which is mapped into $r$, while $\chi_{[r_j,
    r_{j+1}]}$ is the indicator function and thus equal to one in the
interval $[r_j, r_{j+1}]$ and zero elsewhere. Expanding $\pmb{F}^t$
over the Lagrange basis of $S_j$
\begin{equation}
  \big[\pmb{F}^t \circ \varphi_j\big](x) = \sum_{k=0}^D \pmb{F}^t_{jk}
  l_k(x)\,,
\end{equation}
we obtain a set of coupled ordinary differential equations for the
coefficients $\pmb{F}^t_{ji}$
\begin{equation}
\label{eq:dg.fully.discrete}
\begin{split}
  r_{ji}^2 \partial_t \pmb{F}^t_{ji} = \big( & r^2 X \pmb{F}^r,
  \partial_r l_i \big)_{j,D} \\ & - \frac{1}{w_i |\varphi_j'|} \Big[
  r^2\, X\, \pmb{\mathcal{F}}^r l_i \Big]_{r_j}^{r_{j+1}} + r_{ji}^2
  \pmb{s}_{ji}\,,
\end{split}
\end{equation}
where $r_{ji} = \varphi_j(x_i)$ and we used the fact that
$(l_i,l_k)_{j,D} = \delta_{ik}$. An explicit example of
(\ref{eq:dg.fully.discrete}) for the continuity equation is given in
the Appendix \ref{sec:example}. Hereafter we will consider only affine
maps $\varphi_j$, so that the transformation between the reference
element $T$ and the finite element $S$ is given by
\begin{equation}
\label{eq:affine_transf}
  \varphi_j(x) = \left(\frac{x+1}{2}\right) r_{j+1}  - 
  \left(\frac{x-1}{2}\right) r_{j} \,,
\end{equation}
and $|\varphi_j'|=|r_{j+1}-r_{j}|/2$. 

Figure~\ref{fig:edges.grid} offers a schematic representation of the
spacetime grid structure in \texttt{EDGES}. The $D+1$ collocation
nodes (\ie the four filled squares for a polynomial of degree three)
are generated and stored on the reference element, $T$, and then
mapped onto the finite element $S_j$ with an affine transformation,
$\varphi_j$.  Each point on the boundary of an element is associated
with two distinct degrees of freedom, so that the corresponding values
of the functions can be different there. 

This system of equations is coupled with the equations for the
evolution of the metric quantities, described in Sect.
\ref{sec:coupling.spacetime}, and closed with an EOS. In particular we
use an ideal-fluid (or $\Gamma$-law) EOS
\begin{equation}\label{eq:eos}
  p = (\Gamma-1)\, \rho\, \varepsilon\,.
\end{equation}

Expression (\ref{eq:dg.fully.discrete}) is clearly a collocation scheme
for $\mathcal{D}, S$ and $\tau$ on the grid illustrated in
Fig.~\ref{fig:edges.grid}, since we can interpret the expansion
coefficients as collocated values, \ie $f_{ji} = f(r_{ji})$ for any
function $f$. In our code we evaluate the fluxes, $\pmb{\mathcal{F}}^r$,
using the relativistic HLLE approximate Riemann solver and evolve
numerically the resulting set of equations using a second-order strongly
stability preserving (SSP) Runge-Kutta method \cite{gottlieb2009}.  We
note that, as expected, if $D=0$, then $l_j={\rm const.}$, the first term
on the right-hand side vanishes and we are left with a standard FV
scheme.

A final note concerns the CFL condition needed to ensure the linear
stability of the scheme. The linear stability of Legendre pseudospectral
methods for hyperbolic equations has been studied
in~\cite{gottlieb_1991_ccs}, where it has been shown that $L^2$-stability
can be obtained if $\Delta t \sim D^{-2}$. As a result, in our code, we
use a timestep given by
\begin{equation}
  \Delta t = \frac{C_{\text{CFL}}}{(D+1)^2}\, \Delta x\,,
\end{equation}
where $\Delta x$ is the size of the smallest element, the
$C_{\text{CFL}}$ is a coefficient that is reminiscent (but distinct
from) the traditional ``CFL factor'' and has to be determined
empirically.  This coefficient is usually taken to be
$C_{\text{CFL}}\sim 0.2-0.3$, but our numerical experience (at least
in 1D) seems to suggest that this condition is overly restrictive as
stable evolutions can be obtained with $C_{\text{CFL}}\sim 1$ in most
situations.

\subsection{Coupling with the spacetime}
\label{sec:coupling.spacetime}

As discussed in Sect.~\ref{sec:spherical.symmetry}, in spherical
symmetry and with the gauge chosen, the Einstein equations are simple
constraints on each spatial slice and thus ODEs in the form
\begin{equation}
  \partial_r u(t,r) = f(t,r)\,,
\end{equation}
which could be easily integrated to very high accuracy. However,
instead of using a standard Runge-Kutta method for their integration,
we found that a more efficient and accurate approach to solve
equations (\ref{eq:hamiltonian.constraint}) and
(\ref{eq:slicing.condition}) consists in using an implicit
discontinuous spectral ODE solver which makes use of the same grid as
the hydrodynamical variables. Such approach has the advantage of
obtaining a numerical solution with a degree of accuracy which is of
the same order or higher with respect to the one attained for the
hydrodynamical variables, without requiring a very small step, as
would have been the case for a Runge-Kutta method.

In order to implement this approach it is necessary to allow the
solution to be discontinuous across the elements, while imposing the
continuity using an interior penalty technique~\cite{arnold_1982_ipf,
  arnold_2002_uad}. In particular the discontinuous Galerkin
formulation that we use reads
\begin{equation}
\label{eq:dg.ode}
  \sum_{j=1}^N \Bigg[ \int_{r_j}^{r_{j+1}} \partial_r u\, \phi\, \dd r -
  \mu_j \llbracket u \rrbracket_j\, \phi_j \Bigg] = \sum_{j=1}^N
  \int_{r_j}^{r_{j+1}} f\, \phi\, \dd r\,,
\end{equation}
where $\phi_j \equiv \phi(r_j^+)$, $\llbracket u \rrbracket_j \equiv
u(t,r_j^-) - u(t,r_j^+)$ is the jump term, and $\mu_j \sim 1 / \Delta
r_j$ is the penalization coefficient. It is straightforward to see that
this penalization term, which we refer to as ``upwind penalization''
in contrast to the usual symmetric penalization term $\sum_{j=1}^N
\mu_j \llbracket u \rrbracket_j \llbracket \phi \rrbracket_j$, has the
effect of enforcing $u(t, r_j^+) = u(t, r_j^-)$ without affecting the
equation for the collocation value in $r_j^-$.

As customary in the context of DG-SEM-NI methods, we approximate
$(\ref{eq:dg.ode})$ as
\begin{equation}
  (\partial_r u, \phi)_D - \sum_{j=1}^N \mu_j \llbracket u \rrbracket_j\,
  \phi_j = (f, \phi)_D\,,
\end{equation}
so that if we arrange the values of $u(t,r)$ and $f(t,r)$ on the
collocation points in two arrays $\pmb{u}(t)$ and $\pmb{f}(t)$, the
previous can be written as
\begin{equation}
  \pmb{A}\, \pmb{u}(t) = \pmb{f}(t)\,,
\end{equation}
where $\pmb{A}$ is a large-sparse matrix. As $\pmb{A}$ does not depend
on $t$, in \texttt{EDGES} this matrix is pre-computed, stored and
pre-factorized using \texttt{UMFPACK} (see \eg~\cite{saad_96}). At
each step, then, we have simply to compute $\pmb{f}(t)$ and use the
factorized version of $\pmb{A}$ to efficiently compute $\pmb{u}(t) =
\pmb{A}^{-1} \pmb{f}(t)$.

\subsection{Limiters, spectral viscosity and spectral filtering}

It is well known that high-order numerical methods suffer from numerical
oscillations in the presence of discontinuities (Gibbs phenomenon). If
these oscillations are not suitably handled, they tend to grow out of
control and destabilize the method. To overcome this difficulty several
different ``flattening techniques'' have been developed in the context of
FD, FV and spectral methods to artificially lower the order of the
methods in the presence of shock waves. Some examples are artificial
viscosity, flux limiting, PPM or ENO/WENO reconstruction. Many of these
techniques are implemented in \texttt{EDGES} and can be activated during
the evolution.

\subsubsection{Complications of spherical symmetry}
\label{complicationsSS}

While discontinuous Galerkin methods can in general be used in
combination with the large majority of flattening techniques, this is
not the case in spherical symmetry. The reason for this is that if we
consider a function $u$ and interpret any of the flatting methods
above as the application of an operator $\mathcal{L}$ to $u$, it can be
shown that, for the above mentioned flattening techniques,
\begin{equation}\label{eq:filter.conservation.1d}
  \int_{\Omega_j} u\, \dd x = \int_{\Omega_j} \mathcal{L} (u) \, \dd x\,,
\end{equation}
which is the desired behaviour for a conservative scheme in
1D. However, in the case of spherical symmetry, the conservation
property that we should satisfy reads
\begin{equation}
\label{eq:filter.conservation}
  \int_{\Omega_j} u\, r^2\, \dd r = 
\int_{\Omega_j} \mathcal{L} (u)\, r^2\, \dd r\,.
\end{equation}
Unfortunately, the property~\eqref{eq:filter.conservation} does not hold
for almost all of the stabilization techniques which will be discussed in
Sect.~\ref{sec:stabilization} and which are generally coordinate
dependent (the only exception being given by the spectral viscosity
method which is inherently a differential operator). In the case of FV
codes this is not a problem because the reconstruction is only used in
the computation of the fluxes and the volume averages are evolved using
the correct cell volume, but in a scheme that works with the actual
point-wise value of the solution, this leads to unacceptable variations
of the volume integral of $u$ (\eg of the total ``mass'') in the elements
with $r\ll 1$ and $r\gg 1$, which in turn leads to the development of
large numerical errors and/or instabilities. For this reason it is
necessary to introduce a correction to the flattening procedure in the
spherically symmetric case.

A naive way to obtain the wanted result would be to modify the
operator $\mathcal{L}$ as 
\begin{equation}
  \tilde{\mathcal{L}}(u) = \frac{1}{r^2} \mathcal{L} (u r^2)\,,
\end{equation}
so that $\tilde{\mathcal{L}}$ now satisfies
(\ref{eq:filter.conservation}). In practice, however, this strategy
results, as can be easily foreseen, in large numerical errors near $r
= 0$ and is this of little use. For this reason we had to adopt a more
radical approach and add a ``correction step'' after the application
of $\mathcal{L}$ to enforce (\ref{eq:filter.conservation}). In
particular we implemented three different strategies which we discuss
below.

The first one, which we refer to as ``dummy'' correction, consists in
simply adding to $\mathcal{L}(u)$ a function $C$, constant on every
element, such that (\ref{eq:filter.conservation}) is satisfied. This
is a very simple approach and has the advantage of not increasing the
total variation of $\mathcal{L}(u)$ inside the single elements, as
conservation is basically obtained at the expense of the amplification
of the jumps of $\mathcal{L}(u)$ across the elements. In this way the
additional total variation generated by the corrective procedure is
concentrated over the grid points which constitute the ``finite volume
part'' of the method.

The second one, which we refer to as ``bubble'' correction, consists
in adding to $\mathcal{L}(u)$ a function $b$, which is a bubble
function over each element, \ie a function which is zero at the edges
of the element, such as
\begin{equation}
  \big[b \circ \varphi_j\big](x) = K_j (1-x^2)\,,
\end{equation}
where the $K_j$'s are chosen in each element so that
(\ref{eq:filter.conservation}) is satisfied. This approach has the
advantage of avoiding the creation of artificial discontinuities at
the boundary of the elements, at the price of a small increase in the
total variation of $\mathcal{L}(u)$.

Finally the third one, which we refer to ``intrinsic'' correction,
consists in modifying the action of $\mathcal{L}$ so that $L_q =
\mathcal{L} (L_q)$ for $q\leq 2$. In this way both
(\ref{eq:filter.conservation.1d}) and (\ref{eq:filter.conservation})
are satisfied. In practice this is obtained by overwriting the
low-order coefficients of the discrete Legendre transform (DLT) of
$\mathcal{L}(u)$ with the ones of $u$. This method has the advantage
of not introducing any unwanted extra total variation and, for this
reason, is the one that has the best mathematical properties. The only
limitation is that it effectively weakens $\mathcal{L}$ and could thus
fail to completely removing the Gibbs oscillations.

All these three techniques can be basically used in combination with
any of the stabilization methods which are described in the following
Section.

\subsubsection{Stabilization techniques}
\label{sec:stabilization}

The most commonly used stabilization technique in discontinuous
Galerkin methods is based on slope limiting. These methods were
originally developed for FV schemes~\cite{Leveque92}, but are easily
modified to work with discontinuous Galerkin schemes as done by
Cockburn and Shu~\cite{cockburn_2001_rkd}, who introduced the
``$\Lambda\Pi_1$'' limiter based on a generalized version of the
``minmod limiter''. In our code we implemented a refined version of
this limiter originally proposed by~\cite{biswas_1994_paf} and
subsequently improved and extended
by~\cite{krivodonova_2007_lho}. This essentially works by recursively
limiting the coefficients of the spectral representation of the
solution on the various elements. The main advantage of this technique
is that it does not require tuning and it is usually very reliable,
while its main limitation is that its use often results in excessive
flattening of the solution in the presence of
discontinuities. Moreover we found that all these methods perform
rather poorly in conjunction with the correction techniques outlined
above and for this reason we have rarely used them.

Another method implemented in \texttt{EDGES} is the ``spectral
viscosity'' method first proposed by Maday et
al.~\cite{maday_1993_lpv} in the context of Legendre pseudospectral
methods for conservation laws. This method consists in the addiction
to the right hand side of (\ref{eq:dg.fully.discrete}) of a term
\begin{equation}
\label{eq:spectral.viscosity.1d}
  - \varepsilon_D\, (Q\, \partial_x \pmb{F}^t, \partial_x \phi)_D\,,
\end{equation}
where $\varepsilon_D$ is a coefficient depending on the number of grid
points and $Q$ is a viscosity kernel whose action on a scalar function
$u$ reads, in every element,
\begin{equation}
  \big[Q u\big](x) = \sum_{k=0}^D \hat{Q}_k \hat{u}_k L_k(x)\,.
\end{equation}
Here, $\hat{u}_k$ are the coefficients of the DLT of $u$ and
$\hat{Q}_k$ are real numbers such that
\begin{subequations}
\begin{align}
&  \hat{Q}_k = 0\,,  
&&\textrm{for} \quad k \leq m_D\,,\\
&  1 - \bigg(\frac{m_D}{k}\bigg)^4  \leq \hat{Q}_k \leq 1 \,,
&&\textrm{for} \quad k > m_D\,.
\end{align}
\end{subequations}
Note that $m_D$ effectively plays the role of the cut-off frequency of
the filter. Meday et al.~\cite{maday_1993_lpv}, were able to show
that, in the context of scalar conservation laws in 1D and using only
one domain, this method is able to stabilize the scheme, without
spoiling its spectral accuracy if
\begin{equation}
  \varepsilon_D \sim D^{-\theta}\,,
  \quad \textrm{for} \quad D \gg 1\,, 
\end{equation}
where $theta$ is just an exponent such that $0 <\theta \leq 1$ and
\begin{equation}
  m_D \sim D^{q/4}\,,
  \quad \textrm{for} \quad 0 < q < \theta \leq 1\,.
\end{equation}
In our code we have set
\begin{equation}
  \varepsilon_D = \mu \frac{\Delta t}{N} \frac{1}{D}
\end{equation}
and 
\begin{subequations}
\begin{align}
&  \hat{Q}_k = f\,,  
&&\textrm{for} \quad k \leq m_D\,,\\
&  \hat{Q}_k = f + 1 - \bigg(\frac{m_D}{k}\bigg)^4\,, 
&&\textrm{for} \quad k > m_D\,,
\end{align}
\end{subequations}
where $\mu$, $f$ and $m_D$ are set by the user (standard reference
values are $\mu=1, f=0$ and $m_D=0,1$).

Besides being particularly dissipative, this method has the important
advantage that it can be adapted to spherical symmetry by modifying
(\ref{eq:spectral.viscosity.1d}) as
\begin{equation}
\label{eq:spectral.viscosity}
  - \varepsilon_D\, (r^2\, Q\, \partial_r \pmb{F}^t, \partial_r
  \phi)_D\,.
\end{equation}
In this way, the condition~(\ref{eq:filter.conservation}) is satisfied
without the need of adding corrective terms. In \texttt{EDGES}, the
term (\ref{eq:spectral.viscosity}) is added using an
operator-splitting technique, so that its use corresponds to the
application of a linear filter. This is discretized with two different
techniques: locally on each element, or globally, using an interior
penalty technique. We call the latter approach ``interior-penalty
spectral-viscosity'' (IPSV) stabilization. While the first approach is
completely local, the second one is able to relax also the jump terms
between the elements. The main drawback of this method, however, is
that we have found that the quality of the results can be quite
sensitive to the tuning of $\mu$ and $m_D$, which are necessarily
problem dependent.

The third method implemented in \texttt{EDGES} is usually referred to
as ``spectral filtering'' (see
\eg~\cite{Canuto-Hussaini-Quarteroni-Zang:pseudospectral}). The idea
behind this technique is to filter the numerical solution, or its time
derivatives, with a low-pass filter in order to remove high-frequency
components and keep the Gibbs oscillations under control.  In the
context of Legendre pseudospectral methods, this filtering, which we
indicate with $\mathcal{F}_D$, reads
\begin{equation}
  \big[\mathcal{F}_D u\big](x) = \sum_{k=0}^D \sigma\bigg(\frac{k}{D}\bigg) \,
  \hat{u}_k \, L_k(x)\,,
\end{equation}
where $\hat{u}_k$ are the coefficients of the Legendre expansion of $u$
and $\sigma(\eta)$ is a \emph{filter function} of order $p$ in the
Vandeven's sense, that is a function $\sigma\in
C^p\big(\mathbb{R}^+;[0,1]\big)$ such that 
\begin{align}
  &\cdot~ \sigma(0) = 1\,; & \\
  &\cdot~ \sigma^{(k)}(0) = 0\,, &\textrm{for~}& k = 1,2,\ldots, p-1\,;\\
  &\cdot~ \sigma(\eta) = 0\,,   &\textrm{for~}& \eta \geq 1\,; \\
  &\cdot~ \sigma^{(k)}(1) = 0\,, &\textrm{for~}& k = 1,2,\ldots, p-1\,; 
\end{align}
where $f^{(p)}(x)$ denotes the $p$-th derivative of $f$ in $x$. These
schemes were studied by Hesthaven and Kirby~\cite{hesthaven_2008_fls},
who showed that if $\sigma(\cdot)$ is a filter of order $p$, and $u
\in C^p(-1,1)$, then
\begin{equation}
  \big|u(x) - \mathcal{F}_Du(x)\big| \leq D^{1-p} \| u^{(p)}
  \|_{L^2(-1,1)}\,,
\end{equation}
so that, if $p \sim D$, the filter does not spoil the spectral
accuracy of the method.

The effects of filtering on the stability of the numerical method can be
intuitively understood from the fact that some filters, in particular the
exponential filter of order $p$,
\begin{equation}\label{eq:exp.filter}
  \sigma(\eta) = \exp\Big[-\mu (D+1) \frac{\Delta t}{\Delta x}
  \eta^p\Big]\,,
\end{equation}
can be interpreted as equivalent to the use of numerical diffusion of
order $p$ and strength $\mu$~\cite{meister_2009_sfd}. At present,
however, a mathematical understanding of the impact of filtering on
the accuracy and the stability of Legendre pseudospectral methods is
still lacking~\cite{hesthaven_2008_fls}.

\texttt{EDGES} provides a number of different filter functions, but the
most used are the exponential filter (\ref{eq:exp.filter}), which is only
an approximate filter, but which is very popular due to its flexibility
\cite{Canuto-Hussaini-Quarteroni-Zang:pseudospectral}, and the ``Erfc-Log
filter'' proposed by Boyd~\cite{boyd_1996_efa},
\begin{equation}
  \sigma(\eta) = \frac{1}{2} \mathrm{erfc} \Bigg\{ 2 p^{1/2}
  \bigg(|\eta|-\frac{1}{2}\bigg)
  \sqrt{\frac{-\log[1-4(\eta-1/2)^2]}{4(\eta-1/2)^2}} \Bigg\}\,,
\end{equation}
which is a semi-analytic approximation of the Vandeven's filter,
\begin{equation}
  \sigma(\eta) = 1-\frac{\Gamma(2p)}{\Gamma(p)^2} \int_0^\eta \big[ t
  (1-t) \big]^{p-1} \dd t\,,
\end{equation}
where $\Gamma$ is the Euler Gamma function. Note that it is customary
to use filters of order $p \sim 2D$ or larger when expecting regular
solutions and $2D+1$ is a reasonable value for a strong filter such as
Erfc-Log.

The main limitation of spectral filtering is that, being a high-order
linear method, it cannot completely remove the Gibbs
oscillations. This is a consequence of the well-known Godunov theorem
stating that no linear monotonicity-preserving method exists of
second-order or higher (see \eg~\cite{Leveque92}). The idea is,
instead, to allow for small oscillations of the numerical solution at
the location of shocks, while preventing them from growing out of
control. In our numerical experience we found that spectral filtering
is a robust and viable alternative to slope limiting in the shock-tube
case, where its use often results in a sharper resolution of the
discontinuities, and, at the same time, it is well-behaved in the
spherically symmetric case. Spectral filtering seems also to be much
less sensitive than the spectral viscosity methods to the tuning of
the parameters of the filter functions. Finally, we point out that, in
contrast to IPSV and slope limiters, this technique is completely
local in the sense that its application in a given element does not
require the knowledge of the solution elsewhere. For these reasons,
most of the results that we will show in the next Sections were
obtained with the use of spectral filtering.

\begin{figure}
\begin{center}
  \includegraphics[width=8.0cm]{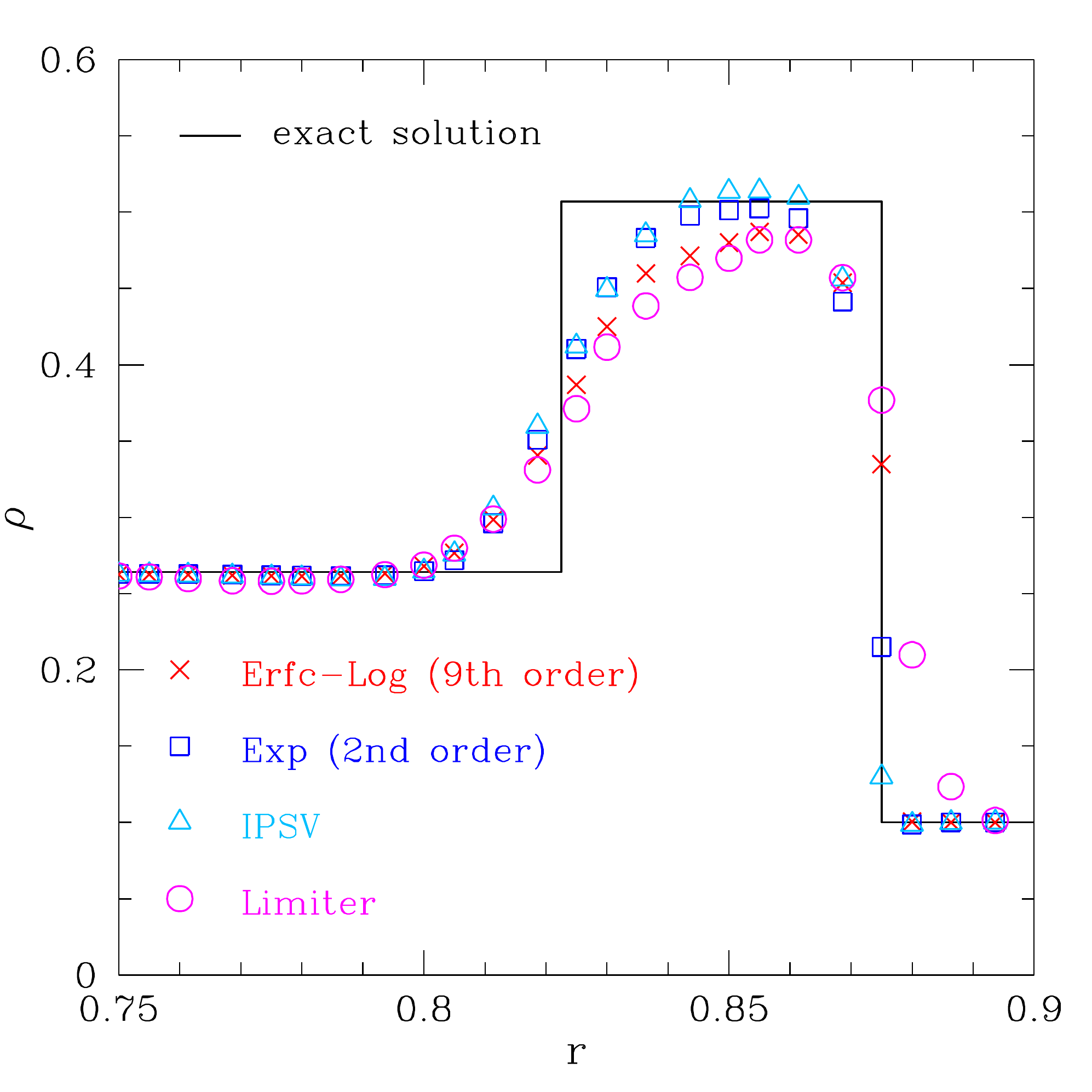}
  \caption{\label{fig:filters} Solution for the rest-mass density of a
    relativistic shock tube with different stabilization
    techniques. The solid black line represents the exact, analytic
    solution, while the red crosses and the blue squares refer to the
    use of the two different DLT filters, \ie the Erfc-Log and the
    exponential filter, respectively. The blue triangles and the
    violet circles refer instead to the use of the IPSV filter and of
    high-order slope limiters, respectively. To avoid excessive
    cluttering, only a data point every five is shown. The results were
    obtained on a grid of $N=200$ elements with a polynomial of degree
    $D=3$.}
\end{center}
\end{figure}

Before doing that, however, and to illustrate the difference between
the various stabilization techniques, we show in
Fig.~\ref{fig:filters} a comparison of the results obtained using
different filters, while keeping all the other discretization
parameters fixed. In particular, we show the results obtained in the
case of a relativistic shock tube in flat spacetime. The details of
the physical setup will be given in Sect.~\ref{sec:shock_tubes}, here
instead we focus on the effects of the different stabilization
techniques. As anticipated we find that slope-limiting, even in their
high-order variant, result in a much larger smearing of the shock
front, then all the other methods. The 9th order Erfc-Log filter
yields a much sharper resolution of the shock and a stable evolution,
even tough, as discussed before, filtering does not have such a strong
mathematical basis, while slope-limiting is known to yield a
total-variation diminishing in mean (TVDM)
scheme~\cite{cockburn_2001_rkd}. The best overall results are the ones
obtained with the second-order exponential filter, for which we used a
small strength factor, $\mu = 1$, and the IPSV technique with $\mu =
1$ and $\hat{Q}_k = 1$. We remark that the results obtained with these
last two methods are very similar since, as discussed before, the
action of an exponential filter is roughly equivalent to that of a
spectral viscosity. The results in Fig.~\ref{fig:filters} obtained on
a grid of $N=200$ elements with a polynomial of degree $D=3$.

\subsection{Treatment of low-density regions}

The treatment of interfaces between vacuum region and fluid regions is
one of the most challenging problems in Eulerian (relativistic)
hydrodynamics codes (see \eg~\cite{galeazzi_master, kastaun_2006_hrs,
Millmore2010}). The most commonly used approach to treat vacuum regions
is to fill them with a low-density fluid, such that if a fluid element is
evolved to have a rest-mass density below a certain threshold, it is set
to have a floor value and zero coordinate
velocity~\cite{Font02c,Baiotti04}. This approach works reasonably well
and has been adopted by the vast majority of the
relativistic-hydrodynamics codes. Nevertheless, the introduction of this
``atmosphere'' creates several numerical difficulties. First of all, this
approach is gauge dependent and, for example, a star can accrete or lose
mass from/to the atmosphere due to oscillations in the coordinate
variables. Secondly, this approach often results in the introduction of
errors at the surface of the star, which, on the one hand, usually have
small influence on the overall dynamics because of the small momenta
involved, but which, on the other hand are relatively large if compared
with the magnitude of the involved quantities at the surface. Finally,
and most importantly, it is possible that the algorithm governing the
evolution of the floor can couple with the Gibbs oscillations, leading
quickly to their amplification and destabilizing the scheme (After-all,
the introduction of an atmosphere treatment is de-facto equivalent to the
use of a boundary condition and this can very well lead to an unstable
algorithm).  For this reason this procedure could not be used in a
straightforward way in our code, but required a particular care.

Other solutions, such as the use of the equations in Lagrangian form,
\eg~\cite{Gabler:2009yt}, the use of moving grids techniques
\eg~\cite{gourgoulhon_1991_seg}, or the use of suitable limiters at the
surface~\cite{galeazzi_master}, do not suffer from these issues, but are
restricted to the spherically symmetric case. As our code is meant to be
a prototype code to study the viability of DG methods for relativistic
hydrodynamics, the use of these techniques would have defeated our
purpose.

As a result, several different approaches were implemented and tested
in \texttt{EDGES} to overcome the difficulties with the atmosphere
discussed above. The main idea behind all these approaches was to use
stronger stabilization techniques at the interfaces between fluid and
vacuum regions. Unfortunately all these techniques performed quite
poorly because they resulted in an unacceptable lowering of the
resolution at the surface and thus in large numerical errors.
Moreover, in the spherically symmetric case, the situation is greatly
worsened by the fact that these errors tend to build up coherently
while traveling towards the center.  For example, in the case of a
neutron star, they tend to be amplified by a factor $R / r \sim 10^3$,
$R \sim 10$ being the radius of the star and $r \sim 0.01$ being the
location of the closest grid point to the center.

\begin{figure}
\begin{center}
  \includegraphics[width=8.0cm]{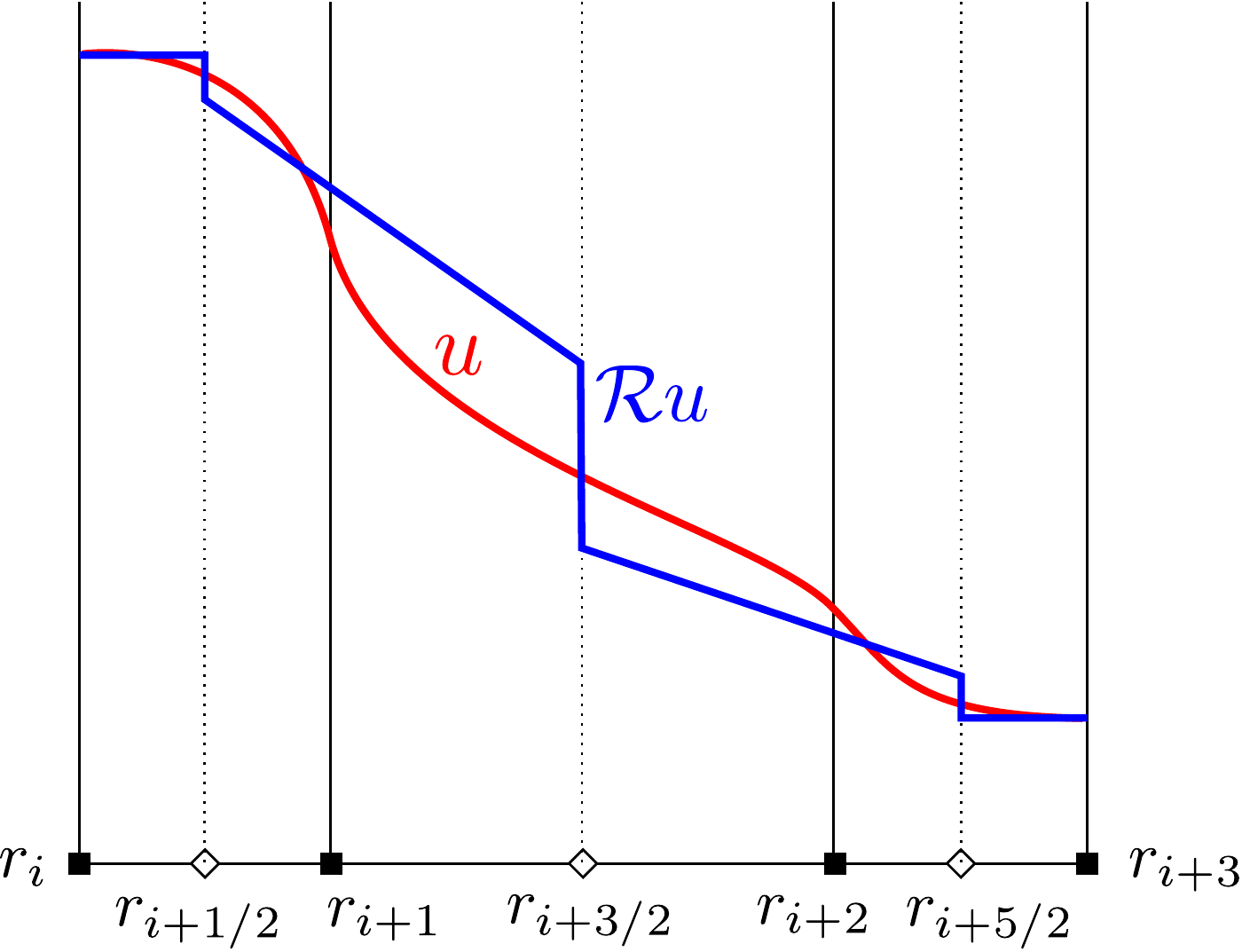}
  \caption{\label{fig:spectral.volumes} Division of an element into
    control volumes for the spectral-volume method applied at the
    fluid-vacuum interface. The values of the solution $u$ on the
    collocation points (filled squares) are interpreted as
    volume-averaged values of the solution in appropriate control
    volumes (hollow diamonds denote the boundaries). These values are
    then used to generate a reconstructed solution $\mathcal{R}u$
    using an adapted slope-limiter method.}
\end{center}
\end{figure}

Within \texttt{EDGES} a solution to this problem was eventually found
with the use of a method that is, at the same time, oscillation-free
and capable of obtaining high-enough resolution at the surface. In
particular, we derived a ``sub-element method'' approach that we later
discovered to be very similar to the spectral volumes (SV) method
already suggested by Wang~\cite{wang_2002_sfv}. In this method, we
first flag those elements in which the rest mass density falls below a
certain threshold. Secondly we interpret the value of the solution on
the collocation points of those elements as being the volume-averaged
value of the solution on appropriated \emph{control volumes}. Finally
these values are evolved as in a FV scheme: a linear reconstruction
with slope limiting is used to compute the value of the primitive
variables at the interface of the control volumes and the HLLE
approximate Riemann solver is used to compute the fluxes.

The structure of the grid used for this spectral volume method is
shown in Fig.~\ref{fig:spectral.volumes}. The control volumes are
simply obtained by considering the points of the dual mesh as the cell
interfaces, while their volume has to be corrected to take into
account the weights of the Gaussian quadrature associated with the
primal grid.  The linear reconstruction can be performed only for the
interior control volumes, while for the control volumes at the
boundary of each element we already have the value of the solution at
one side of the control volume, thus we are forced to reconstruct the
solution there as a constant. In our tests we found that the
``superbee''~\cite{Font08} limiter is the one which guarantees the best
results among the ones that we tested. For this reason, all the
results shown in the following make use of this limiter.

\begin{figure}
\begin{center}
  \includegraphics[width=8.0cm]{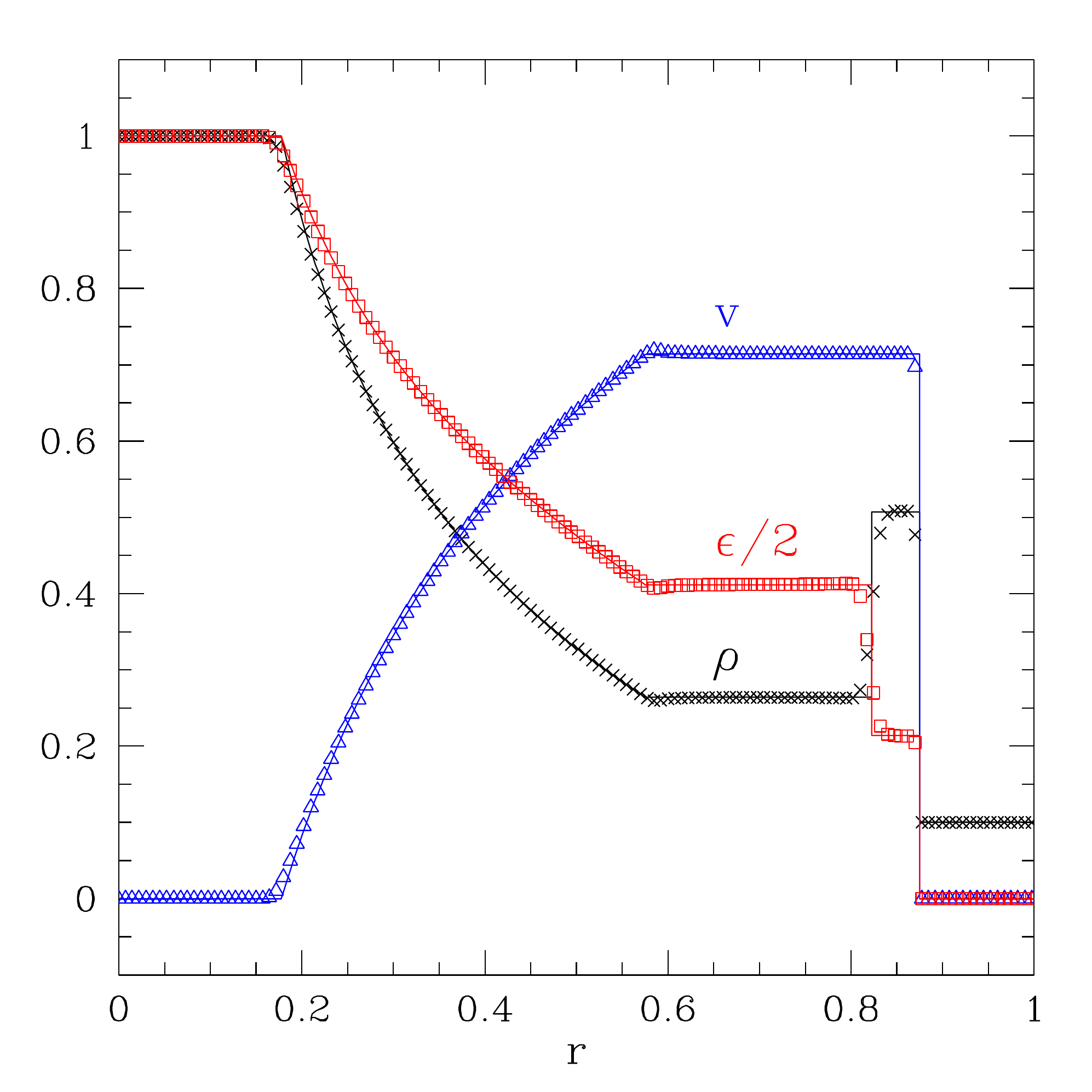}
  \caption{\label{fig:shock.tube.1} First shock-tube test. Shown with
    solid lines are the exact solution for the rest-mass density
    (black line), the velocity (blue line), and the specific internal
    energy (red line). The corresponding numerical solution is
    represented by the black crosses, the blue triangles, and the red
    squares, respectively. To avoid excessive cluttering we show only
    one point every $15$ of the numerical solution. The results were
    obtained on a grid of $N=500$ elements with a polynomial of degree
    $D=3$.}
\end{center}
\end{figure}

The important advantage of this approach, with respect to a more
traditional approach in which the troublesome elements are split into
equal-size cells and a classical FV method is used, see
\eg~\cite{touil_2007_tdh}, is that no interpolation is required in the
switching from/to the discontinuous Galerkin method. The coupling
between the two methods is also very natural and is done through the
Riemann solvers between the elements. Again no special treatment is
required to handle different type of elements: the DG ones and the SV
ones. The main limitation of this approach is that, as we do not
increase the number of degrees of freedom in the flagged elements, it
has the effect of reducing the accuracy to second order where it is
used. As it will be shown in Sect.~\ref{sec:tov}, this seems not to be
limiting the accuracy of our code, probably because the use of the SV
method is confined to small regions containing low-density fluid. In
the case of a neutron star, for example, the use of the SV method is
typically limited to one element containing the surface of the star.

As a final remark we note that, as we are working with a nonuniform
grid, the slope limiting method in its standard form, also used by us,
is not guaranteed to be TVD and/or second
order~\cite{berger_2005_asl}.  Again this seems not to be a problem in
practice.

\section{Numerical tests}\label{sec:results}

In what follows we present the results obtained from some of the tests
performed with \texttt{EDGES}. These tests have been chosen because
they highlight the capabilities of our code in idealized settings,
such as in the simulation of shock tubes and shock waves, and in more
``astrophysically motivated'' settings, such as in the simulation of
the dynamics of spherical stars and of the gravitational collapse to
black hole. In all cases considered, the evolution has been made
employing an ideal-fluid EOS to take into account for non-isentropic
transformations, such as shock-heating effects.

\subsection{Shock tubes}
\label{sec:shock_tubes}

The first tests performed are shock-tube tests and, more specifically,
we first present the results obtained in the case of two standard
benchmarks for relativistic-hydrodynamics codes described
in~\cite{Marti03}. These are referred to as ``blast-wave'' problems
$1$ and $2$ in~\cite{Marti03}, and are essentially
one-dimensional flat-space Riemann problems.

\begin{figure}
\begin{center}
  \includegraphics[width=8.0cm]{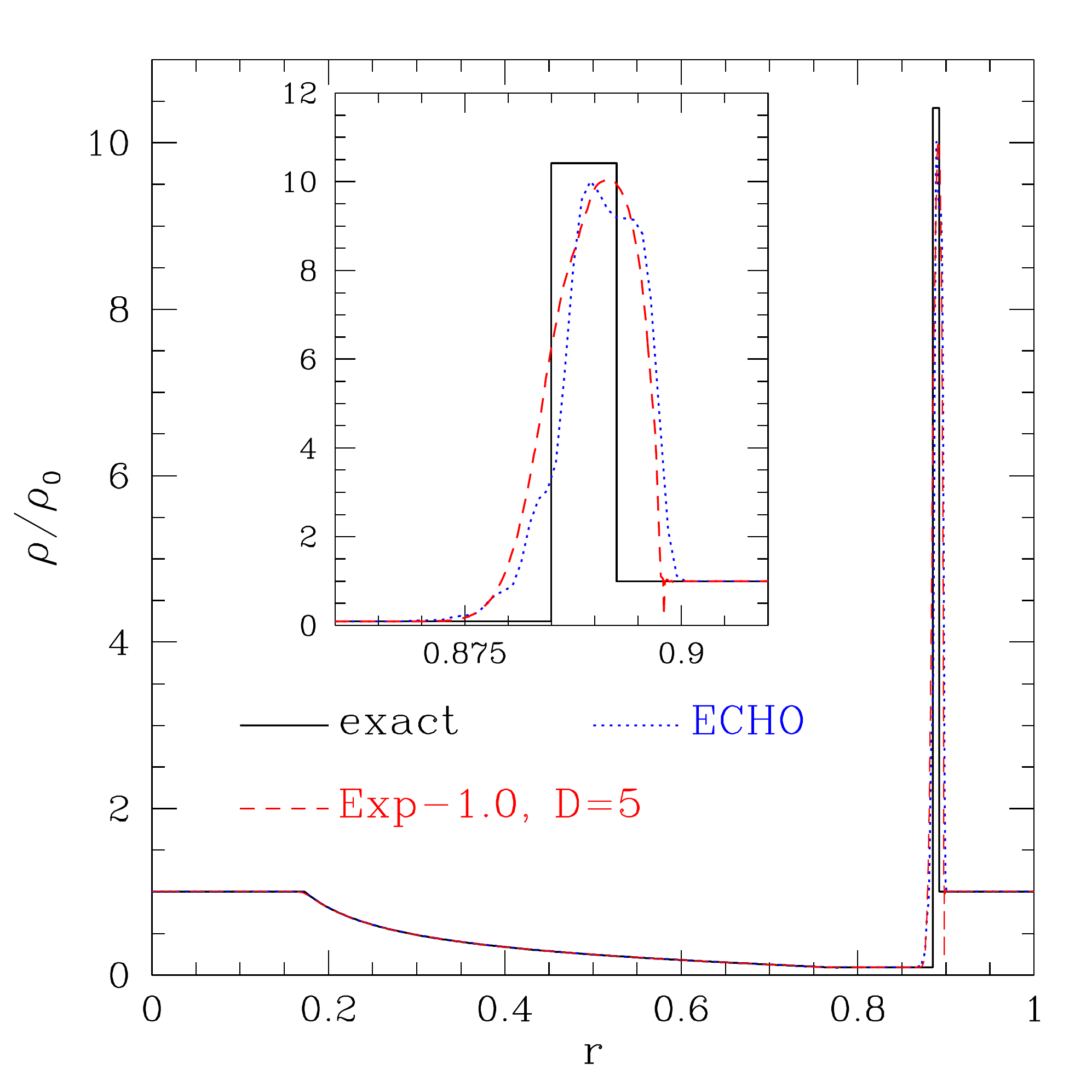}
  \caption{\label{fig:shock.tube.2} Rest-mass density profile for the
    second shock-tube test. Shown with different lines are the exact
    solution (solid black), the solution obtained with the
    \texttt{ECHO} code~\cite{DelZanna2007} (dotted blue). Both codes
    use $1000$ cells/elements and algorithms at fifth order;
    \texttt{EDGES} also uses a exponential filter with $\mu=1, p=2$.}
\end{center}
\end{figure}

The first problem describes the propagation of a relativistic blast
wave through a low-density medium. The solution obtained with
\texttt{EDGES}, as well as the analytic solution, are shown in Fig.
\ref{fig:shock.tube.1}, where the solution is computed using
polynomials of degree three (\ie $D=3$) and $500$ elements (\ie
$N=500$). The stabilization is obtained with the use of an Erfc-Log
filter of 9th order, which has been applied directly to the conserved
variables in the regions where they fail to be monotone at every
timestep. As it can be seen from the figure, the stabilization is
strong enough to eliminate any oscillation, thus reducing the order of
the method to the expected first-order near the discontinuity, yet
with a very small smearing of the shock.

\begin{figure}
\begin{center}
  \includegraphics[width=8.0cm]{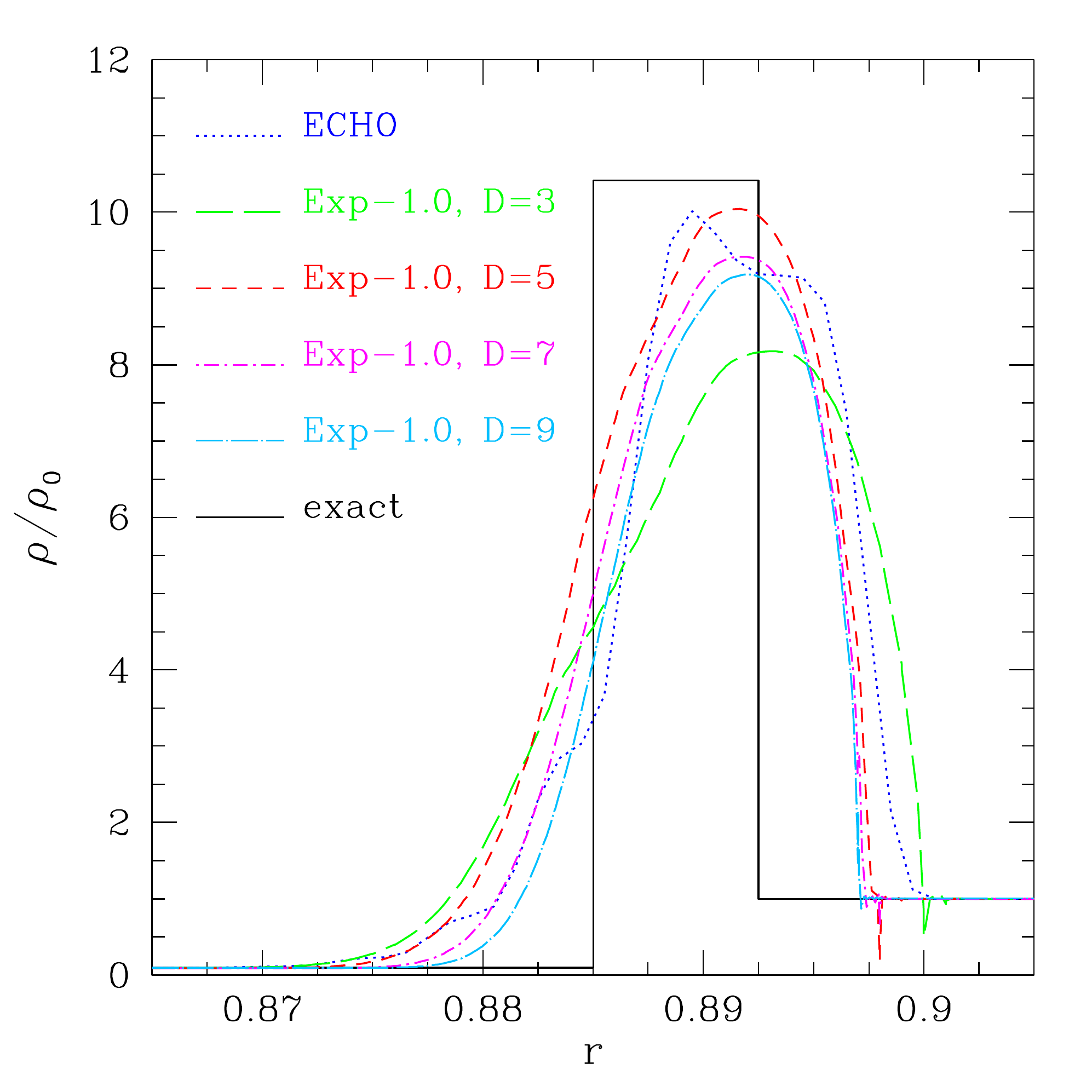}
  \caption{\label{fig:shock.tube.2bis} The same as in
  Fig.~\ref{fig:shock.tube.2}, but with a magnification of the shock and
  a comparison of the results at different orders $D$ of the polynomial
  representation of the solution. Note that the accuracy with which the
  shock location is founded increases with $D$}
\end{center}
\end{figure}

The second shock-tube problem is similar to the first one, but is much
more extreme. This benchmark is considered to be a challenging test
also for modern HRSC codes~\cite{Marti03}. The main difficulty is to
handle the very high compression ratios, typical of relativistic
hydrodynamics, produced in this case. In Fig.~\ref{fig:shock.tube.2}
we show a comparison of the results obtained with our code when using
the exponential filter (\ref{eq:exp.filter}) at second-order (\ie
$p=2$) with $\mu = 1.0$, which is employed in the same way as the
Erfc-Log filter in the previous test. Also shown for a comparison is
the solution obtained with the high-order HRSC code
\texttt{ECHO}~\cite{DelZanna2007}, which was kindly provided by Olindo
Zanotti. The \texttt{ECHO} code uses HLLE Riemann solver and
monotonicity-preserving reconstruction at fifth
order~\cite{suresh_1997_amp}, while for \texttt{EDGES} we employed
fifth-order polynomials\footnote{Note that the solution of the
  \texttt{ECHO} code used here does not make employ the high-order DER
  flux reconstruction discussed in the Appendix of
  Ref.~\cite{DelZanna2007} and which would remove some of the
  oscillations.}. Furthermore, the comparison is made when using the
same number, $1000$, of cells (for \texttt{ECHO}) and of elements (for
\texttt{EDGES}); for both codes, in fact, we expect a convergence
order (on smooth solutions) of the type $\Delta x^{-p}$, with $\Delta
x$ being the width of each cell/element and $p$ being the order of the
scheme. Note, however, that, while in \texttt{ECHO} high order is
obtained with the use of data across multiple cells, in \texttt{EDGES}
this is attained with the use of a polynomial representation of the
solution in each element. Thus the \texttt{EDGES} code actually uses
around six times more grid points than the \texttt{ECHO} code.
Nevertheless, we argue that this is the correct comparison because, as
mentioned before, the convergence properties of the two codes once the
reconstruction procedure and the order of the polynomial
representation is fixed, scale with the width of the
cells/elements. This kind of comparison, where results obtained with
discontinuous Galerkin and finite volume methods are obtained using
the same number of elements/cells, but different number of degrees of
freedom has already been discussed in the literature, see \eg
\cite{zhou_2001_ncw}.

As can be seen from Fig.~\ref{fig:shock.tube.2} the quality of the
solution obtained by the two codes is comparable, with the solution
obtained by the \texttt{EDGES} code being slightly closer to the exact
position of the shock, but also showing signs of Gibbs oscillations ahead
of the shock. Overall, the exponential filter can prevent the
oscillations from growing and yields the best results when compared to
the other flattening techniques discussed above. A more quantitative
estimate of the quality of the numerical solution can be obtained by
looking at the ratio $\sigma/\sigma_{\mathrm{exact}}$, between the
observed compression ratio and the analytic one~\cite{Marti03}. We obtain
a value of $0.96$ with both \texttt{ECHO} and \texttt{EDGES}. At lower
resolution, \ie using only $400$ elements, we obtain, for \texttt{EDGES},
a value of $0.60$ which is close to the ones reported for PPM at the same
resolution: \ie $\sigma/\sigma_{\mathrm{exact}} \simeq 0.54 -
0.69$~\cite{Marti03}. We conclude therefore that in the case of
discontinuous solutions, DG methods behave similarly to FV methods with
the same number of elements/regions.

It is worth stressing that while for \texttt{ECHO} the fifth order
represents the largest achievable, the \texttt{EDGES} code can perform
this very severe test also with higher-order polynomials. This is
shown in Fig.~\ref{fig:shock.tube.2bis}, where we report again the
exact solution shown in Fig.~\ref{fig:shock.tube.2} and compare it
with the numerical one for $D=3, 5, 7$ and $D=9$. Of course the
mathematical convergence in all of the different solutions is still
only at first order, but it is remarkable to see that the code can
handle such large shocks even when $D=9$ and that the position of the
shock and of the contact discontinuity become increasingly accurate as
$D$ is increased.  Overall, the solution with $D=5$ is the one
yielding the closest match with the density right behind the shock and
we have not considered orders higher than this in most of the
subsequent tests. As a final remark we note that, to the best of our
knowledge, this is the first time that a shock-tube test in
relativistic hydrodynamics has been computed using a method of this
order.

\subsection{Spherical shock reflection}

\begin{figure}
\begin{center}
  \includegraphics[width=8.0cm]{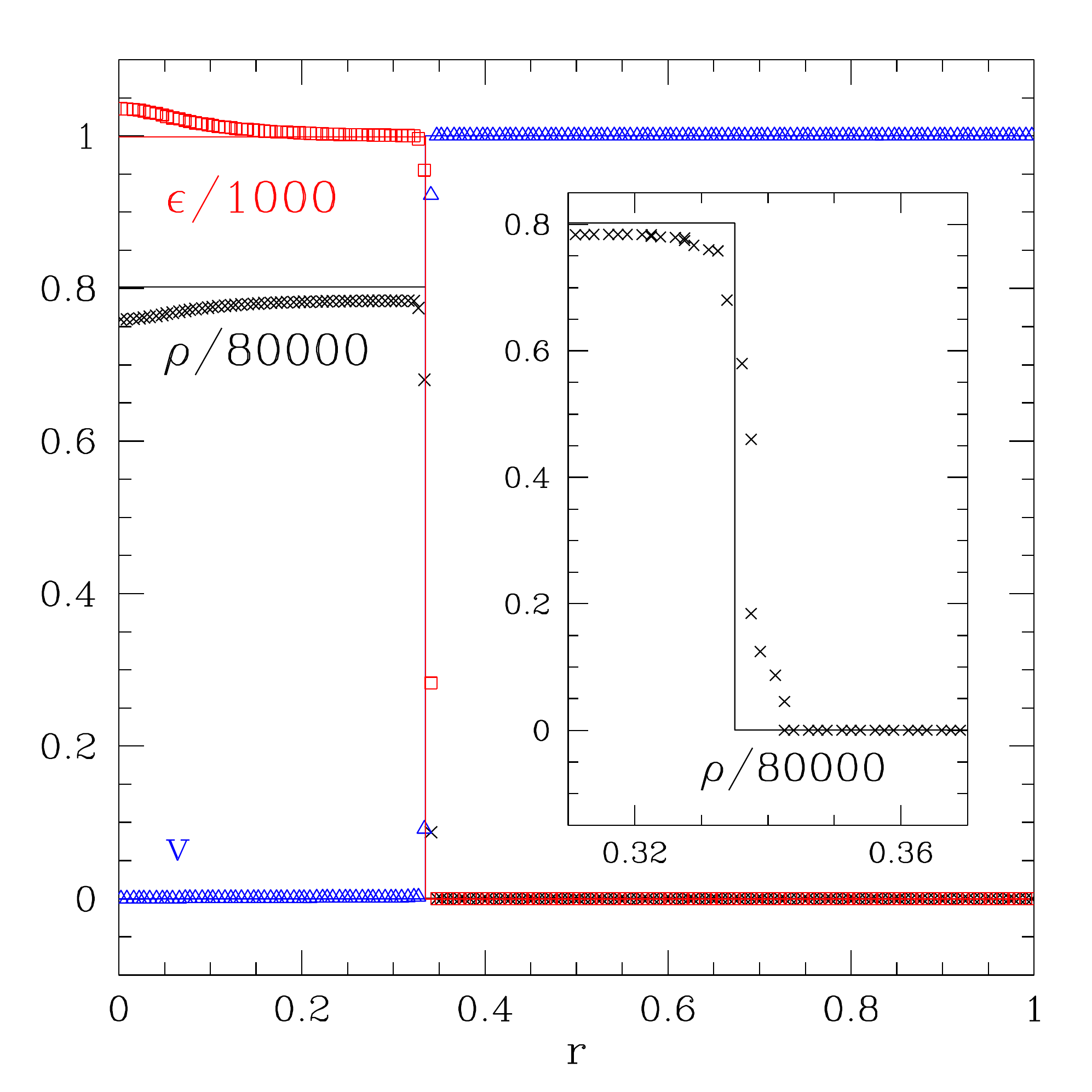}
  \caption{\label{fig:shock.reflection} The same as
    Fig.~\ref{fig:shock.tube.1}, but for the spherical
    shock-reflection test with a Lorentz factor of $W = 1000$. To
    avoid excessive cluttering in the main frame of the picture we
    plot only a point every five of the numerical solution, while the
    inset shows the solution on all the collocation points.}
\end{center}
\end{figure}

Another classical benchmark for relativistic-hydrodynamics codes is
the spherical shock-reflection test. This is the relativistic version
of the classical Noh test and its setup is described in detail
in~\cite{Marti03}. The initial data consist in a cold fluid converging
at the center of the domain. As the fluid flows towards the center, a
hot dense shell of matter is formed and a shock wave is propagated
outwards. In this situation it is possible to generate arbitrarily
high compression ratios by increasing the Lorentz factor of the
ingoing flow. This is a peculiarity of relativistic hydrodynamics and,
again, the key point of the test is to assess the capability of the
numerical methods to handle such strong density contrasts.

This problem is also interesting to assess the quality of the
correction procedure for the filtering in spherical symmetry. In
particular, the quality of the results depends on the used correction
procedure: they are of comparable quality when using the ``dummy'' and
the ``bubble correction'' procedures (the former being slightly
better), while the ``intrinsic'' correction yields a filtering which
is too weak to guarantee stability. This is not a surprise since with
the intrinsic procedure we are not able to filter the coefficients of
order less than two in the Legendre expansion, so that we cannot
reduce the order of the method below the second even with a strong
targeted filter.

\begin{figure}
\begin{center}
  \includegraphics[width=8.0cm]{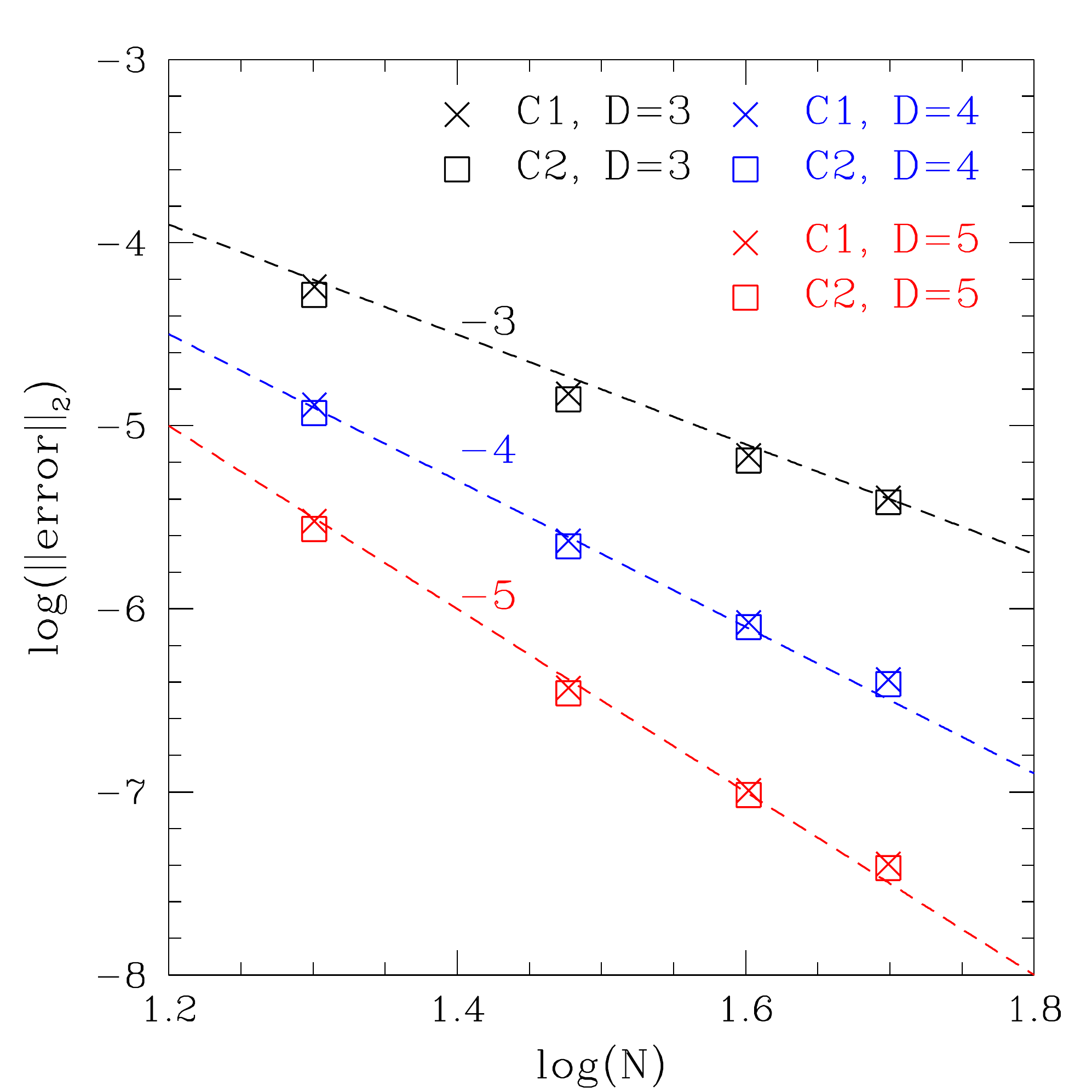}
  \caption{\label{fig:michel.hconv} $L^2$-error in the spherical
    accretion test as a function of the number of elements
    $N$. Crosses and squares refer to the errors on $C_1$ and $C_2$,
    respectively, while the black, blue and red colors denote refer to
    the simulations using third, fourth, and fifth-order polynomials.
    Finally the dashed lines show the slopes associated with third,
    fourth, and fifth-order convergence.}
\end{center}
\end{figure}

In Fig.~\ref{fig:shock.reflection} we present the results obtained for
a flow with $W = 1\, 000$ solved using polynomials of degree three, on
$200$ elements and with an Erfc-Log filter with $p=4$ and a dummy
correction. As shown in the figure, the filtering procedure is able to
suppress any spurious oscillation and the shock front is captured
within two elements, \ie eight collocation points. The average
relative error on the compression level is of $1.4\,\%$, comparable
with the $2.2\,\%$ reported by Romero et al.~\cite{Romero96} using FV
with a minmod reconstruction on a grid with the same number of
cells. Furthermore, as in Romero et al.~\cite{Romero96}, the error in
$\epsilon$ and $\rho$ increases in the inner part of the
solution. This is due to the well known wall-heating effect and is due
to the small but nonzero dissipative and dispersive features of the
numerical methods (see~\cite{noh_1987_ecs, rider_2000_rwh} for an
extended discussion of this effect).

\subsection{Spherical accretion onto a Schwarzschild black hole}

Having verified the capability of our code to handle shock waves, we
next present some results concerning its accuracy in the case of
regular solutions. In particular, we consider the case of spherical
accretion onto a Schwarzschild black hole in the Cowling approximation
(\ie with a fixed spacetime). An analytic solution exists in the case
of stationary flows and was first presented by Michel~\cite{Michel72}
and later used as a numerical test for other numerical codes, for
example by \cite{Romero96}. This solution can be described in term of
two constants:
\begin{equation}
  C_1 \equiv \sqrt{-g}\, J^r \,,
\end{equation}
and
\begin{equation}
  C_2 \equiv \sqrt{-g}\, T^r_{\phantom{r}t}\,.
\end{equation}

\begin{figure}
\begin{center}
  \includegraphics[width=8.0cm]{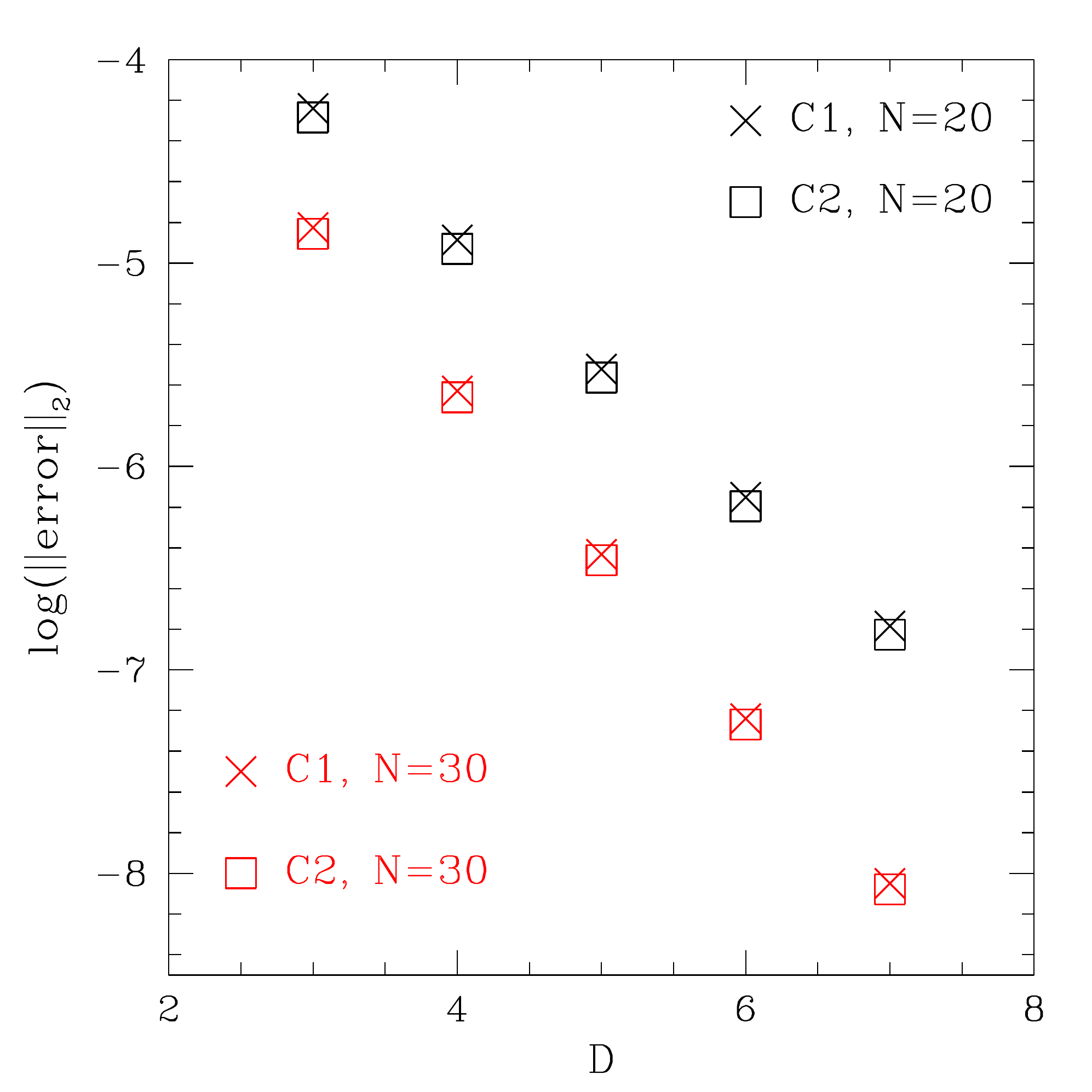}
  \caption{\label{fig:michel.pconv} The same as in
    Fig.~\ref{fig:michel.hconv}, but when the error is shown as a
    function of the polynomial degree when the number of elements $N$
    is kept fixed (black for $N=20$ and red for $N=30$). Note that an
    exponential convergence is measured.}
\end{center}
\end{figure}

In our simulation we consider a spherical shell with extent $3-N/40
\leq r \leq 20$ in the spacetime metric of a Schwarzschild black hole
of unit mass, while the initial conditions for the hydrodynamical
variables describe a low-density fluid at rest with respect to an
observer at infinity. At $t=0$ we start injecting higher-density fluid
from the outer boundary and after a short transient the solution
reaches stationarity, allowing us to measure $C_1$ and $C_2$ and to
compare them with the analytic values fixed by the outer boundary
condition.

In Fig.~\ref{fig:michel.hconv} we show the $L^2$-norm of the error for
$C_1$ and $C_2$ as a function of the number of elements $N$, and for
different orders $D$ of the polynomials used for the representation of
the numerical solution over the single elements. The stabilization was
obtained using an exponential DLT filter of order $2D$ and strength
$\mu = 1.0$, corrected using the ``dummy'' strategy. Clearly, the
errors computed from $C_1$ and from $C_2$ are basically
identical. More importantly the convergence order is the one that one
would expect from the theory: third, fourth, and fifth-order using
third, fourth, and fifth-order polynomials, respectively.

A complementary measure of the convergence properties of the code is
shown in Fig.~\ref{fig:michel.pconv}, where we report the $L^2$-norm
of the error obtained after keeping fixed the number of elements, $N =
20,30$, but changing the order of the polynomials used for the
representation of the solution. As expected, SDGM schemes behave in
this case as multi-domain spectral methods~\cite{canuto_2008_sme},
with the error decreasing exponentially with the polynomial
degree. This is indeed what we observe in Fig.~\ref{fig:michel.pconv},
where the errors decrease exponentially by almost three orders of
magnitude for both $C_1$ and $C_2$, and at both resolutions. No sign
of saturation appears in the error and we find it remarkable that the
same method that was able to capture so sharply the discontinuous
solutions in the previous tests, is also able to attain exponential
convergence in smooth flows.

\subsection{Linear oscillations of spherical stars}\label{sec:tov}

The results presented so far refer to situations that are somehow
idealized and are meant mainly as a way to highlight the code's
properties with respect to shock capturing, filtering correction and
accuracy on smooth solutions. We next present the results obtained in
tests describing systems of more direct physical interest. Furthermore,
in contrast to the previous tests, the matter will not be considered as a
test-fluid and the spacetime will be properly evolved.

A first interesting test is the study of linear oscillations of
spherical stars. The initial data considered describes the equilibrium
configuration of a self-gravitating fluid sphere, described by a
polytropic equation of state, $p = K \rho^\Gamma$, and is obtained by
integrating the Tolman-Oppenheimer-Volkoff (TOV) equations, as
described by \cite{tooper_1965_afs}, and interpolating the result on
the computational grid used by \texttt{EDGES}. The system is then
evolved under the sole effects of numerical perturbations, mainly due
to the interpolation errors in the generation of the initial data and
to the interaction between the star and the atmosphere. Note that in
order to avoid the presence of collocation points at $r=0$, we stagger
our numerical grid so that the $r=0$ point falls at the centre of the
first element and we use only polynomials of odd degree.

\begin{figure}
\begin{center}
  \includegraphics[width=8.0cm]{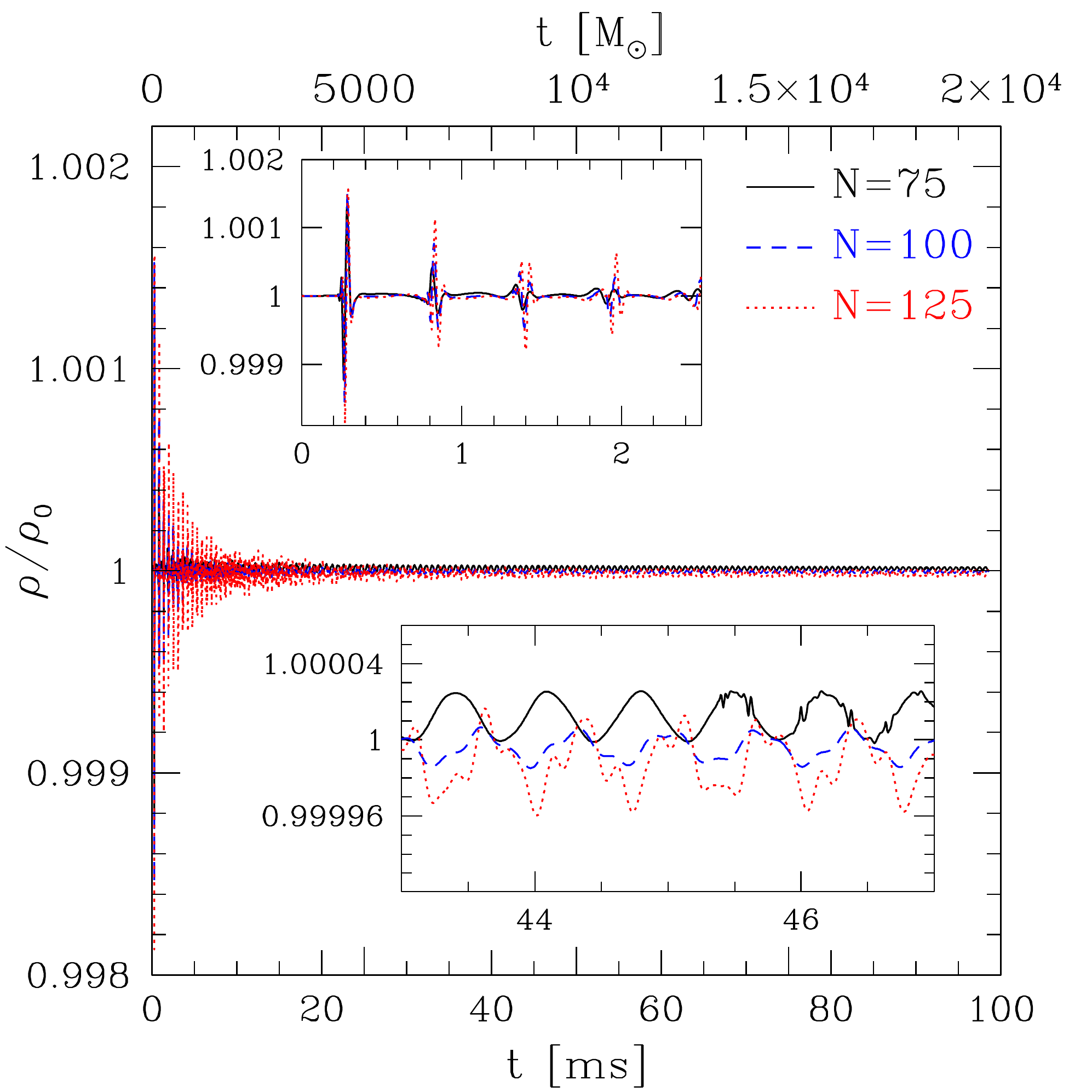}
  \caption{\label{fig:tov.rhoc} Evolution of the central rest-mass
    density normalized to its initial value for a stable TOV and
    evolved with a polynomial of degree five. Different lines refer to
    different resolutions: solid black, dashed blue and dotted red for
    $N=75\,,100$ and $125$, respectively. Shown in the insets are the
    evolution at the initial time (\ie $t\lesssim 2$ ms) and when the
    error at the surface is released triggering new small-scale
    oscillations (\ie $t\sim 45$ ms).}
\end{center}
\end{figure}

More specifically, we have considered a TOV constructed with a
polytropic EOS having $K = 100$ and $\Gamma = 2$ and whose initial
central density is $\rho_c = 1.28 \times 10^{-3}$. This could be taken
to represent a stable, nonrotating, neutron star with gravitational
mass $M = 1.4\,M_{\odot}$ and areal radius $R = 9.6\,M_{\odot} \simeq
14.2$ km. The degree of the polynomial basis is five and an Erfc-Log
filter of order eleven is used to ensure a stable evolution. This
filter is corrected with the ``intrinsic'' procedure outlined in
Sect.~\ref{complicationsSS} and is applied to the right-hand-side of
the time stepping scheme. This results in a very weak stabilization
algorithm whose main effect is to diffuse back into the atmosphere the
numerical errors that would otherwise accumulate at the surface of the
star and destabilize high-resolution runs. The latter, in fact, have
very low intrinsic viscosity and run for many more timesteps on
timescales of $100$ ms or longer (we recall that the dynamical
timescale for this star is of the order of $0.7$ ms).

In Fig.~\ref{fig:tov.rhoc} we show the evolution of the central
density for three of these runs using different resolutions. They span
a time frame of $20\,000\,M_{\odot} \sim 100\ \mathrm{ms}$, with the
grid extending over the range $0 \leq r \leq 15$ and being composed by
$N=50, 75, 100, 125$ and $150$ elements.

\begin{figure}
\begin{center}
  \includegraphics[width=8.0cm]{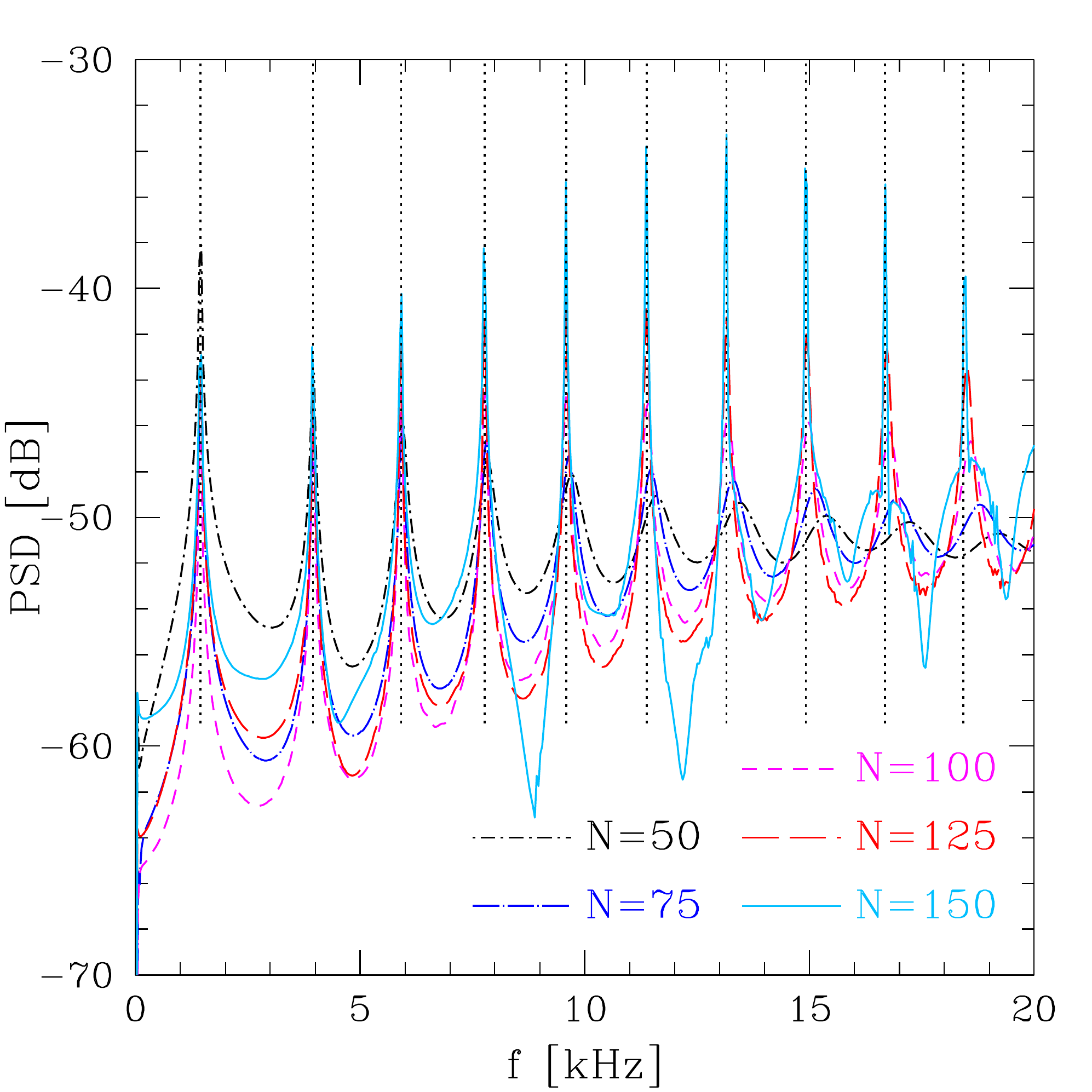}
  \caption{\label{fig:tov.spectrum} Power spectral density computed
    from the first $5000\,M_{\odot}$ of the evolution of the central
    rest-mass density shown in Fig.~\ref{fig:tov.rhoc}. Different line
    types and colors refer to the different resolutions used (see
    legend). Shown as vertical dotted lines are the eigenfrequencies
    computed from linear perturbation theory.}
\end{center}
\end{figure}

The dynamics is similar among the three runs, with the central density
first exhibiting a ``burst'' with a variation of the order of $0.12\,
\%$, which is due to the fact that the star is ``cut'' by our
interpolation algorithm of the initial data there where the density
falls below the atmosphere threshold. This phase is magnified by the
inset in the top of the figure. In a second stage the ``burst'' is
quickly damped and the star starts to vibrate radially with its
characteristic eigenfrequencies and eigenmodes under the effects of
numerical perturbations, mainly at the surface of the star.  In a
third phase these oscillations are damped by the numerical viscosity,
which depends on the resolution, and by the crude treatment used to
represent the surface of the star. High-frequency modes are damped
quickly while low-frequency modes have longer damping times. As a
result, towards the end of the simulation we are left only with
slowly-damped sinusoidal oscillations associated with the
$F$-mode. This is particularly evident in the $N = 75$ run, which has
the largest numerical viscosity.

The evolution is stable for all the resolutions that we have
considered, but a careful examination of the behaviour of the central
density reveals the existence of a fourth stage of the dynamics in
some of the runs with lower resolution. More specifically, it is
possible to note that during the third phase of the dynamics numerical
errors accumulate at the surface of the star, producing small
variations in the density profile near the atmosphere.  When these
variations are large enough so that they cannot be controlled by the
employed flattening methods, they are ``released'' and new energy is
pumped into the high-frequency modes, starting a new phase in the
dynamics. This happens at around $45$ ms for the $N = 75$ run in
Fig.~\ref{fig:tov.rhoc}, when a small high frequency component
modulated by the $F$-mode suddenly appears, as shown by the inset at
the bottom of the figure. The same happens around $80$ ms for the $N =
100$ run, while this phenomenon is not observed within the simulation
time for higher resolution runs. Although the onset of this ``energy
release'' at the surface can be delayed with the use of stronger
filtering, it occurs only on a secular timescale and it does not
affect the stability of the star. As a result, the central density
keeps oscillating with a very small amplitude, of the order of
$10^{-3}\, \%$, even if, as discussed in
Sect.~\ref{sec:stabilization}, perturbations at the surface tend to be
greatly amplified when they reach the center as an artifact of the
spherical symmetry. For this reason we believe that the energy release
does not impact the quality of our simulations.

\begin{figure}
\begin{center}
  \includegraphics[width=8.0cm]{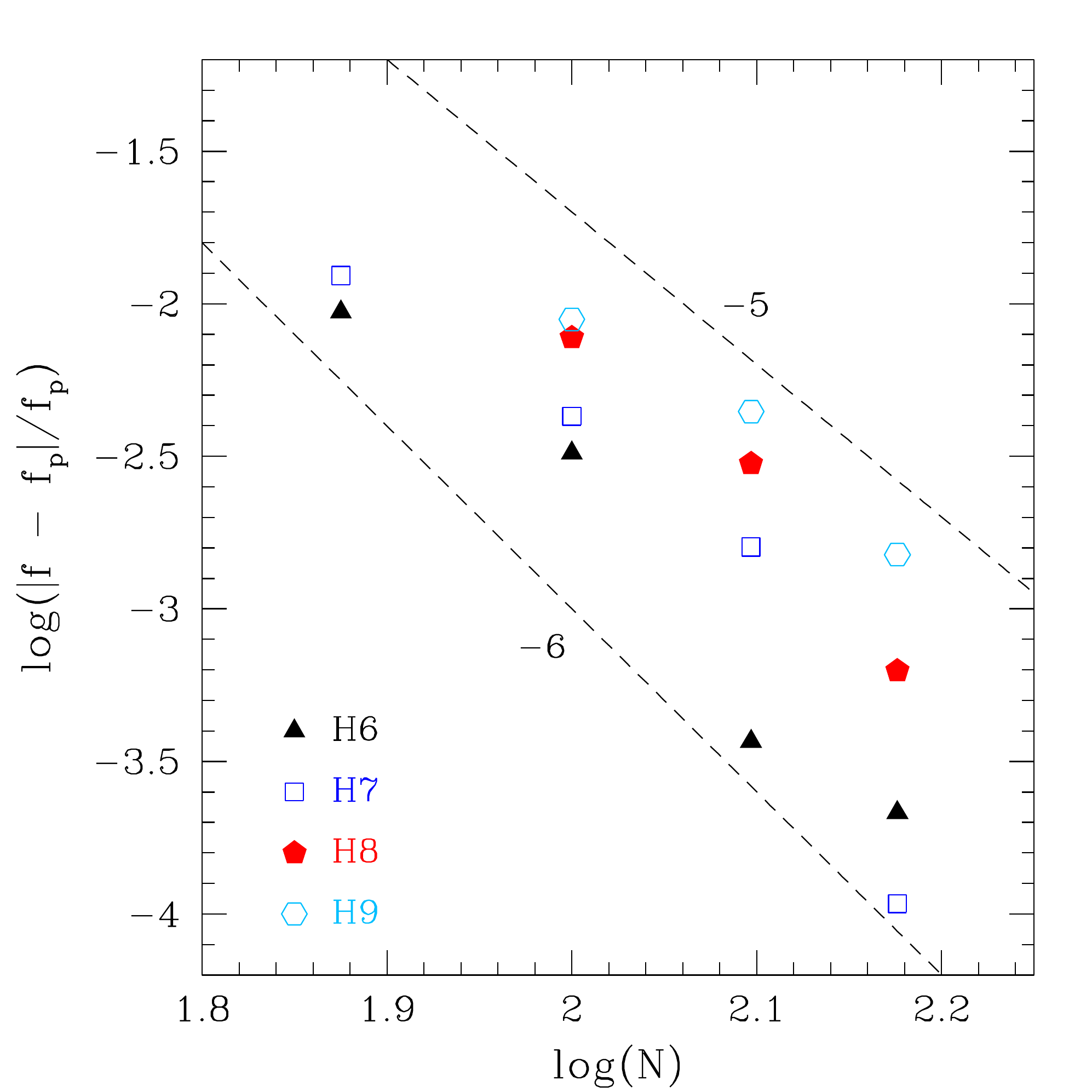}
  \caption{\label{fig:tov.frequencies} Absolute relative difference
    between the estimated eigenfrequencies and the ones computed from
    linear perturbation theory, shown for different modes and as a
    function of the resolution. The lower-order modes are not
    considered because their error is smaller than the nominal one of
    the power spectral density and we report only the values for which
    a reliable measure of the error was possible. Finally, indicated
    as black dashed lines are the slopes associated with fifth and
    sixth-order convergence.}
\end{center}
\end{figure}

A traditional measure of the accuracy of general-relativistic
hydrodynamics codes is the comparison of the power spectrum of the
oscillations of TOVs against the values provided by linear perturbation
theory. In Fig.~\ref{fig:tov.spectrum} we make this comparison by showing
the power spectrum of the first $5\, 000\,M_\odot$ of the evolution of
the central density for different resolutions. The vertical dotted lines
represent the eigenfrequencies computed from perturbation theory, which
were kindly provided us by Kentaro Takami and computed using the method
described in~\cite{Yoshida01}. We note that, even at the lowest
resolution, \texttt{EDGES} shows a perfect agreement between the observed
proper frequencies and the perturbative ones for the $F$-mode and the
first four overtones, $H1, H2, H3$ and $H4$. Furthermore, as the
resolution increases we are able to match more and more modes to the
point that, with the $N = 150$ run, we match the first ten modes to a
very good precision. We also note that the $N = 150$ run gives evidence
of some nonlinear features in the spectrum, which shows considerable
power also in its high-frequency part.

A more quantitative measure of the convergence of the power spectrum
is shown in Fig.~\ref{fig:tov.frequencies}, where we report as a
function of the resolution the relative difference between the
measured frequencies for the overtones $H6, H7, H8$ and $H9$, and the
ones computed from perturbation theory. More specifically, the
numerical frequencies were computed from the data during the first
$5\, 000\,M_{\odot}$ by using the procedure proposed
in~\cite{agrez_2007_dfe}. Namely, we computed the discrete Fourier
transform (DFT) of the data with an Hanning window and used a three
point interpolation of the power spectrum to correct for the
incoherency error and determine the correct eigenfrequencies (this
procedure is conceptually equivalent to a Lorentzian fit of the peaks
of the DFT). Note that the lower-order modes are not shown because
their precision is such that the error on those frequencies is well
below the nominal uncertainty of the DFT and is basically dominated by
the error on our measure.  (The values of the error is not shown for
those resolutions/frequencies that could not be determined in a
reliable way: namely, the $H8$ and $H9$ modes for the $N = 75$
run.). Overall, we find that the measured convergence rate depends
somehow on the specific mode, but it is compatible with a fifth-order
convergence rate (\cf the fifth and sixth-order convergence rates
which are shown as dashed lines).  As we will comment later on, this
results indicates that our largest source of error on the frequencies
is not dominated by the low-order FV approach used at the surface of
the star.  Rather, the behaviour in Fig.~\ref{fig:tov.frequencies}
shows that in the case of global quantities such as the oscillation
frequencies of a TOV, the treatment of the surface is not the most
critical element of the ``error budget'' in
\texttt{EDGES}\footnote{Note that the rapid decrease of the error in
  the estimate of $H6$ and $H7$ at high resolution is most probably
  due to the nonlinear effects mentioned above which concentrate power
  in these higher-order modes, making them sharper and better
  resolved.}.

\begin{figure}
\begin{center}
  \includegraphics[width=8.0cm]{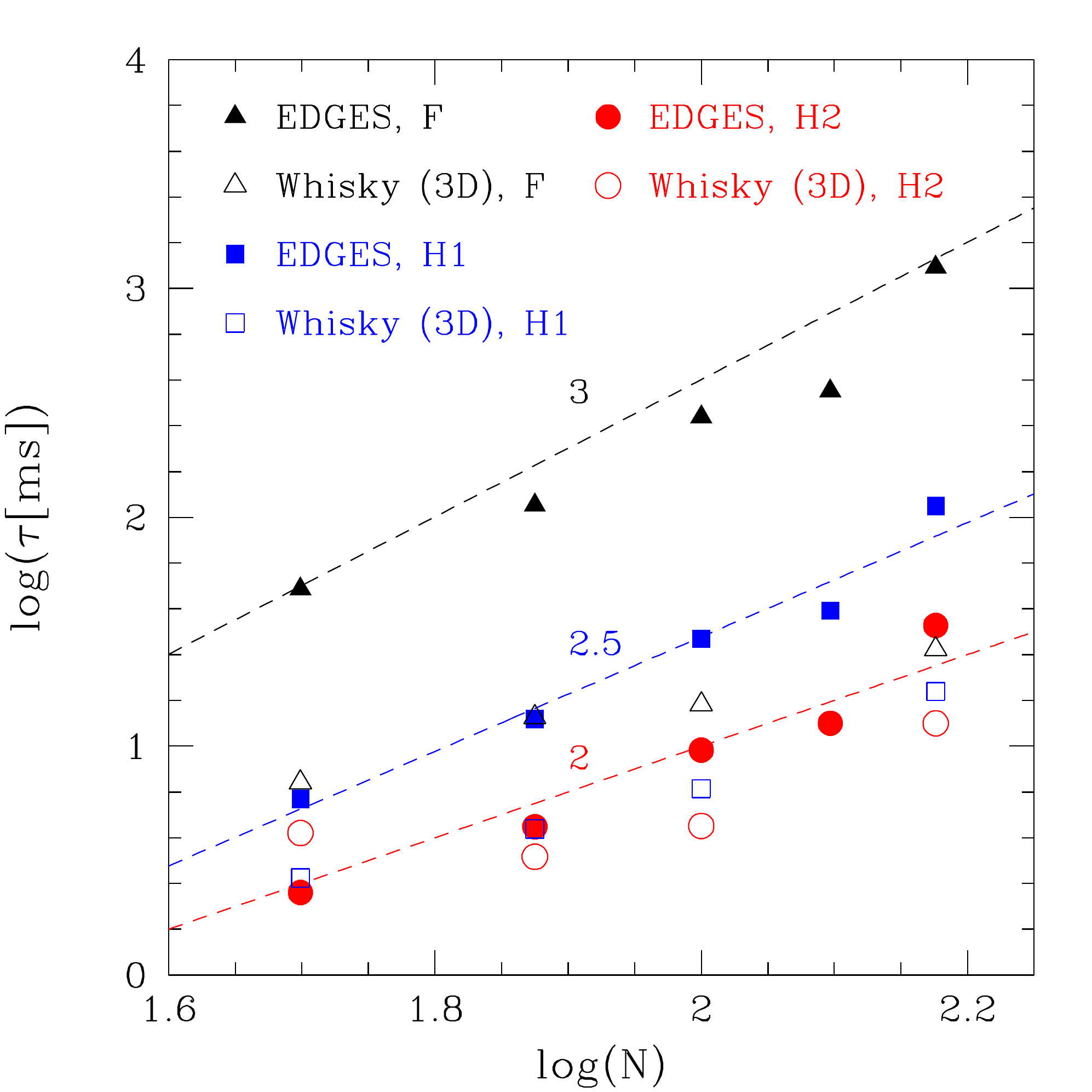}
  \caption{\label{fig:tov.damping} Estimated damping rates for various
    eigenmodes as a function of the resolution as computed by
    \texttt{EDGES} in 1D and by \texttt{Whisky} in 3D. Black
    triangles, blue squares and red circles refer respectively to the
    $F\,, H1$ and $H2$ modes, while filled/hollow points distinguish
    the estimates with \texttt{EDGES} from those with
    \texttt{Whisky}. Finally the dashed lines show the estimated
    convergence order.}
\end{center}
\end{figure}

As a concluding consideration we note that in a recent work
Cerd\'a-Dur\'an~\cite{cerda-duran_2009_nvh} has proposed to measure
the numerical bulk viscosity of general-relativistic hydrodynamical
codes by looking at the damping time of the $F$-mode in the case of
oscillating TOVs. In particular, the rate of change with resolution of
the damping time (and which clearly increases with resolution) can be
used as a measure of the convergence rate of the code. To explore this
interesting suggestion we have computed the damping time by analysing
more systematically the evolution of the central density. More
specifically, using a sampling frequency of $\approx 1\,M_{\odot}$, we
have built a discrete signal which was then divided into chunks of
$512$ points with an overlap of $128$ points. A DFT of each chunk was
then computed using an Hanning window and the power of the signal at
the frequency associated with each mode was computed with a linear
interpolation of the absolute value of the DFT. In this way we
obtained an estimate of the energy in the mode for each time
window. Finally we performed a least square fit of an exponential
function to determine the damping time. (We have estimated the
accuracy of this procedure to be better then $10\%$, on the basis of
tests performed with signals of known spectral properties.)

In Fig.~\ref{fig:tov.damping} we show the results of this measure when
made on different modes. The data for the $N = 50\,,75$ and $N = 100$
runs have been obtained by windowing the evolution in the interval $0
\leq t \leq 5\, 000$ to avoid spurious values due to the fourth phase of
the dynamics described above, while for the $N = 125\,, 150$ runs it was
necessary to use the full data in the interval $0 \leq t \leq 20\, 000
\approx 100\ \mathrm{ms}$ to obtain a reliable estimate of the damping
time for the $F$-mode. As a comparison, we also report the results
obtained in 3D using the~\texttt{Whisky} code~\cite{Baiotti04,
Giacomazzo:2007ti, Pollney:2007ss, Schnetter-etal-03b}, when computed
using a similar procedure. The simulations with \texttt{Whisky} were done
using a PPM reconstruction and the HLLE Riemann solver for the
hydrodynamics and using a fourth-order FD scheme for the evolution of the
spacetime on a uniform grid also covering $0 \leq r \leq
15$\footnote{Note that, since \texttt{Whisky} and \texttt{EDGES} use
different gauges, the two resolutions are only roughly equivalent.}.

We find that the convergence order of the damping time of the $F$-mode
with \texttt{Whisky} is one, in agreement with the results reported
by~\cite{cerda-duran_2009_nvh} for the \texttt{COCONUT}
code~\cite{Dimmelmeier02a, Dimmelmeier05a}.
In~\cite{cerda-duran_2009_nvh} it was argued that the reason why the
order reduces to first is that the damping is active mainly at the
surface of the star, where the numerical methods are only first order.
The results found with \texttt{EDGES} show however a different behaviour,
with the order of the damping being $3, 2.5$ and $2$ for the $F$, $H1$
and $H2$ modes, respectively. Because the treatment of the surface in
\texttt{EDGES} can be seen as a variant of the FV method used by both
\texttt{Whisky} and \texttt{COCONUT} and is therefore only first-order
accurate, our results suggest that the coefficient of the first-order
(and surface-induced) error is much smaller than the coefficient of the
error coming from the high-order filtering procedure, which then becomes
the dominant source of damping. This is in agreement with the high
convergence order found in the measurement of the eigenfrequencies (\cf
Fig.~\ref{fig:tov.frequencies}).

\begin{figure}
\begin{center}
  \includegraphics[width=8.0cm]{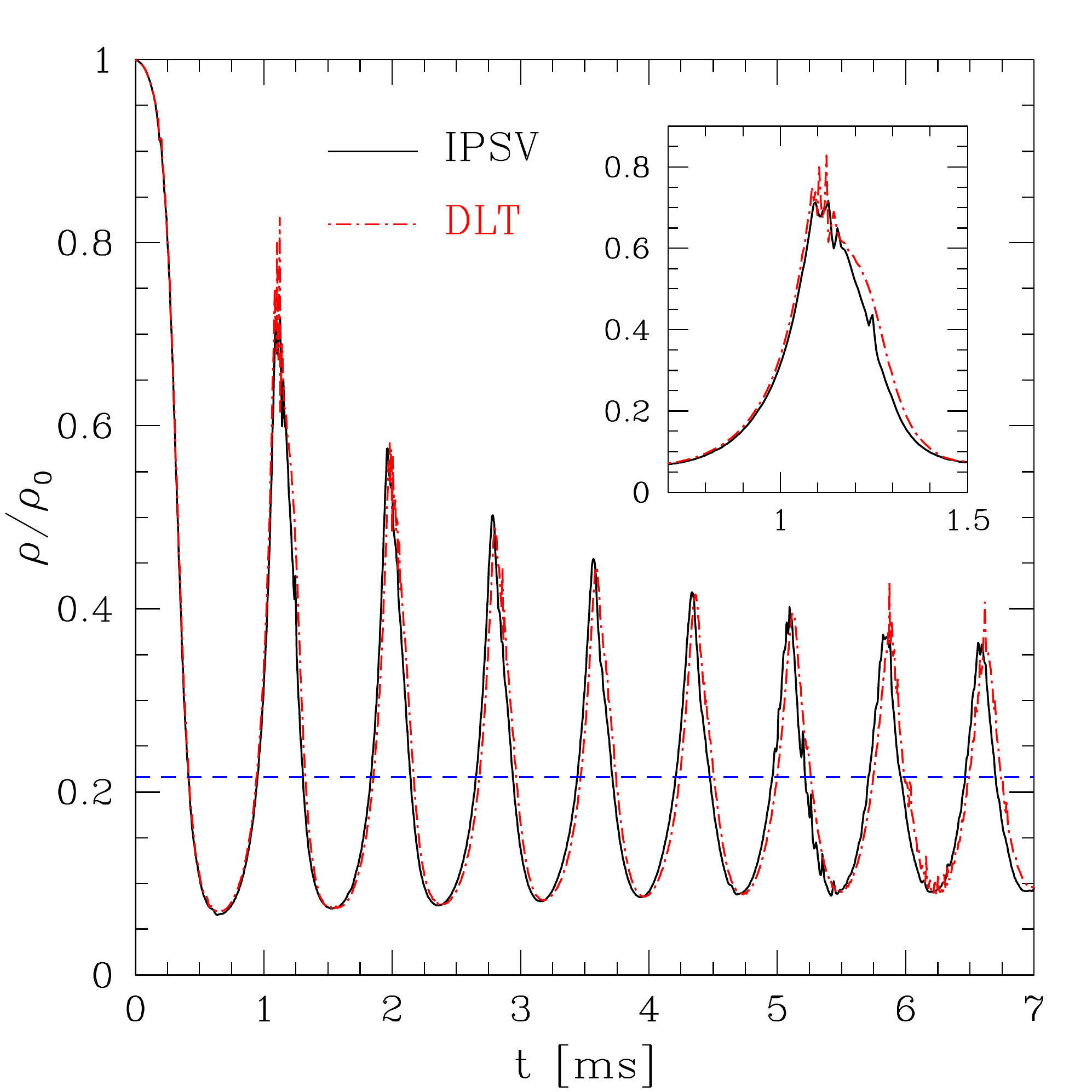}
  \caption{\label{fig:tov.migration.rho} Evolution of the normalized
    central rest-mass density of an unstable TOV migrating towards the
    stable branch for a resolution of $N=150$.  The black solid line
    refers to a run employing an IPSV stabilization, while the red
    dashed line refers to a run employing a DLT filter; the blue
    horizontal line denotes the central rest-mass density of the
    stable model associated with the initial configuration. The inset
    shows a magnification of the dynamics around $1 \, \mathrm{ms}$,
    at the first peak in the central density.}
\end{center}
\end{figure}

As a side comment we want to point out that we are able to attain
higher-than-second order convergence in our results, because the
time-step that we use is small enough so that the errors in the spatial
discretization are dominating over the errors due to the time evolution,
which is only second order.

\subsection{Nonlinear oscillations of spherical stars: the migration test}

As a direct extension of the analysis carried in the previous Section on
linear oscillations, we next study large nonlinear oscillations which are
produced as an equilibrium star model on the unstable branch of
equilibria models migrates to the stable branch. This process has been
used as a numerical test in 3D codes (see,
\eg~\cite{Font02c,Baiotti03a,Baiotti04,Cordero2009}), has been analyzed
extensively in the past~\cite{gourgholon_1992_nra,
novak_2001_vic,noble_2003_nsr} and has gained special interest recently
when it was shown that it exhibits a critical
behaviour~\cite{liebling_2010_emr, Radice:10}.

\begin{figure}
\begin{center}
  \includegraphics[width=8.0cm]{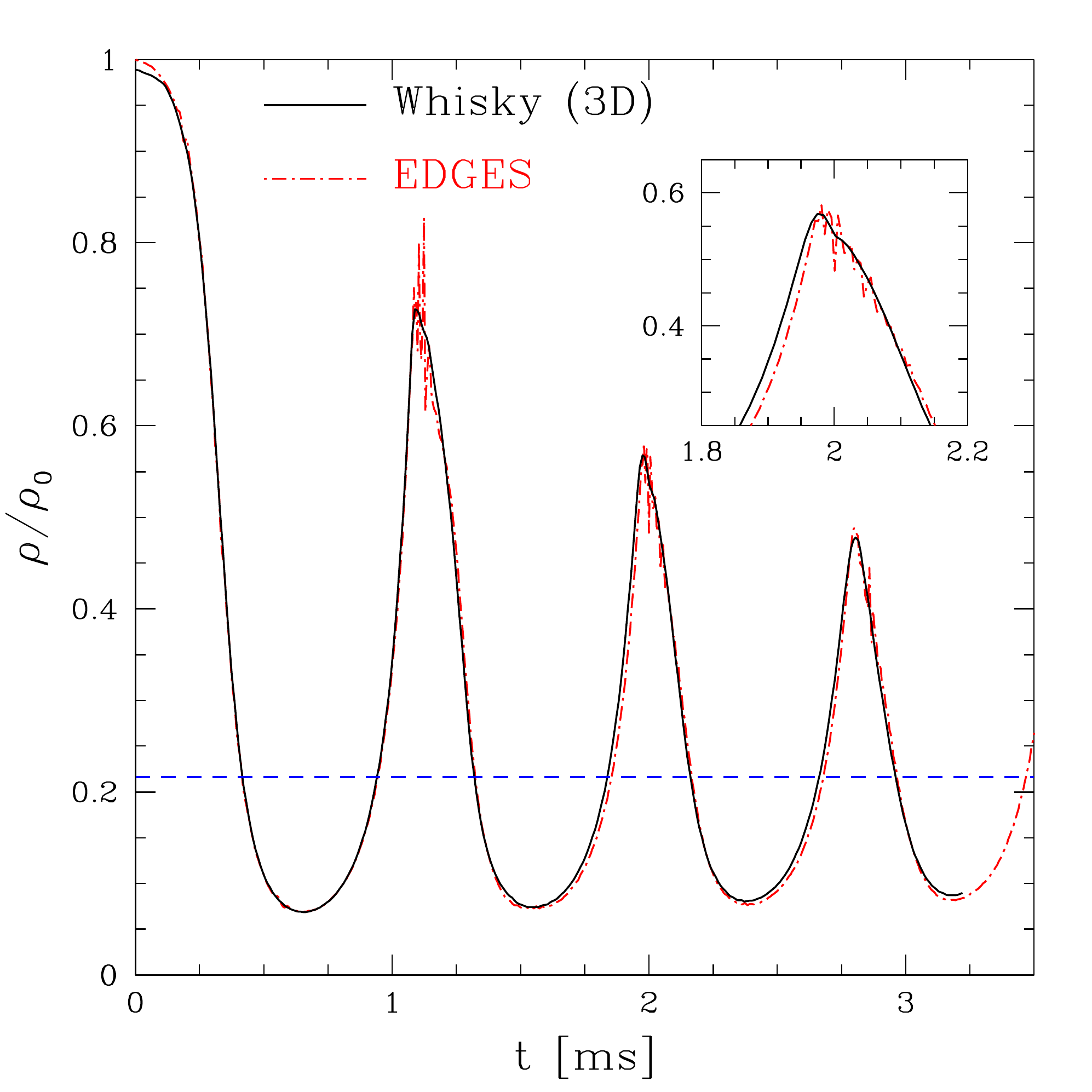}
  \caption{\label{fig:tov.migration.edges.vs.whisky} Comparison in the
    evolution of the normalized central rest-mass density of the
    migrating star between a simulation in 1D with \texttt{EDGES}
    (dot-dashed red line), and a simulation in 3D with \texttt{Whisky}
    (solid black line). Also in this case, the blue dashed line
    denotes the central rest-mass density of the stable model
    associated with the initial configuration and the inset shows a
    magnification of the second peak (see main text for details).}
\end{center}
\end{figure}

Here, in particular, we have considered a TOV constructed with a
polytropic EOS having $K = 100$ and $\Gamma = 2$, central density
$\rho_c = 7.0 \times 10^{-3}$, gravitational mass $M =
1.49\,M_{\odot}$ and areal radius $R = 6\,M_{\odot} \simeq 8.8$ km.
The evolution, on the other hand, was made with an ideal-fluid EOS to
properly take into account shock-heating effects. The numerical grid
covers the region $0 \leq r \leq 30$ and the simulations reported used
a polynomial representation of the solution of degree five. Two
different stabilization techniques were used. A first one employed an
exponential filter of order six with $\mu = 40$, applied to the DLTs
of the conserved variables and corrected with the ``intrinsic''
procedure. A second one used instead an IPSV stabilization with
$\hat{Q}_k = 1 - \delta_{k0}$ and strength $\mu = 1$.

To trigger the migration on the stable branch, the star is perturbed
with an \textit{outgoing} velocity perturbation of the form
\begin{equation}\label{eq:vel.perturbation}
  v(x) = \frac{A}{2} | x^3 - 3 x |\,,\quad x = \frac{r}{R}\,,
\end{equation}
where $A = 0.01$. Under the effect this perturbation the star exhibits
a violent expansion and migrates towards a new stable equilibrium
configuration with a series of large-amplitude oscillations. During
these violent oscillations the exterior layers of the star tend to
infall with higher velocity then the interior layers and this leads to
the formation of shock waves that heat the neutron star matter and
result in the ejection of a small portion of the material of the star.

In Fig.~\ref{fig:tov.migration.rho} we show the evolution of the
central density, normalized to its initial value, for two runs
employing $150$ elements and different stabilization
techniques. Because this test does not have an analytic counterpart,
we have compared it the corresponding evolution performed with the
\texttt{Whisky} code in a 3D simulation having $N = 100$ grid cells in
each direction, PPM reconstruction and the HLLE Riemann solver. 

\begin{figure}
\begin{center}
  \includegraphics[width=8.0cm]{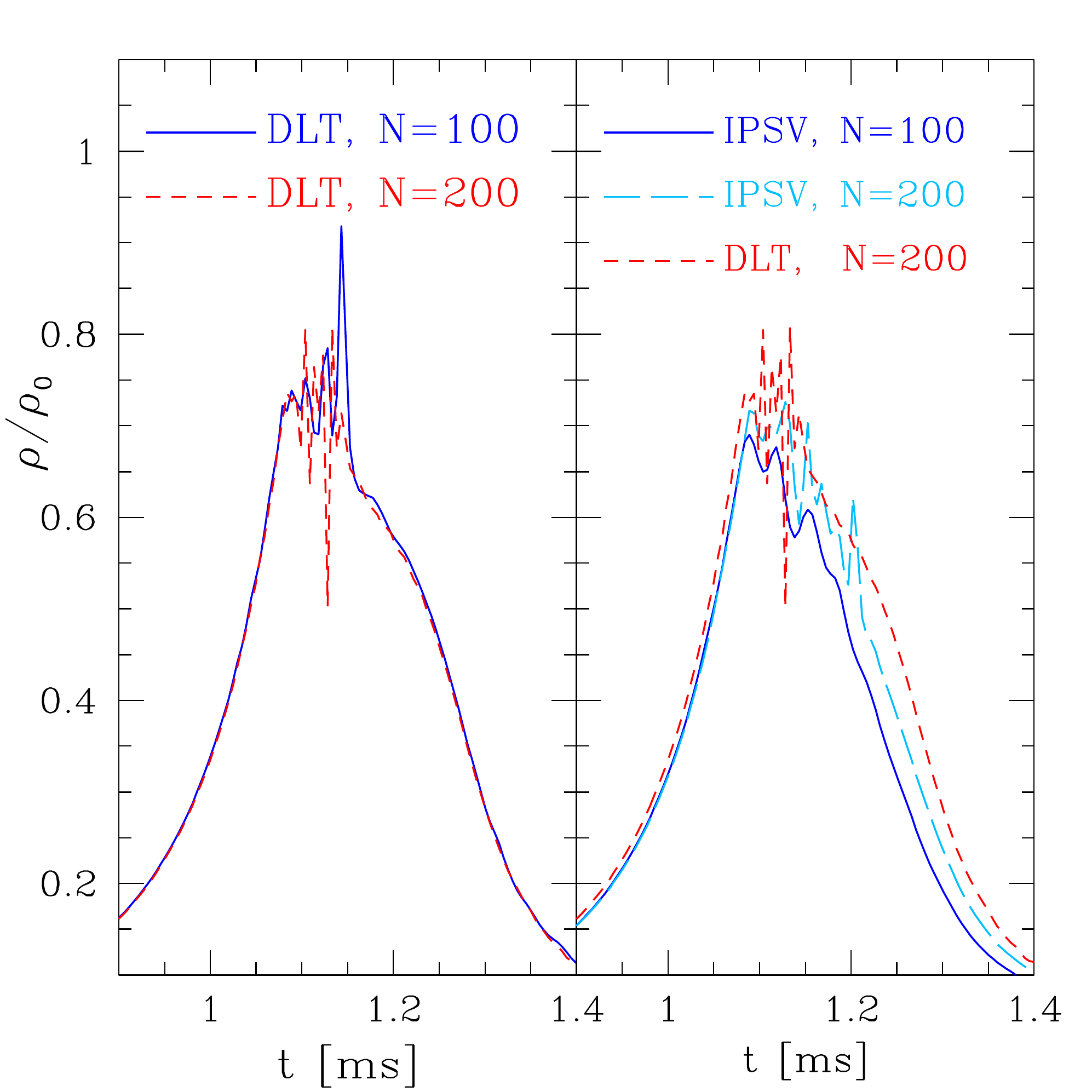}
  \caption{\label{fig:tov.migration.ipsv.vs.exp} Comparison of
    different stabilization techniques in the migration test. Shown in
    the two panels is the evolution of the central rest-mass density
    around the first peak. The left panel refers to a DLT filter on
    grids of $N=100$ and $N=200$ elements (blue solid and dashed red
    lines), while the right panel refers the IPSV stabilization
    technique with $N=100$ and $N=200$ elements (blue solid and
    cyan long-dashed lines). Note that the DLT-stabilized run reaches
    convergences with a smaller number of elements.}
\end{center}
\end{figure}

The comparison is offered in Fig.~\ref{fig:tov.migration.edges.vs.whisky}
and shows an extremely good agreement. We note that because the two codes
have intrinsically different initial truncation errors, the expansion
phase in \texttt{Whisky} is slightly delayed with respect to the dynamics
produced by \texttt{EDGES}, and we account for this difference by
shifting the data obtained by \texttt{Whisky} in order to obtain an
approximate alignment at the first minimum of the density.
Notwithstanding the very good agreement, the main difference between the
two solutions is the presence of high-frequency oscillations in the
central density computed by \texttt{EDGES}, and in particular for the DLT
runs and near the maximum value of the density. These ``spikes'' can be
tracked down to the propagation of shock waves which are formed in the
outer layers of the star during the collapse phase. They have initially a
small compression factor, but they also tend to sum up coherently as they
travel towards the center, thanks to the assumption of spherical
symmetry, so that they result in strong variation of the density near the
center of the star. This phenomenon has also been observed in other works
in spherical symmetry, employing standard finite-volume schemes, see
\eg~\cite{Cordero2009}, but is not usually observed in 2D or 3D
simulations and, indeed, it is not present in the results obtained with
\texttt{Whisky}. In particular \texttt{Whisky} employs Cartesian
coordinates, which inevitably introduce preferred directions in the grid,
thus preventing the spherical waves produced at the surface from
interfering constructively and focussing towards the center. Another
reason why these spikes are not observed in multi-dimensional solutions
could be related to the larger numerical viscosity due to the necessity
of using a much coarser resolution than in the 1D case. This is confirmed
by the fact that the use of the IPSV stabilization techniques, which
allows us to introduce a much larger numerical dissipation, is able to
greatly suppress this phenomenon.

\begin{figure}
\begin{center}
  \includegraphics[width=8.0cm]{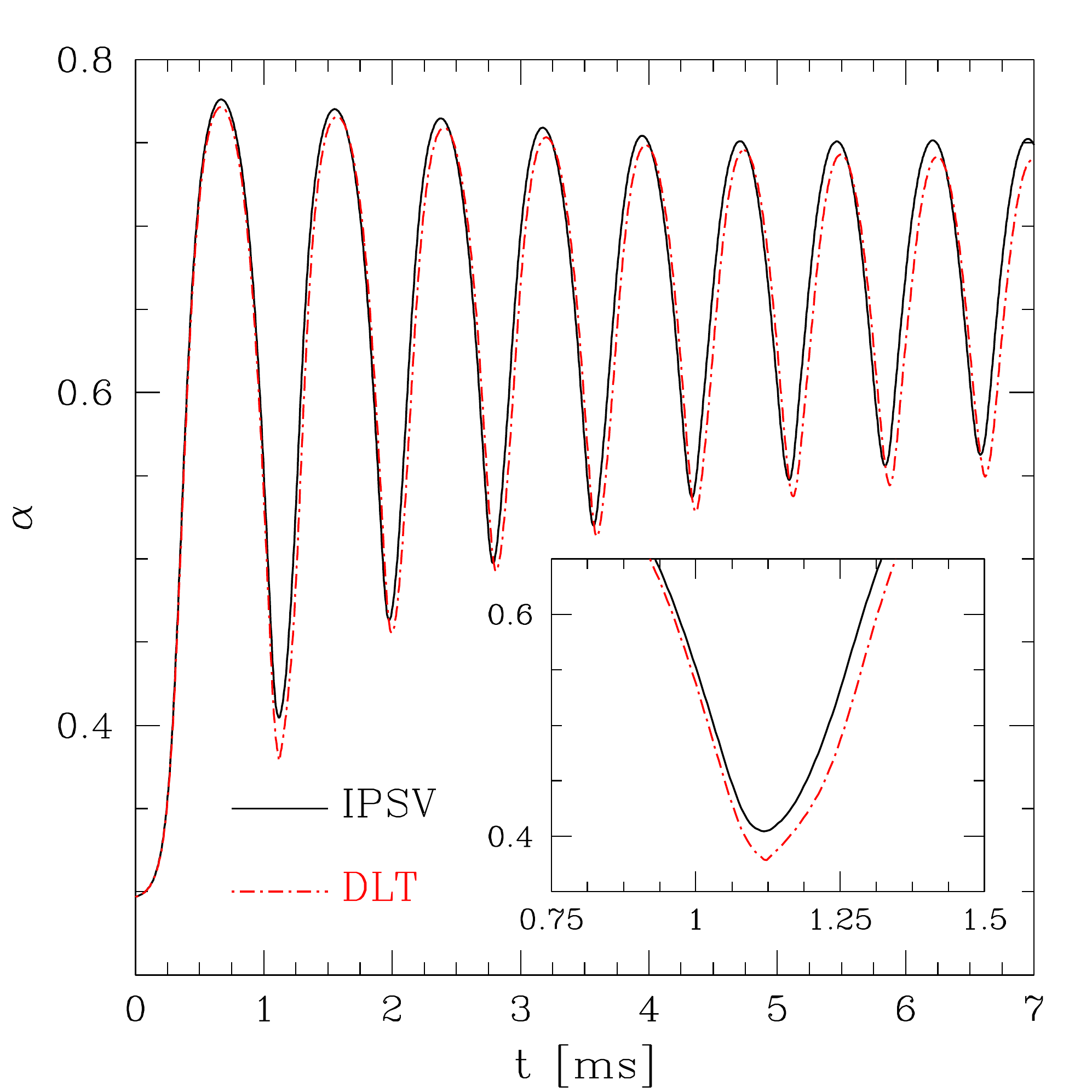}
  \caption{\label{fig:tov.migration.alp} Evolution of the lapse
    function at the center of an unstable TOV migrating towards the
    stable branch for two different runs using different stabilization
    techniques (black solid line for the IPSV stabilization, and red
    dashed line for the DLT filter) and $N=150$. The inset shows a
    magnification of the dynamics around the first peak at $1 \,
    \mathrm{ms}$ and highlights that the solution is always smooth.}
\end{center}
\end{figure}

A more detailed comparison between the results obtained with the two
stabilization techniques is shown in
Fig.~\ref{fig:tov.migration.ipsv.vs.exp}, where we offer a comparison
between the evolution of the central density around the first bounce
obtained with the two approaches and with different resolutions. Note
that the DLT runs show signs of spikes in the density for all the
resolutions (left panel), while the IPSV runs are much smoother (right
panel). On the other hand, the results obtained with the DLT filters
are already in a convergent regime and in fact the results obtained
with $N=100$ elements are very similar to those obtained by doubling
the resolution. This is not the case for the solutions obtained with
the IPSV filter, that seem to be only slowly approaching the DLT ones
(\cf note in the right panel that the IPSV solutions approaches the
DLT one as $N$ goes from $100$ to $200$). The evidence that the
simulations with the DLT filtering are already in a convergent regime
and yet show spikes in the evolution can be interpreted, therefore, as
a confirmation that the latter are not a numerical effect, but rather
an artifact of the symmetry, which leads to a focusing of the waves
travelling towards the center.

Fortunately, because these perturbations actually carry only a very
small energy and are amplified by the symmetry, the evolution of the
spacetime variables is totally unaffected.  This is shown in
Fig.~\ref{fig:tov.migration.alp}, where we report the evolution of a
particularly representative metric quantity, namely, the lapse
function at the center. As can be seen from the figure, the lapse
function shows no spikes or spurious oscillation and appears to be
smooth with both the DLT and the IPSV filters, the main difference
being the value of the minimum attained during the first bounce with
the former stabilization technique.

\subsection{Gravitational collapse of unstable spherical stars}

As a final test we consider another classical testbed in
general-relativistic hydrodynamics: namely, the gravitational collapse to
black hole of an unstable TOV star. The problem of the gravitational
collapse to a black hole has been already studied in great detail in a
number of different conditions involving 1D, 2D and 3D simulations, as
well as different physical conditions (see,
\eg~\cite{gourgholon_1992_nra, Romero96, novak_2001_vic, noble_2003_nsr,
Baiotti04, Duez:2005cj, Baiotti06, Baiotti07, Noble08a, Cordero10,
OConnor10, Thierfelder10}), and has become a standard test of
general-relativistic codes. For this reason, we will not discuss the
details of the dynamics of the collapse and concentrate instead on the
quality of the results obtained with \texttt{EDGES}.

For the tests considered here we have evolved an unstable TOV built
with a polytropic EOS with $K=100$ and $\Gamma = 2$, having central
density $\rho_c = 4.5\times 10^{-3}$, gravitational mass $M =
1.6\,M_{\odot}$ and areal radius $R = 6.9\,M_{\odot} \simeq 10.2$
km. In contrast to the migration test, the collapse is triggered by
introducing an \textit{ingoing} velocity perturbation of the type
(\ref{eq:vel.perturbation}) with $A=-0.01$. The evolution of this
system is studied on a grid covering $0 \leq r \leq 15$, staggered
about the origin, using polynomials of degree five and employing an
exponential filter of order six, with intrinsic correction and
strength $\mu = 40$, applied directly on the conserved variables at
every time step.

\begin{figure}
\begin{center}
  \includegraphics[width=8.0cm]{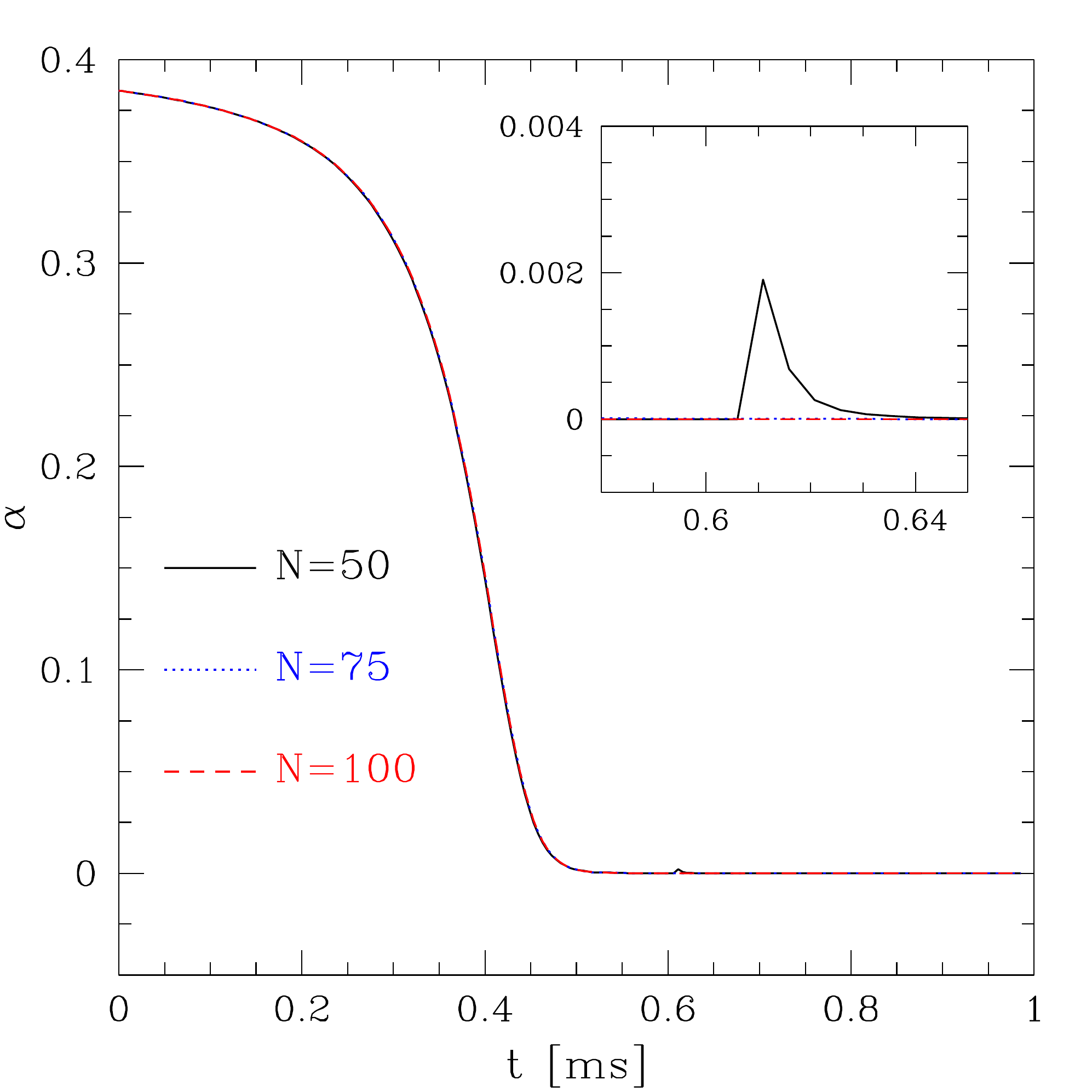}
  \caption{\label{fig:tov.collapse} Evolution of the lapse function at
    the center of an unstable TOV collapsing to a black-hole for
    different resolutions on. The solid black line, the dotted blue
    line and the dashed red line are associated with runs employing a
    grid of $50$, $75$ and $100$ elements, respectively. The inset
    shows a magnification of the dynamics around $0.6\,\mathrm{ms}$,
    highlighting some post-collapse dynamics in the lapse for the
    low-resolution run, which is however absent at higher
    resolutions.}
\end{center}
\end{figure}

During the evolution and as expected, the central density of the star
shows an exponential increase, which halts when the lapse function
collapses to zero, signalling the formation of the black hole and
``freezing'' the evolution in the inner regions of the star. The
maximum value attained by the central density does not have any
physical meaning, partly because it depends entirely on the gauge
conditions and partly because the gauge used will prevent its growth
past a certain value. For this reason it is more interesting to look
at the evolution of the lapse function at the center of the star, and
which we show in Fig.~\ref{fig:tov.collapse}. Note that for all the
considered resolutions, the lapse function quickly collapses to zero
and shows no appreciable subsequent evolution, with the exception of
the lowest resolution case, for which a small growth appears again
around $0.6\ \mathrm{ms}$. This sudden evolution is the result of
errors coming from the surface of the frozen star, where the spatial
slice is stretched due to the fact that we fixed our shift gauge
condition to zero and, thus, the metric functions present large
spatial gradients. These errors, which cannot be compensated by simply
increasing the resolution and are a shortcoming of the gauge used,
propagate towards the inner regions of the frozen star and can induce
a growth of the lapse, restarting the dynamics. This is clearly a
resolution-dependent effect which disappears quickly by increasing the
resolution (see inset in Fig.~\ref{fig:tov.collapse}). Apart from this
late-time, and low-resolution dynamics, the three curves are on top of
each other, signaling a convergent regime, with a rate we measure to
be slightly above the fourth order before the collapse of the lapse.
Afterwards the convergence order starts to slowly decay, due to the
fact that, as the lapse collapses to zero, the equations are only
weakly hyperbolic. At the end of the simulation the convergence order
reaches a value which is $\gtrsim 1.5$.

In summary, the solution of the gravitational collapse to a black hole
with \texttt{EDGES} has been straightforward and indeed a test which
is much less demanding that the migration one, at least in terms of
the hydrodynamical variables.  The only difficulty encountered is a
well known one and has to do with the fact that with such a gauge the
collapse of the lapse is not compensated by any change in the shift
vector, which is identically zero. As a result, as the black hole is
produced it quickly produces a stretching of the coordinates at the
location immediately outside the ``frozen star''. This, in turn,
results in the development of strong gradients in the numerical
solution which cannot be resolved indefinitely without suitable
adaptivity.

\section{Conclusions}
\label{sec:conclusions}

Numerical, relativistic hydrodynamics and MHD have seen a tremendous
growth in the number and the quality of its results after the
introduction of modern high-resolution shock capturing
schemes~\cite{Font08}. These schemes have been proven to be of central
importance in the modelling of complex systems involving strong
gravity and/or high Lorentz factor. Yet, they suffer from important
limitations that ultimately impact the accuracy of the obtained
results~\cite{Baiotti:2009gk}. For this reason the search for better
numerical schemes is still on-going (and will always be).

Discontinuous Galerkin schemes were developed to overcome the
previously mentioned limitations of finite-volume and
finite-difference schemes, while maintaining important properties of
these schemes, such as conservation and shock capturing, that made
them so successful in a number of
applications~\cite{cockburn_2001_rkd}. For this reason they are a
natural candidate as an alternative to more traditional methods also
in relativistic hydrodynamics.

In this paper we have developed the necessary mathematical framework
needed for the application of discontinuous Galerkin schemes to
relativistic hydrodynamics in curved spacetimes. In particular, we
have presented both a manifestly covariant weak formulation of
relativistic hydrodynamics and a more traditional one obtained within
a $3+1$ split. We have then specialized the latter formulation to the
spherically symmetric case and implemented it in a new one-dimensional
relativistic hydrodynamical code, \texttt{EDGES}, which uses a
high-order spectral discontinuous Galerkin method.

The code was tested in a number of situations, including shock waves,
spherical accretion, linear and nonlinear oscillations of relativistic
spherical stars and the gravitational collapse of unstable stars to
black holes. Our results show that discontinuous Galerkin methods are
able to sharply resolve shock waves and, at the same time, attain very
high, spectral, accuracy in the case of smooth solutions. For this
reason they constitute an excellent alternative to classical
finite-volume and finite-difference schemes for relativistic
hydrodynamics, especially in those situations in which shock waves as
well as small-scale features of the flows have to be resolved. In
light of the promising prospects shown with these tests and of their
affinity with a pseudospectral solution of the Einstein equations, we
anticipate that discontinuous Galerkin methods could represent a new
paradigm for the accurate modelling in computational relativistic
astrophysics.

As a final remark we note that the success of these methods in
multi-dimensional implementations will ultimately depend on the
development of techniques such as local timestepping, $hp$-adaptivity
and load balancing~\cite{canuto_2008_sme}, which are needed to take
full advantage of the flexibility of these schemes~\cite{Shu01}. This
will represent the focus of our future work.

\begin{acknowledgments}
We wish to acknowledge Alessandro Proverbio for his work on the
\texttt{MPI-Spectral} library upon which some of the low-level data
structure of \texttt{EDGES} are based. We are also grateful to Olindo
Zanotti for many useful discussions and for providing the numerical
solution to the second shock-tube test obtained with the \texttt{ECHO}
code and to Kentaro Takami for providing accurate perturbative
eigenfrequencies for the TOV model we evolved. It is also a pleasure
to thank Aaryn Tonita, Filippo Galeazzi, Luca Baiotti, Pablo
Cerd\'a-Du\'ran, Toni Font, Bruno, Giacomazzo, Ian Hawke, Jos\'e Mar\'ia
Ib\'a\~nez, and Jos\'e Mar\'ia Marti for useful comments. This work is
supported in part by the DFG grant SFB/Transregio~7 and by
``CompStar'', a Research Networking Programme of the ESF.
\end{acknowledgments}

\appendix

\section{Causal triangulations}
\label{sec:caus_triang}

In what follows we provide some basic definitions related with the causal
structure of the triangulation. 

For every \emph{open set} $A$, we introduce the notation $\partial^+
A$, to indicate the \emph{future boundary} of $A$, that is the set of
all the points, $p \in \partial A$, such that
\begin{equation}
  J^+(p) \cap A = \emptyset\,,
\end{equation}
$J^+(p)$ being the causal future of $p$. Analogously we call \emph{past
boundary} of $A$, $\partial^- A$, the set of all the points $p \in
\partial A$ for which there exist an open neighborhood, $U$, such that
\begin{equation}
  J^+(p) \cap U \subset \overline{A}\,,
\end{equation}
where we have indicated with $\overline{A}$ the closure of $A$.
Finally we will write $\partial^\times A$ for $\partial A \setminus
[\partial^+ A \cup \partial^- A]$. 

\begin{figure}
\begin{center}
  \includegraphics[width=8.0cm]{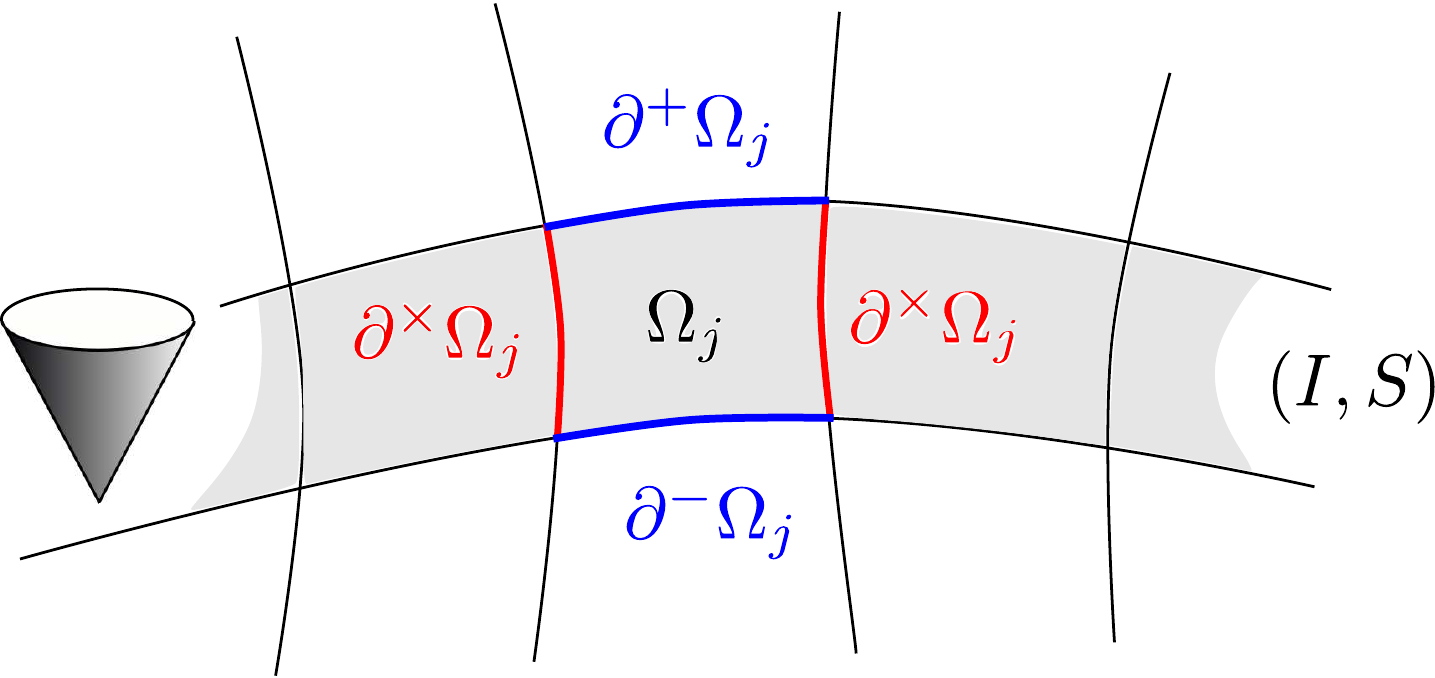}
  \caption{\label{fig:spacetime.dg} Schematic representation of a causal
  slice $(I,S)$ containing the element $\Omega_j$. The future and past
  boundaries $\partial^{\pm} \Omega_j$, as well as the time-like or
  null-like boundaries $\partial^{\times} \Omega_j$ of $\Omega_j$
  are indicated with blue and red lines, respectively. Finally, the
  shaded region represent the causal slice.}
\end{center}
\end{figure}

For a \emph{generic set} $E$, instead, the above definitions are
modified by considering the interior of the set, $\mathring{E}$. In
the case of an element of the grid, $\Omega_j$, these definitions are
useful to identify the regions of the boundary of $\Omega_j$ where the
characteristics are always ingoing, $\partial^- \Omega_j$, always
outgoing, $\partial^+ \Omega_j$, or of mixed nature, $\partial^\times
\Omega_j$. These different parts of the causal slice $(I,S)$
containing the element $\Omega_j$ are shown in
Fig.~\ref{fig:spacetime.dg}.

We also define as \emph{slice of the triangulation}, or simply slice,
any tuple $(I,S)$ where $I\subset\{1,2,\ldots,N\}$ and $S = \bigcup_{j
  \in I} \Omega_j$ is connected. We will say that a slice $(I,S)$ is a
\emph{causal slice} if
\begin{equation}
  \partial^\times S \subset \partial \Omega \,.
\end{equation}

An example of such a slice of the triangulation is given by the shaded
region in Fig.~\ref{fig:spacetime.dg}: a causal slice is basically a
slice whose timelike spatial boundaries $\partial^+$ and $\partial^-$
are parts of a Cauchy foliation of the spacetime. A causal slice
$(I,S)$ is said to be a \emph{minimal causal slice} if it also
satisfies the CFL-like condition
\begin{equation}
  \Bigg[\bigcup_{j\in I} \Big(\partial^\times \Omega_j
    \cup \partial^- \Omega_j\Big) \setminus \partial \Omega_k \Bigg]
    \cap J^+(\Omega_k) = \emptyset, \ \ \forall k \in I,
\end{equation}
that is, if the characteristics originating in each element, $\Omega_k$,
only intersect the boundaries of the element itself, $\partial\Omega$, or
the future boundaries of the other elements in the slice.

Finally we say that a triangulation \emph{follows the causal structure
  of the spacetime} if it can be written as union of minimal causal
slices.  Intuitively a triangulation that follows the causal structure
of the spacetime is one that can be sliced in minimal causal slices,
which are grids associated with a Cauchy foliation of the spacetime
and which satisfy a CFL condition on each hyper-surface of this
foliation.

If the triangulation $\{\Omega_j\}_{j=1}^N$ follows the causal
structure of the spacetime, in the sense defined above, then we can
introduce a family of successive slices $\{(I_k,S_k)\}_{k=0}^Q$,
parametrized by a discrete time $k$, such that $\partial^- S_{k+1}
\subset \partial^+ S_k \cup \partial \Omega$ and $\Omega = \bigcup_{k}
S_k$. We can then construct a globally explicit, locally implicit,
scheme by solving (\ref{eq:spacetime.dg}) on the successive
slices. The reason is that, for all $j \in I_{k+1}$, the fluxes across
$\partial^-\Omega_j$ and $\partial^\times\Omega_j$ are completely
determined by the data on $S_k$ and the boundary conditions, while the
fluxes across $\partial^+\Omega_j$ do not couple $\Omega_j$ with other
elements within $S_{k+1}$.

\section{The continuity equation in the $D=3$ case}
\label{sec:example}

We next present an example in which we show in the explicit form the
discontinuous Galerkin scheme for the simplest of the equations of
relativistic hydrodynamics, namely the continuity equation, and with a
polynomial of order $D=3$. In particular we will show the explicit
form of the semi-discrete equations for the collocated values in a
given element $S_j$. In the \texttt{EDGES} code, nodes, weights and
interpolating polynomials are always computed numerically. However,
for clarity, we are going to present, here, the analytic expressions
for the considered case.

The Legendre polynomial of degree three in $[-1,1]$ is given by
\[
  L_3(x) = \frac{1}{2}(5x^3 - 3x),
\]
so that the Gauss-Legendre-Lobatto quadrature points, the zeros of
$(1-x^2){\dd L_3(x)}/{\dd x}$, are
\[
  \{x_i\}_{i=0}^3 = \{-1,\, -1/\sqrt{5},\, 1/\sqrt{5}, 1\}.
\]
The Lagrange polynomials associated with the quadrature points are
\begin{align*}
  l_0(x) &= \frac{1}{8} (1-x) (-1 + 5 x^2)\,,\\
  l_1(x) &= \frac{5}{8} (-1+x)(1+x)(-1 + \sqrt{5} x)\,, \\
  l_2(x) &= -\frac{5}{8} (-1+x)(1+x)(1+\sqrt{5}x)\,, \\
  l_3(x) &= \frac{1}{8} (1+x) (-1 + 5 x^2)\,,
\end{align*}
and the associated weights, $w_i = \int_{-1}^1 l_i(x)\, \dd x$,
\[
  \{w_i\}_{i=0}^3 = \{1/6,\, 5/6,\, 5/6,\, 1/6\}\,.
\]

Dropping now the element index $j$ in the notation of
(\ref{eq:dg.fully.discrete}), so that for any function $f$, the symbol
$f_i$ will denote $f\big(\varphi_j(x_i)\big)$, and restricting
ourselves to the continuity equation in $S_j$ we find
\begin{equation}
\begin{split}
  r_i^2 \frac{\dd {\cal D}_i}{\dd t} = &\sum_{k=0}^4 w_k r_k^2 X_k v_k
  {\cal D}_k \frac{\dd l_i(x_k)}{\dd x} -\\ &\frac{2}{w_i |r_3-r_0|}
  \Big[ r_0^2 X_0 \mathcal{F}^-_{\cal D} \delta_{0i} - r_3^2 X_3
  \mathcal{F}^+_{\cal D} \delta_{3i} \Big]\,,
\end{split}
\end{equation}
where $\mathcal{F}_{\cal D}^\pm$ denotes the fluxes of ${\cal D}$,
between $S_j$ and $S_{j\pm 1}$, as obtained by the HLLE solver, \ie
\begin{equation}
\begin{split}
  \mathcal{F}^-_{\cal D} = \frac{1}{\lambda^+ - \lambda^-} 
  \Big[ \lambda^+ (&v_{j-1, 3} {\cal D}_{j-1,  3}) - 
    \lambda^- (v_{j, 0} {\cal D}_{j, 0})\\ 
  & + \lambda^+ \lambda^- ({\cal D}_{j-1, 3} - {\cal D}_{j, 0})
  \Big]\,,
\end{split}
\end{equation}
and
\begin{equation}
\begin{split}
  \mathcal{F}^+_{\cal D} = \frac{1}{\lambda^+ - \lambda^-} \Big[
  \lambda^+ (&v_{j+1, 0} {\cal D}_{j+1, 0}) - \lambda^- (v_{j, 3}
  {\cal D}_{j, 3})\\ 
  & + \lambda^+ \lambda^- ({\cal D}_{j+1, 0} - {\cal D}_{j, 3})
  \Big]\,,
\end{split}
\end{equation}
where $v_{j, i}\,,{\cal D}_{j, i}$ are the values of $v$ and ${\cal
  D}$ at the $i$-th collocation point in the $j$-th element and
$\lambda^+$ and $\lambda^-$ are the maximum and minimum characteristic
speeds, respectively.


\begin{thebibliography}{96}%
\makeatletter
\providecommand \@ifxundefined [1]{%
 \@ifx{#1\undefined}
}%
\providecommand \@ifnum [1]{%
 \ifnum #1\expandafter \@firstoftwo
 \else \expandafter \@secondoftwo
 \fi
}%
\providecommand \@ifx [1]{%
 \ifx #1\expandafter \@firstoftwo
 \else \expandafter \@secondoftwo
 \fi
}%
\providecommand \natexlab [1]{#1}%
\providecommand \enquote  [1]{``#1''}%
\providecommand \bibnamefont  [1]{#1}%
\providecommand \bibfnamefont [1]{#1}%
\providecommand \citenamefont [1]{#1}%
\providecommand \href@noop [0]{\@secondoftwo}%
\providecommand \href [0]{\begingroup \@sanitize@url \@href}%
\providecommand \@href[1]{\@@startlink{#1}\@@href}%
\providecommand \@@href[1]{\endgroup#1\@@endlink}%
\providecommand \@sanitize@url [0]{\catcode `\\12\catcode `\$12\catcode
  `\&12\catcode `\#12\catcode `\^12\catcode `\_12\catcode `\%12\relax}%
\providecommand \@@startlink[1]{}%
\providecommand \@@endlink[0]{}%
\providecommand \url  [0]{\begingroup\@sanitize@url \@url }%
\providecommand \@url [1]{\endgroup\@href {#1}{\urlprefix }}%
\providecommand \urlprefix  [0]{URL }%
\providecommand \Eprint [0]{\href }%
\providecommand \doibase [0]{http://dx.doi.org/}%
\providecommand \selectlanguage [0]{\@gobble}%
\providecommand \bibinfo  [0]{\@secondoftwo}%
\providecommand \bibfield  [0]{\@secondoftwo}%
\providecommand \translation [1]{[#1]}%
\providecommand \BibitemOpen [0]{}%
\providecommand \bibitemStop [0]{}%
\providecommand \bibitemNoStop [0]{.\EOS\space}%
\providecommand \EOS [0]{\spacefactor3000\relax}%
\providecommand \BibitemShut  [1]{\csname bibitem#1\endcsname}%
\let\auto@bib@innerbib\@empty
\bibitem [{\citenamefont {May}\ and\ \citenamefont {White}(1966)}]{May66}%
  \BibitemOpen
  \bibfield  {author} {\bibinfo {author} {\bibfnamefont {M.~M.}\ \bibnamefont
  {May}}\ and\ \bibinfo {author} {\bibfnamefont {R.~H.}\ \bibnamefont
  {White}},\ }\href@noop {} {\bibfield  {journal} {\bibinfo  {journal} {Phys.
  Rev.}\ }\textbf {\bibinfo {volume} {141}},\ \bibinfo {pages} {1232} (\bibinfo
  {year} {1966})}\BibitemShut {NoStop}%
\bibitem [{\citenamefont {Wilson}(1972)}]{wilson_1972_nsf}%
  \BibitemOpen
  \bibfield  {author} {\bibinfo {author} {\bibfnamefont {J.~R.}\ \bibnamefont
  {Wilson}},\ }\href {\doibase 10.1086/151434} {\bibfield  {journal} {\bibinfo
  {journal} {\apj}\ }\textbf {\bibinfo {volume} {173}},\ \bibinfo {pages} {431}
  (\bibinfo {year} {1972})}\BibitemShut {NoStop}%
\bibitem [{\citenamefont {Font}(2008)}]{Font08}%
  \BibitemOpen
  \bibfield  {author} {\bibinfo {author} {\bibfnamefont {J.~A.}\ \bibnamefont
  {Font}},\ }\href {http://www.livingreviews.org/lrr-2008-7} {\bibfield
  {journal} {\bibinfo  {journal} {Living Reviews in General Relativity}\ }
  (\bibinfo {year} {2008})}\BibitemShut {NoStop}%
\bibitem [{\citenamefont {Mart{\'\i}}\ \emph {et~al.}(1991)\citenamefont
  {Mart{\'\i}}, \citenamefont {Ib{\'a}{\~n}ez},\ and\ \citenamefont
  {Miralles}}]{Marti91}%
  \BibitemOpen
  \bibfield  {author} {\bibinfo {author} {\bibfnamefont {J.~M.}\ \bibnamefont
  {Mart{\'\i}}}, \bibinfo {author} {\bibfnamefont {J.~M.}\ \bibnamefont
  {Ib{\'a}{\~n}ez}}, \ and\ \bibinfo {author} {\bibfnamefont {J.~A.}\
  \bibnamefont {Miralles}},\ }\href@noop {} {\bibfield  {journal} {\bibinfo
  {journal} {Phys. Rev. D}\ }\textbf {\bibinfo {volume} {43}},\ \bibinfo
  {pages} {3794} (\bibinfo {year} {1991})}\BibitemShut {NoStop}%
\bibitem [{\citenamefont {Banyuls}\ \emph {et~al.}(1997)\citenamefont
  {Banyuls}, \citenamefont {Font}, \citenamefont {Ib{\'a}{\~n}ez},
  \citenamefont {Mart{\'\i}},\ and\ \citenamefont {Miralles}}]{Banyuls97}%
  \BibitemOpen
  \bibfield  {author} {\bibinfo {author} {\bibfnamefont {F.}~\bibnamefont
  {Banyuls}}, \bibinfo {author} {\bibfnamefont {J.~A.}\ \bibnamefont {Font}},
  \bibinfo {author} {\bibfnamefont {J.~M.}\ \bibnamefont {Ib{\'a}{\~n}ez}},
  \bibinfo {author} {\bibfnamefont {J.~M.}\ \bibnamefont {Mart{\'\i}}}, \ and\
  \bibinfo {author} {\bibfnamefont {J.~A.}\ \bibnamefont {Miralles}},\
  }\href@noop {} {\bibfield  {journal} {\bibinfo  {journal} {Astrophys. J.}\
  }\textbf {\bibinfo {volume} {476}},\ \bibinfo {pages} {221} (\bibinfo {year}
  {1997})}\BibitemShut {NoStop}%
\bibitem [{\citenamefont {Leveque}(1992)}]{Leveque92}%
  \BibitemOpen
  \bibfield  {author} {\bibinfo {author} {\bibfnamefont {R.~J.}\ \bibnamefont
  {Leveque}},\ }\href@noop {} {\emph {\bibinfo {title} {Numerical Methods for
  Conservation Laws}}}\ (\bibinfo  {publisher} {Birkhauser Verlag},\ \bibinfo
  {address} {Basel},\ \bibinfo {year} {1992})\BibitemShut {NoStop}%
\bibitem [{\citenamefont {Toro}(1999)}]{Toro99}%
  \BibitemOpen
  \bibfield  {author} {\bibinfo {author} {\bibfnamefont {E.~F.}\ \bibnamefont
  {Toro}},\ }\href@noop {} {\emph {\bibinfo {title} {Riemann Solvers and
  Numerical Methods for Fluid Dynamics}}}\ (\bibinfo  {publisher}
  {Springer-Verlag},\ \bibinfo {year} {1999})\BibitemShut {NoStop}%
\bibitem [{\citenamefont {Giacomazzo}\ \emph {et~al.}(2011)\citenamefont
  {Giacomazzo}, \citenamefont {Rezzolla},\ and\ \citenamefont
  {Baiotti}}]{Giacomazzo:2010}%
  \BibitemOpen
  \bibfield  {author} {\bibinfo {author} {\bibfnamefont {B.}~\bibnamefont
  {Giacomazzo}}, \bibinfo {author} {\bibfnamefont {L.}~\bibnamefont
  {Rezzolla}}, \ and\ \bibinfo {author} {\bibfnamefont {L.}~\bibnamefont
  {Baiotti}},\ }\href {\doibase 10.1103/PhysRevD.83.044014} {\bibfield
  {journal} {\bibinfo  {journal} {Phys. Rev. D}\ }\textbf {\bibinfo {volume}
  {83}},\ \bibinfo {pages} {044014} (\bibinfo {year} {2011})}\BibitemShut
  {NoStop}%
\bibitem [{\citenamefont {{Rezzolla}}\ \emph {et~al.}(2011)\citenamefont
  {{Rezzolla}}, \citenamefont {{Giacomazzo}}, \citenamefont {{Baiotti}},
  \citenamefont {{Granot}}, \citenamefont {{Kouveliotou}},\ and\ \citenamefont
  {{Aloy}}}]{Rezzolla:2011}%
  \BibitemOpen
  \bibfield  {author} {\bibinfo {author} {\bibfnamefont {L.}~\bibnamefont
  {{Rezzolla}}}, \bibinfo {author} {\bibfnamefont {B.}~\bibnamefont
  {{Giacomazzo}}}, \bibinfo {author} {\bibfnamefont {L.}~\bibnamefont
  {{Baiotti}}}, \bibinfo {author} {\bibfnamefont {J.}~\bibnamefont {{Granot}}},
  \bibinfo {author} {\bibfnamefont {C.}~\bibnamefont {{Kouveliotou}}}, \ and\
  \bibinfo {author} {\bibfnamefont {M.~A.}\ \bibnamefont {{Aloy}}},\
  }\href@noop {} {\bibfield  {journal} {\bibinfo  {journal} {Astrophys. Journ.
  Lett.}\ }\textbf {\bibinfo {volume} {732}},\ \bibinfo {pages} {L6} (\bibinfo
  {year} {2011})}\BibitemShut {NoStop}%
\bibitem [{\citenamefont {Johnsen}\ \emph {et~al.}(2010)\citenamefont
  {Johnsen}, \citenamefont {Larsson}, \citenamefont {Bhagatwala}, \citenamefont
  {Cabot}, \citenamefont {Moin}, \citenamefont {Olson}, \citenamefont {Rawat},
  \citenamefont {Shankar}, \citenamefont {Sj{\"{o}}green},\ and\ \citenamefont
  {Yee}}]{johnsen_2010_ahr}%
  \BibitemOpen
  \bibfield  {author} {\bibinfo {author} {\bibfnamefont {E.}~\bibnamefont
  {Johnsen}}, \bibinfo {author} {\bibfnamefont {J.}~\bibnamefont {Larsson}},
  \bibinfo {author} {\bibfnamefont {A.~V.}\ \bibnamefont {Bhagatwala}},
  \bibinfo {author} {\bibfnamefont {W.~H.}\ \bibnamefont {Cabot}}, \bibinfo
  {author} {\bibfnamefont {P.}~\bibnamefont {Moin}}, \bibinfo {author}
  {\bibfnamefont {B.~J.}\ \bibnamefont {Olson}}, \bibinfo {author}
  {\bibfnamefont {P.~S.}\ \bibnamefont {Rawat}}, \bibinfo {author}
  {\bibfnamefont {S.~K.}\ \bibnamefont {Shankar}}, \bibinfo {author}
  {\bibfnamefont {B.}~\bibnamefont {Sj{\"{o}}green}}, \ and\ \bibinfo {author}
  {\bibfnamefont {H.}~\bibnamefont {Yee}},\ }\href {\doibase
  10.1016/j.jcp.2009.10.028} {\bibfield  {journal} {\bibinfo  {journal}
  {Journal of Computational Physics}\ }\textbf {\bibinfo {volume} {229}},\
  \bibinfo {pages} {1213} (\bibinfo {year} {2010})}\BibitemShut {NoStop}%
\bibitem [{\citenamefont {Baiotti}\ \emph {et~al.}(2009)\citenamefont
  {Baiotti}, \citenamefont {Giacomazzo},\ and\ \citenamefont
  {Rezzolla}}]{Baiotti:2009gk}%
  \BibitemOpen
  \bibfield  {author} {\bibinfo {author} {\bibfnamefont {L.}~\bibnamefont
  {Baiotti}}, \bibinfo {author} {\bibfnamefont {B.}~\bibnamefont {Giacomazzo}},
  \ and\ \bibinfo {author} {\bibfnamefont {L.}~\bibnamefont {Rezzolla}},\
  }\href@noop {} {\bibfield  {journal} {\bibinfo  {journal} {Class. Quantum
  Grav.}\ }\textbf {\bibinfo {volume} {26}},\ \bibinfo {pages} {114005}
  (\bibinfo {year} {2009})}\BibitemShut {NoStop}%
\bibitem [{\citenamefont {Meier}(1999)}]{meier_1999_mas}%
  \BibitemOpen
  \bibfield  {author} {\bibinfo {author} {\bibfnamefont {D.~L.}\ \bibnamefont
  {Meier}},\ }\href {\doibase 10.1086/307292} {\bibfield  {journal} {\bibinfo
  {journal} {The Astrophysical Journal}\ }\textbf {\bibinfo {volume} {518}},\
  \bibinfo {pages} {788} (\bibinfo {year} {1999})}\BibitemShut {NoStop}%
\bibitem [{\citenamefont {Gourgoulhon}(1991)}]{gourgoulhon_1991_seg}%
  \BibitemOpen
  \bibfield  {author} {\bibinfo {author} {\bibfnamefont {E.}~\bibnamefont
  {Gourgoulhon}},\ }\href@noop {} {\bibfield  {journal} {\bibinfo  {journal}
  {Astronomy and Astrophysics}\ }\textbf {\bibinfo {volume} {252}},\ \bibinfo
  {pages} {651} (\bibinfo {year} {1991})}\BibitemShut {NoStop}%
\bibitem [{\citenamefont {Grandcl{\'{e}}ment}\ and\ \citenamefont
  {Novak}(2009)}]{grandclement_2009_smn}%
  \BibitemOpen
  \bibfield  {author} {\bibinfo {author} {\bibfnamefont {P.}~\bibnamefont
  {Grandcl{\'{e}}ment}}\ and\ \bibinfo {author} {\bibfnamefont
  {J.}~\bibnamefont {Novak}},\ }\href {http://www.livingreviews.org/lrr-2009-1}
  {\bibfield  {journal} {\bibinfo  {journal} {Living Reviews in Relativity}\
  }\textbf {\bibinfo {volume} {12}} (\bibinfo {year} {2009})}\BibitemShut
  {NoStop}%
\bibitem [{\citenamefont {Bonazzola}\ \emph {et~al.}(1997)\citenamefont
  {Bonazzola}, \citenamefont {Gourgoulhon},\ and\ \citenamefont
  {Marck}}]{Bonazzola97}%
  \BibitemOpen
  \bibfield  {author} {\bibinfo {author} {\bibfnamefont {S.}~\bibnamefont
  {Bonazzola}}, \bibinfo {author} {\bibfnamefont {E.}~\bibnamefont
  {Gourgoulhon}}, \ and\ \bibinfo {author} {\bibfnamefont {J.~A.}\ \bibnamefont
  {Marck}},\ }\href@noop {} {\bibfield  {journal} {\bibinfo  {journal} {Phys.
  Rev. D}\ }\textbf {\bibinfo {volume} {56}},\ \bibinfo {pages} {7740}
  (\bibinfo {year} {1997})}\BibitemShut {NoStop}%
\bibitem [{\citenamefont {Ansorg}\ \emph {et~al.}(2002)\citenamefont {Ansorg},
  \citenamefont {Kleinw{\"a}chter},\ and\ \citenamefont {Meinel}}]{Ansorg01b}%
  \BibitemOpen
  \bibfield  {author} {\bibinfo {author} {\bibfnamefont {M.}~\bibnamefont
  {Ansorg}}, \bibinfo {author} {\bibfnamefont {A.}~\bibnamefont
  {Kleinw{\"a}chter}}, \ and\ \bibinfo {author} {\bibfnamefont
  {R.}~\bibnamefont {Meinel}},\ }\href@noop {} {\bibfield  {journal} {\bibinfo
  {journal} {Astron. Astrophys.}\ }\textbf {\bibinfo {volume} {381}},\ \bibinfo
  {pages} {L49} (\bibinfo {year} {2002})},\ \bibinfo {note}
  {astro-ph/0111080}\BibitemShut {NoStop}%
\bibitem [{\citenamefont {{Dumbser}}\ and\ \citenamefont
  {{Zanotti}}(2009)}]{Dumbser2009}%
  \BibitemOpen
  \bibfield  {author} {\bibinfo {author} {\bibfnamefont {M.}~\bibnamefont
  {{Dumbser}}}\ and\ \bibinfo {author} {\bibfnamefont {O.}~\bibnamefont
  {{Zanotti}}},\ }\href {\doibase 10.1016/j.jcp.2009.06.009} {\bibfield
  {journal} {\bibinfo  {journal} {Journal of Computational Physics}\ }\textbf
  {\bibinfo {volume} {228}},\ \bibinfo {pages} {6991} (\bibinfo {year}
  {2009})}\BibitemShut {NoStop}%
\bibitem [{\citenamefont {Canuto}\ \emph {et~al.}(2008)\citenamefont {Canuto},
  \citenamefont {Hussaini}, \citenamefont {Quarteroni},\ and\ \citenamefont
  {Zang}}]{canuto_2008_sme}%
  \BibitemOpen
  \bibfield  {author} {\bibinfo {author} {\bibfnamefont {C.}~\bibnamefont
  {Canuto}}, \bibinfo {author} {\bibfnamefont {M.}~\bibnamefont {Hussaini}},
  \bibinfo {author} {\bibfnamefont {A.}~\bibnamefont {Quarteroni}}, \ and\
  \bibinfo {author} {\bibfnamefont {T.}~\bibnamefont {Zang}},\ }\href@noop {}
  {\emph {\bibinfo {title} {Spectral Methods: Evolution to Complex Gemetries
  and Applications to Fluid Dynamics}}}\ (\bibinfo  {publisher} {Springer},\
  \bibinfo {year} {2008})\BibitemShut {NoStop}%
\bibitem [{\citenamefont {Cockburn}\ and\ \citenamefont
  {Shu}(2001)}]{cockburn_2001_rkd}%
  \BibitemOpen
  \bibfield  {author} {\bibinfo {author} {\bibfnamefont {B.}~\bibnamefont
  {Cockburn}}\ and\ \bibinfo {author} {\bibfnamefont {C.-W.}\ \bibnamefont
  {Shu}},\ }\href@noop {} {\bibfield  {journal} {\bibinfo  {journal} {Journal
  of Scientific Computing}\ }\textbf {\bibinfo {volume} {16}},\ \bibinfo
  {pages} {173} (\bibinfo {year} {2001})}\BibitemShut {NoStop}%
\bibitem [{\citenamefont {Patera}(1984)}]{patera_1984_sem}%
  \BibitemOpen
  \bibfield  {author} {\bibinfo {author} {\bibfnamefont {A.}~\bibnamefont
  {Patera}},\ }\href {\doibase 10.1016/0021-9991(84)90128-1} {\bibfield
  {journal} {\bibinfo  {journal} {Journal of Computational Physics}\ }\textbf
  {\bibinfo {volume} {54}},\ \bibinfo {pages} {468} (\bibinfo {year}
  {1984})}\BibitemShut {NoStop}%
\bibitem [{\citenamefont {Cockburn}\ \emph {et~al.}(2000)\citenamefont
  {Cockburn}, \citenamefont {Karniadakis},\ and\ \citenamefont
  {Shu}}]{cockburn_2000_dg}%
  \BibitemOpen
  \bibfield  {author} {\bibinfo {author} {\bibfnamefont {B.}~\bibnamefont
  {Cockburn}}, \bibinfo {author} {\bibfnamefont {G.~E.}\ \bibnamefont
  {Karniadakis}}, \ and\ \bibinfo {author} {\bibfnamefont {C.-W.}\ \bibnamefont
  {Shu}},\ }\href@noop {} {\emph {\bibinfo {title} {Discontinuous Galerkin
  Methods: Theory, Computation and Applications}}},\ Lacture Notes on
  Computational Science and Engineering\ (\bibinfo  {publisher} {Springer},\
  \bibinfo {year} {2000})\BibitemShut {NoStop}%
\bibitem [{\citenamefont {Zumbusch}(2009)}]{zumbusch_2009_fed}%
  \BibitemOpen
  \bibfield  {author} {\bibinfo {author} {\bibfnamefont {G.}~\bibnamefont
  {Zumbusch}},\ }\href {\doibase 10.1088/0264-9381/26/17/175011} {\bibfield
  {journal} {\bibinfo  {journal} {Classical and Quantum Gravity}\ }\textbf
  {\bibinfo {volume} {26}},\ \bibinfo {pages} {175011} (\bibinfo {year}
  {2009})}\BibitemShut {NoStop}%
\bibitem [{\citenamefont {{Field}}\ \emph {et~al.}(2010)\citenamefont
  {{Field}}, \citenamefont {{Hesthaven}}, \citenamefont {{Lau}},\ and\
  \citenamefont {{Mroue}}}]{field10}%
  \BibitemOpen
  \bibfield  {author} {\bibinfo {author} {\bibfnamefont {S.~E.}\ \bibnamefont
  {{Field}}}, \bibinfo {author} {\bibfnamefont {J.~S.}\ \bibnamefont
  {{Hesthaven}}}, \bibinfo {author} {\bibfnamefont {S.~R.}\ \bibnamefont
  {{Lau}}}, \ and\ \bibinfo {author} {\bibfnamefont {A.~H.}\ \bibnamefont
  {{Mroue}}},\ }\href {\doibase 10.1103/PhysRevD.82.104051} {\bibfield
  {journal} {\bibinfo  {journal} {\prd}\ }\textbf {\bibinfo {volume} {82}},\
  \bibinfo {pages} {104051} (\bibinfo {year} {2010})}\BibitemShut {NoStop}%
\bibitem [{\citenamefont {Sopuerta}\ \emph {et~al.}(2006)\citenamefont
  {Sopuerta}, \citenamefont {Sun}, \citenamefont {Laguna},\ and\ \citenamefont
  {Xu}}]{Sopuerta:2005rd}%
  \BibitemOpen
  \bibfield  {author} {\bibinfo {author} {\bibfnamefont {C.~F.}\ \bibnamefont
  {Sopuerta}}, \bibinfo {author} {\bibfnamefont {P.}~\bibnamefont {Sun}},
  \bibinfo {author} {\bibfnamefont {P.}~\bibnamefont {Laguna}}, \ and\ \bibinfo
  {author} {\bibfnamefont {J.}~\bibnamefont {Xu}},\ }\href {\doibase
  10.1088/0264-9381/23/1/013} {\bibfield  {journal} {\bibinfo  {journal}
  {Class. Quant. Grav.}\ }\textbf {\bibinfo {volume} {23}},\ \bibinfo {pages}
  {251} (\bibinfo {year} {2006})}\BibitemShut {NoStop}%
\bibitem [{\citenamefont {Sopuerta}\ and\ \citenamefont
  {Laguna}(2006)}]{Sopuerta:2005gz}%
  \BibitemOpen
  \bibfield  {author} {\bibinfo {author} {\bibfnamefont {C.~F.}\ \bibnamefont
  {Sopuerta}}\ and\ \bibinfo {author} {\bibfnamefont {P.}~\bibnamefont
  {Laguna}},\ }\href {\doibase 10.1103/PhysRevD.73.044028} {\bibfield
  {journal} {\bibinfo  {journal} {Phys. Rev.}\ }\textbf {\bibinfo {volume}
  {D73}},\ \bibinfo {pages} {044028} (\bibinfo {year} {2006})}\BibitemShut
  {NoStop}%
\bibitem [{\citenamefont {{Dimmelmeier}}\ \emph {et~al.}(2005)\citenamefont
  {{Dimmelmeier}}, \citenamefont {{Novak}}, \citenamefont {{Font}},
  \citenamefont {{Ib{\'a}{\~n}ez}},\ and\ \citenamefont
  {{M{\"u}ller}}}]{Dimmelmeier05a}%
  \BibitemOpen
  \bibfield  {author} {\bibinfo {author} {\bibfnamefont {H.}~\bibnamefont
  {{Dimmelmeier}}}, \bibinfo {author} {\bibfnamefont {J.}~\bibnamefont
  {{Novak}}}, \bibinfo {author} {\bibfnamefont {J.~A.}\ \bibnamefont {{Font}}},
  \bibinfo {author} {\bibfnamefont {J.~M.}\ \bibnamefont {{Ib{\'a}{\~n}ez}}}, \
  and\ \bibinfo {author} {\bibfnamefont {E.}~\bibnamefont {{M{\"u}ller}}},\
  }\href {\doibase 10.1103/PhysRevD.71.064023} {\bibfield  {journal} {\bibinfo
  {journal} {Phys. Rev. D}\ }\textbf {\bibinfo {volume} {71}},\ \bibinfo
  {pages} {064023} (\bibinfo {year} {2005})}\BibitemShut {NoStop}%
\bibitem [{\citenamefont {Duez}\ \emph {et~al.}(2008)\citenamefont {Duez} \emph
  {et~al.}}]{Duez:2008rb}%
  \BibitemOpen
  \bibfield  {author} {\bibinfo {author} {\bibfnamefont {M.~D.}\ \bibnamefont
  {Duez}} \emph {et~al.},\ }\href {\doibase 10.1103/PhysRevD.78.104015}
  {\bibfield  {journal} {\bibinfo  {journal} {Phys. Rev. D}\ }\textbf {\bibinfo
  {volume} {78}},\ \bibinfo {pages} {104015} (\bibinfo {year}
  {2008})}\BibitemShut {NoStop}%
\bibitem [{\citenamefont {Chen}\ \emph {et~al.}(2009)\citenamefont {Chen},
  \citenamefont {Ziemer},\ and\ \citenamefont {Torres}}]{chen_2009_ggt}%
  \BibitemOpen
  \bibfield  {author} {\bibinfo {author} {\bibfnamefont {G.-Q.}\ \bibnamefont
  {Chen}}, \bibinfo {author} {\bibfnamefont {W.~P.}\ \bibnamefont {Ziemer}}, \
  and\ \bibinfo {author} {\bibfnamefont {M.}~\bibnamefont {Torres}},\ }\href
  {\doibase 10.1002/cpa.20262} {\bibfield  {journal} {\bibinfo  {journal}
  {Communications on Pure and Applied Mathematics}\ }\textbf {\bibinfo {volume}
  {62}},\ \bibinfo {pages} {242} (\bibinfo {year} {2009})}\BibitemShut
  {NoStop}%
\bibitem [{\citenamefont {Chen}\ and\ \citenamefont
  {Frid}(2003)}]{chen_2003_edm}%
  \BibitemOpen
  \bibfield  {author} {\bibinfo {author} {\bibfnamefont {G.-Q.}\ \bibnamefont
  {Chen}}\ and\ \bibinfo {author} {\bibfnamefont {H.}~\bibnamefont {Frid}},\
  }\href {\doibase 10.1007/s00220-003-0823-7} {\bibfield  {journal} {\bibinfo
  {journal} {Communications in Mathematical Physics}\ }\textbf {\bibinfo
  {volume} {236}},\ \bibinfo {pages} {251} (\bibinfo {year}
  {2003})}\BibitemShut {NoStop}%
\bibitem [{\citenamefont {Palaniappan}\ \emph {et~al.}(2004)\citenamefont
  {Palaniappan}, \citenamefont {Haber},\ and\ \citenamefont
  {Jerrard}}]{palaniappan_2004_sdg}%
  \BibitemOpen
  \bibfield  {author} {\bibinfo {author} {\bibfnamefont {J.}~\bibnamefont
  {Palaniappan}}, \bibinfo {author} {\bibfnamefont {R.~B.}\ \bibnamefont
  {Haber}}, \ and\ \bibinfo {author} {\bibfnamefont {R.~L.}\ \bibnamefont
  {Jerrard}},\ }\href {\doibase 10.1016/j.cma.2004.01.028} {\bibfield
  {journal} {\bibinfo  {journal} {Comput. Methods Appl. Mech. Engrg.}\ }\textbf
  {\bibinfo {volume} {193}},\ \bibinfo {pages} {3607} (\bibinfo {year}
  {2004})}\BibitemShut {NoStop}%
\bibitem [{\citenamefont {Quarteroni}\ and\ \citenamefont
  {Valli}(1997)}]{quarteroni_1997_nap}%
  \BibitemOpen
  \bibfield  {author} {\bibinfo {author} {\bibfnamefont {A.}~\bibnamefont
  {Quarteroni}}\ and\ \bibinfo {author} {\bibfnamefont {A.}~\bibnamefont
  {Valli}},\ }\href@noop {} {\emph {\bibinfo {title} {Numerical Approximation
  of Partial Differential Equations}}}\ (\bibinfo  {publisher} {Springer},\
  \bibinfo {year} {1997})\BibitemShut {NoStop}%
\bibitem [{\citenamefont {Papadopoulos}\ and\ \citenamefont
  {Font}(1999)}]{Papadopoulos-Font-1999}%
  \BibitemOpen
  \bibfield  {author} {\bibinfo {author} {\bibfnamefont {P.}~\bibnamefont
  {Papadopoulos}}\ and\ \bibinfo {author} {\bibfnamefont {J.~A.}\ \bibnamefont
  {Font}},\ }\href {\doibase 10.1103/PhysRevD.61.024015} {\bibfield  {journal}
  {\bibinfo  {journal} {Phys. Rev. D}\ }\textbf {\bibinfo {volume} {61}},\
  \bibinfo {pages} {024015} (\bibinfo {year} {1999})}\BibitemShut {NoStop}%
\bibitem [{\citenamefont {{Rezzolla}}\ and\ \citenamefont
  {{Miller}}(1994)}]{Rezzolla1994}%
  \BibitemOpen
  \bibfield  {author} {\bibinfo {author} {\bibfnamefont {L.}~\bibnamefont
  {{Rezzolla}}}\ and\ \bibinfo {author} {\bibfnamefont {J.~C.}\ \bibnamefont
  {{Miller}}},\ }\href {\doibase 10.1088/0264-9381/11/7/018} {\bibfield
  {journal} {\bibinfo  {journal} {Classical and Quantum Gravity}\ }\textbf
  {\bibinfo {volume} {11}},\ \bibinfo {pages} {1815} (\bibinfo {year}
  {1994})}\BibitemShut {NoStop}%
\bibitem [{\citenamefont {Musco}\ \emph {et~al.}(2005)\citenamefont {Musco},
  \citenamefont {Miller},\ and\ \citenamefont {Rezzolla}}]{musco05}%
  \BibitemOpen
  \bibfield  {author} {\bibinfo {author} {\bibfnamefont {I.}~\bibnamefont
  {Musco}}, \bibinfo {author} {\bibfnamefont {J.~C.}\ \bibnamefont {Miller}}, \
  and\ \bibinfo {author} {\bibfnamefont {L.}~\bibnamefont {Rezzolla}},\
  }\href@noop {} {\bibfield  {journal} {\bibinfo  {journal} {Class. Quantum
  Grav.}\ }\textbf {\bibinfo {volume} {22}},\ \bibinfo {pages} {1405} (\bibinfo
  {year} {2005})}\BibitemShut {NoStop}%
\bibitem [{\citenamefont {Rosswog}(2010)}]{rosswog_2010_csr}%
  \BibitemOpen
  \bibfield  {author} {\bibinfo {author} {\bibfnamefont {S.}~\bibnamefont
  {Rosswog}},\ }\href {\doibase 10.1016/j.jcp.2010.08.002} {\bibfield
  {journal} {\bibinfo  {journal} {Journal of Computational Physics}\ }\textbf
  {\bibinfo {volume} {229}},\ \bibinfo {pages} {8591} (\bibinfo {year}
  {2010})}\BibitemShut {NoStop}%
\bibitem [{\citenamefont {Siegler}\ and\ \citenamefont
  {Riffert}(2000)}]{Siegler00}%
  \BibitemOpen
  \bibfield  {author} {\bibinfo {author} {\bibfnamefont {S.}~\bibnamefont
  {Siegler}}\ and\ \bibinfo {author} {\bibfnamefont {H.}~\bibnamefont
  {Riffert}},\ }\href {\doibase 10.1086/308482} {\bibfield  {journal} {\bibinfo
   {journal} {The Astrophysical Journal}\ }\textbf {\bibinfo {volume} {531}},\
  \bibinfo {pages} {1053} (\bibinfo {year} {2000})}\BibitemShut {NoStop}%
\bibitem [{\citenamefont {Romero}\ \emph {et~al.}(1996)\citenamefont {Romero},
  \citenamefont {Ib{\'a}{\~n}ez}, \citenamefont {Mart{\'\i}},\ and\
  \citenamefont {Miralles}}]{Romero96}%
  \BibitemOpen
  \bibfield  {author} {\bibinfo {author} {\bibfnamefont {J.~V.}\ \bibnamefont
  {Romero}}, \bibinfo {author} {\bibfnamefont {J.~M.}\ \bibnamefont
  {Ib{\'a}{\~n}ez}}, \bibinfo {author} {\bibfnamefont {J.~M.}\ \bibnamefont
  {Mart{\'\i}}}, \ and\ \bibinfo {author} {\bibfnamefont {J.~A.}\ \bibnamefont
  {Miralles}},\ }\href@noop {} {\bibfield  {journal} {\bibinfo  {journal}
  {Astrophys. J.}\ }\textbf {\bibinfo {volume} {462}},\ \bibinfo {pages} {839}
  (\bibinfo {year} {1996})}\BibitemShut {NoStop}%
\bibitem [{\citenamefont {Noble}(2003)}]{noble_2003_nsr}%
  \BibitemOpen
  \bibfield  {author} {\bibinfo {author} {\bibfnamefont {S.~C.}\ \bibnamefont
  {Noble}},\ }\emph {\bibinfo {title} {A Numerical Study of Relativistic Fluid
  Collapse}},\ \href
  {http://www.citebase.org/abstract?id=oai:arXiv.org:gr-qc/0310116} {Ph.D.
  thesis},\ \bibinfo  {school} {University of Texas at Austin} (\bibinfo {year}
  {2003}),\ \Eprint {http://arxiv.org/abs/gr-qc/0310116v1} {gr-qc/0310116v1}
  \BibitemShut {NoStop}%
\bibitem [{\citenamefont {Yang}(2000)}]{yang_2000_oon}%
  \BibitemOpen
  \bibfield  {author} {\bibinfo {author} {\bibfnamefont {D.}~\bibnamefont
  {Yang}},\ }\href@noop {} {\emph {\bibinfo {title} {C++ and Object-Oriented
  Numeric Computing for Scientists and Engineers}}}\ (\bibinfo  {publisher}
  {Springer-Verlag},\ \bibinfo {year} {2000})\BibitemShut {NoStop}%
\bibitem [{\citenamefont {Veldhuizen}(1998)}]{Veldhuizen98}%
  \BibitemOpen
  \bibfield  {author} {\bibinfo {author} {\bibfnamefont {T.~L.}\ \bibnamefont
  {Veldhuizen}},\ }in\ \href
  {http://portal.acm.org/citation.cfm?id=646894.709708} {\emph {\bibinfo
  {booktitle} {Proceedings of the Second International Symposium on Computing
  in Object-Oriented Parallel Environments}}},\ \bibinfo {series and number}
  {ISCOPE '98}\ (\bibinfo  {publisher} {Springer-Verlag},\ \bibinfo {address}
  {London, UK},\ \bibinfo {year} {1998})\ pp.\ \bibinfo {pages}
  {223--230}\BibitemShut {NoStop}%
\bibitem [{bli()}]{blitz}%
  \BibitemOpen
  \href@noop {} {}\bibinfo {howpublished}
  {\url{http://www.oonumerics.org/blitz/}}\BibitemShut {NoStop}%
\bibitem [{\citenamefont
  {Davis}(2002{\natexlab{a}})}]{Davis-2002a-UMFPACK-report}%
  \BibitemOpen
  \bibfield  {author} {\bibinfo {author} {\bibfnamefont {T.~A.}\ \bibnamefont
  {Davis}},\ }\href@noop {} {\emph {\bibinfo {title} {A column pre-ordering
  strategy for the unsymmetric-pattern multifrontal method}}},\ \bibinfo {type}
  {Tech. Rep.}\ \bibinfo {number} {TR-02-001}\ (\bibinfo  {institution} {Univ.
  of Florida, CISE Dept.},\ \bibinfo {address} {Gainesville, FL},\ \bibinfo
  {year} {2002})\BibitemShut {NoStop}%
\bibitem [{\citenamefont
  {Davis}(2002{\natexlab{b}})}]{Davis-2002b-UMFPACK-report}%
  \BibitemOpen
  \bibfield  {author} {\bibinfo {author} {\bibfnamefont {T.~A.}\ \bibnamefont
  {Davis}},\ }\href@noop {} {\emph {\bibinfo {title} {Algorithm 8xx: {UMFPACK
  V3.2}, an unsymmetric-pattern multifrontal method with a column pre-ordering
  strategy}}},\ \bibinfo {type} {Tech. Rep.}\ \bibinfo {number} {TR-02-002}\
  (\bibinfo  {institution} {Univ. of Florida, CISE Dept.},\ \bibinfo {address}
  {Gainesville, FL},\ \bibinfo {year} {2002})\ \bibinfo {note}
  {(http://www.cise.ufl.edu/tech-reports. Submitted to {\em ACM Trans. Math.
  Softw.})}\BibitemShut {NoStop}%
\bibitem [{\citenamefont {Davis}\ and\ \citenamefont
  {Duff}(1997)}]{Davis-Duff-1997-UMFPACK}%
  \BibitemOpen
  \bibfield  {author} {\bibinfo {author} {\bibfnamefont {T.~A.}\ \bibnamefont
  {Davis}}\ and\ \bibinfo {author} {\bibfnamefont {I.~S.}\ \bibnamefont
  {Duff}},\ }\href@noop {} {\bibfield  {journal} {\bibinfo  {journal} {SIAM J.
  Matrix Anal. Applic.}\ }\textbf {\bibinfo {volume} {18}},\ \bibinfo {pages}
  {140} (\bibinfo {year} {1997})}\BibitemShut {NoStop}%
\bibitem [{\citenamefont {Davis}\ and\ \citenamefont
  {Duff}(1999)}]{Davis-Duff-1999-UMFPACK}%
  \BibitemOpen
  \bibfield  {author} {\bibinfo {author} {\bibfnamefont {T.~A.}\ \bibnamefont
  {Davis}}\ and\ \bibinfo {author} {\bibfnamefont {I.~S.}\ \bibnamefont
  {Duff}},\ }\href@noop {} {\bibfield  {journal} {\bibinfo  {journal} {ACM
  Trans. Math. Softw.}\ }\textbf {\bibinfo {volume} {25}},\ \bibinfo {pages}
  {1} (\bibinfo {year} {1999})}\BibitemShut {NoStop}%
\bibitem [{\citenamefont {Davis}()}]{umfpackweb}%
  \BibitemOpen
  \bibfield  {author} {\bibinfo {author} {\bibfnamefont {T.~A.}\ \bibnamefont
  {Davis}},\ }\href {http://www.cise.ufl.edu/research/sparse/umfpack/}
  {\enquote {\bibinfo {title} {{UMFPACK}: {A} set of routines for solving
  sparse linear systems via {LU} factorization},}\ }\bibinfo {note}
  {{UMFPACK}}\BibitemShut {NoStop}%
\bibitem [{\citenamefont {Gottlieb}\ \emph {et~al.}(2009)\citenamefont
  {Gottlieb}, \citenamefont {Ketcheson},\ and\ \citenamefont
  {Shu}}]{gottlieb2009}%
  \BibitemOpen
  \bibfield  {author} {\bibinfo {author} {\bibfnamefont {S.}~\bibnamefont
  {Gottlieb}}, \bibinfo {author} {\bibfnamefont {D.}~\bibnamefont {Ketcheson}},
  \ and\ \bibinfo {author} {\bibfnamefont {C.-W.}\ \bibnamefont {Shu}},\ }\href
  {http://dx.doi.org/10.1007/s10915-008-9239-z} {\bibfield  {journal} {\bibinfo
   {journal} {Journal of Scientific Computing}\ }\textbf {\bibinfo {volume}
  {38}},\ \bibinfo {pages} {251} (\bibinfo {year} {2009})},\ \bibinfo {note}
  {10.1007/s10915-008-9239-z}\BibitemShut {NoStop}%
\bibitem [{\citenamefont {Gottlieb}\ and\ \citenamefont
  {Tadmor}(1991)}]{gottlieb_1991_ccs}%
  \BibitemOpen
  \bibfield  {author} {\bibinfo {author} {\bibfnamefont {D.}~\bibnamefont
  {Gottlieb}}\ and\ \bibinfo {author} {\bibfnamefont {E.}~\bibnamefont
  {Tadmor}},\ }\href {http://www.jstor.org/stable/2008395} {\bibfield
  {journal} {\bibinfo  {journal} {Mathematics of Computation}\ }\textbf
  {\bibinfo {volume} {56}},\ \bibinfo {pages} {565} (\bibinfo {year}
  {1991})}\BibitemShut {NoStop}%
\bibitem [{\citenamefont {Arnold}(1982)}]{arnold_1982_ipf}%
  \BibitemOpen
  \bibfield  {author} {\bibinfo {author} {\bibfnamefont {D.~N.}\ \bibnamefont
  {Arnold}},\ }\href@noop {} {\bibfield  {journal} {\bibinfo  {journal} {SIAM
  J. Numer. Anal.}\ }\textbf {\bibinfo {volume} {19}},\ \bibinfo {pages} {742}
  (\bibinfo {year} {1982})}\BibitemShut {NoStop}%
\bibitem [{\citenamefont {Arnold}\ \emph {et~al.}(2002)\citenamefont {Arnold},
  \citenamefont {Brezzi}, \citenamefont {Cockburn},\ and\ \citenamefont
  {Marini}}]{arnold_2002_uad}%
  \BibitemOpen
  \bibfield  {author} {\bibinfo {author} {\bibfnamefont {D.~N.}\ \bibnamefont
  {Arnold}}, \bibinfo {author} {\bibfnamefont {F.}~\bibnamefont {Brezzi}},
  \bibinfo {author} {\bibfnamefont {B.}~\bibnamefont {Cockburn}}, \ and\
  \bibinfo {author} {\bibfnamefont {L.~D.}\ \bibnamefont {Marini}},\
  }\href@noop {} {\bibfield  {journal} {\bibinfo  {journal} {SIAM J. Numer.
  Anal.}\ }\textbf {\bibinfo {volume} {39}},\ \bibinfo {pages} {1749} (\bibinfo
  {year} {2002})}\BibitemShut {NoStop}%
\bibitem [{\citenamefont {Saad}(1996)}]{saad_96}%
  \BibitemOpen
  \bibfield  {author} {\bibinfo {author} {\bibfnamefont {Y.}~\bibnamefont
  {Saad}},\ }\href@noop {} {\emph {\bibinfo {title} {Iterative Methods for
  Sparse Linear Systems}}}\ (\bibinfo  {publisher} {SIAM},\ \bibinfo {year}
  {1996})\BibitemShut {NoStop}%
\bibitem [{\citenamefont {Biswas}\ \emph {et~al.}(1994)\citenamefont {Biswas},
  \citenamefont {Devine},\ and\ \citenamefont {Flaherty}}]{biswas_1994_paf}%
  \BibitemOpen
  \bibfield  {author} {\bibinfo {author} {\bibfnamefont {R.}~\bibnamefont
  {Biswas}}, \bibinfo {author} {\bibfnamefont {K.~D.}\ \bibnamefont {Devine}},
  \ and\ \bibinfo {author} {\bibfnamefont {J.~E.}\ \bibnamefont {Flaherty}},\
  }\href@noop {} {\bibfield  {journal} {\bibinfo  {journal} {Applied Numerical
  Mathematics}\ }\textbf {\bibinfo {volume} {14}},\ \bibinfo {pages} {255}
  (\bibinfo {year} {1994})}\BibitemShut {NoStop}%
\bibitem [{\citenamefont {Krivodonova}(2007)}]{krivodonova_2007_lho}%
  \BibitemOpen
  \bibfield  {author} {\bibinfo {author} {\bibfnamefont {L.}~\bibnamefont
  {Krivodonova}},\ }\href {\doibase 10.1016/j.jcp.2007.05.011} {\bibfield
  {journal} {\bibinfo  {journal} {Journal of Computational Physics}\ }\textbf
  {\bibinfo {volume} {226}},\ \bibinfo {pages} {879} (\bibinfo {year}
  {2007})}\BibitemShut {NoStop}%
\bibitem [{\citenamefont {Maday}\ \emph {et~al.}(1993)\citenamefont {Maday},
  \citenamefont {Kaber},\ and\ \citenamefont {Tadmor}}]{maday_1993_lpv}%
  \BibitemOpen
  \bibfield  {author} {\bibinfo {author} {\bibfnamefont {Y.}~\bibnamefont
  {Maday}}, \bibinfo {author} {\bibfnamefont {S.~M.~O.}\ \bibnamefont {Kaber}},
  \ and\ \bibinfo {author} {\bibfnamefont {E.}~\bibnamefont {Tadmor}},\
  }\href@noop {} {\bibfield  {journal} {\bibinfo  {journal} {SIAM J. Numer.
  Anal}\ }\textbf {\bibinfo {volume} {30}},\ \bibinfo {pages} {321} (\bibinfo
  {year} {1993})}\BibitemShut {NoStop}%
\bibitem [{\citenamefont {Canuto}\ \emph {et~al.}(1988)\citenamefont {Canuto},
  \citenamefont {Hussani}, \citenamefont {Quarteroni},\ and\ \citenamefont
  {Zang}}]{Canuto-Hussaini-Quarteroni-Zang:pseudospectral}%
  \BibitemOpen
  \bibfield  {author} {\bibinfo {author} {\bibfnamefont {C.}~\bibnamefont
  {Canuto}}, \bibinfo {author} {\bibfnamefont {M.~Y.}\ \bibnamefont {Hussani}},
  \bibinfo {author} {\bibfnamefont {A.}~\bibnamefont {Quarteroni}}, \ and\
  \bibinfo {author} {\bibfnamefont {T.~A.}\ \bibnamefont {Zang}},\ }\href@noop
  {} {\emph {\bibinfo {title} {Spectral Methods in Fluid Dynamics}}},\ \bibinfo
  {edition} {2nd}\ ed.\ (\bibinfo  {publisher} {Springer-Verlag},\ \bibinfo
  {address} {New York and Berlin},\ \bibinfo {year} {1988})\BibitemShut
  {NoStop}%
\bibitem [{\citenamefont {Hesthaven}\ and\ \citenamefont
  {Kirby}(2008)}]{hesthaven_2008_fls}%
  \BibitemOpen
  \bibfield  {author} {\bibinfo {author} {\bibfnamefont {J.~S.}\ \bibnamefont
  {Hesthaven}}\ and\ \bibinfo {author} {\bibfnamefont {R.~M.}\ \bibnamefont
  {Kirby}},\ }\href {\doibase 10.1090/S0025-5718-08-02110-8} {\bibfield
  {journal} {\bibinfo  {journal} {Math. Comput.}\ }\textbf {\bibinfo {volume}
  {77}},\ \bibinfo {pages} {1425} (\bibinfo {year} {2008})}\BibitemShut
  {NoStop}%
\bibitem [{\citenamefont {Meister}\ \emph {et~al.}(2009)\citenamefont
  {Meister}, \citenamefont {Ortleb},\ and\ \citenamefont
  {Sonar}}]{meister_2009_sfd}%
  \BibitemOpen
  \bibfield  {author} {\bibinfo {author} {\bibfnamefont {A.}~\bibnamefont
  {Meister}}, \bibinfo {author} {\bibfnamefont {S.}~\bibnamefont {Ortleb}}, \
  and\ \bibinfo {author} {\bibfnamefont {T.}~\bibnamefont {Sonar}},\
  }\href@noop {} {\enquote {\bibinfo {title} {On spectral filtering for
  discontinuous {G}alerkin methods on unstructured triangular grids},}\ }
  (\bibinfo {year} {2009}),\ \bibinfo {note}
  {http://cms.uni-kassel.de/unicms/fileadmin/groups/w\_180000/prep/prep0904.pd%
f}\BibitemShut {NoStop}%
\bibitem [{\citenamefont {Boyd}(1996)}]{boyd_1996_efa}%
  \BibitemOpen
  \bibfield  {author} {\bibinfo {author} {\bibfnamefont {J.~P.}\ \bibnamefont
  {Boyd}},\ }in\ \href@noop {} {\emph {\bibinfo {booktitle} {Proceedings of the
  Third International Conference on Spectral and High Order Methods}}},\
  \bibinfo {editor} {edited by\ \bibinfo {editor} {\bibfnamefont {L.~R.~S.}\
  \bibnamefont {A.~V.~Illin}}}\ (\bibinfo  {publisher} {Houston Journal of
  Mathematics},\ \bibinfo {year} {1996})\ pp.\ \bibinfo {pages}
  {267--276}\BibitemShut {NoStop}%
\bibitem [{\citenamefont {Galeazzi}(2008)}]{galeazzi_master}%
  \BibitemOpen
  \bibfield  {author} {\bibinfo {author} {\bibfnamefont {F.}~\bibnamefont
  {Galeazzi}},\ }\emph {\bibinfo {title} {Modelling fluid interfaces in
  numerical relativistic hydrodynamics}},\ \href@noop {} {Master's thesis},\
  \bibinfo  {school} {Universit{\`{a}} degli studi di Padova} (\bibinfo {year}
  {2008})\BibitemShut {NoStop}%
\bibitem [{\citenamefont {Kastaun}(2006)}]{kastaun_2006_hrs}%
  \BibitemOpen
  \bibfield  {author} {\bibinfo {author} {\bibfnamefont {W.}~\bibnamefont
  {Kastaun}},\ }\href {\doibase 10.1103/PhysRevD.74.124024} {\bibfield
  {journal} {\bibinfo  {journal} {Phys. Rev. D}\ }\textbf {\bibinfo {volume}
  {74}},\ \bibinfo {pages} {124024} (\bibinfo {year} {2006})}\BibitemShut
  {NoStop}%
\bibitem [{\citenamefont {{Millmore}}\ and\ \citenamefont
  {{Hawke}}(2010)}]{Millmore2010}%
  \BibitemOpen
  \bibfield  {author} {\bibinfo {author} {\bibfnamefont {S.~T.}\ \bibnamefont
  {{Millmore}}}\ and\ \bibinfo {author} {\bibfnamefont {I.}~\bibnamefont
  {{Hawke}}},\ }\href {\doibase 10.1088/0264-9381/27/1/015007} {\bibfield
  {journal} {\bibinfo  {journal} {Classical and Quantum Gravity}\ }\textbf
  {\bibinfo {volume} {27}},\ \bibinfo {pages} {015007} (\bibinfo {year}
  {2010})}\BibitemShut {NoStop}%
\bibitem [{\citenamefont {Font}\ \emph {et~al.}(2002)\citenamefont {Font},
  \citenamefont {Goodale}, \citenamefont {Iyer}, \citenamefont {Miller},
  \citenamefont {Rezzolla}, \citenamefont {Seidel}, \citenamefont
  {Stergioulas}, \citenamefont {Suen},\ and\ \citenamefont {Tobias}}]{Font02c}%
  \BibitemOpen
  \bibfield  {author} {\bibinfo {author} {\bibfnamefont {J.~A.}\ \bibnamefont
  {Font}}, \bibinfo {author} {\bibfnamefont {T.}~\bibnamefont {Goodale}},
  \bibinfo {author} {\bibfnamefont {S.}~\bibnamefont {Iyer}}, \bibinfo {author}
  {\bibfnamefont {M.}~\bibnamefont {Miller}}, \bibinfo {author} {\bibfnamefont
  {L.}~\bibnamefont {Rezzolla}}, \bibinfo {author} {\bibfnamefont
  {E.}~\bibnamefont {Seidel}}, \bibinfo {author} {\bibfnamefont
  {N.}~\bibnamefont {Stergioulas}}, \bibinfo {author} {\bibfnamefont {W.~M.}\
  \bibnamefont {Suen}}, \ and\ \bibinfo {author} {\bibfnamefont
  {M.}~\bibnamefont {Tobias}},\ }\href@noop {} {\bibfield  {journal} {\bibinfo
  {journal} {Phys. Rev. D}\ }\textbf {\bibinfo {volume} {65}},\ \bibinfo
  {pages} {084024} (\bibinfo {year} {2002})}\BibitemShut {NoStop}%
\bibitem [{\citenamefont {Baiotti}\ \emph {et~al.}(2005)\citenamefont
  {Baiotti}, \citenamefont {Hawke}, \citenamefont {Montero}, \citenamefont
  {L{\"o}ffler}, \citenamefont {Rezzolla}, \citenamefont {Stergioulas},
  \citenamefont {Font},\ and\ \citenamefont {Seidel}}]{Baiotti04}%
  \BibitemOpen
  \bibfield  {author} {\bibinfo {author} {\bibfnamefont {L.}~\bibnamefont
  {Baiotti}}, \bibinfo {author} {\bibfnamefont {I.}~\bibnamefont {Hawke}},
  \bibinfo {author} {\bibfnamefont {P.~J.}\ \bibnamefont {Montero}}, \bibinfo
  {author} {\bibfnamefont {F.}~\bibnamefont {L{\"o}ffler}}, \bibinfo {author}
  {\bibfnamefont {L.}~\bibnamefont {Rezzolla}}, \bibinfo {author}
  {\bibfnamefont {N.}~\bibnamefont {Stergioulas}}, \bibinfo {author}
  {\bibfnamefont {J.~A.}\ \bibnamefont {Font}}, \ and\ \bibinfo {author}
  {\bibfnamefont {E.}~\bibnamefont {Seidel}},\ }\href@noop {} {\bibfield
  {journal} {\bibinfo  {journal} {Phys. Rev. D}\ }\textbf {\bibinfo {volume}
  {71}},\ \bibinfo {pages} {024035} (\bibinfo {year} {2005})}\BibitemShut
  {NoStop}%
\bibitem [{\citenamefont {Gabler}\ \emph {et~al.}(2009)\citenamefont {Gabler},
  \citenamefont {Sperhake},\ and\ \citenamefont {Andersson}}]{Gabler:2009yt}%
  \BibitemOpen
  \bibfield  {author} {\bibinfo {author} {\bibfnamefont {M.}~\bibnamefont
  {Gabler}}, \bibinfo {author} {\bibfnamefont {U.}~\bibnamefont {Sperhake}}, \
  and\ \bibinfo {author} {\bibfnamefont {N.}~\bibnamefont {Andersson}},\ }\href
  {\doibase 10.1103/PhysRevD.80.064012} {\bibfield  {journal} {\bibinfo
  {journal} {Phys. Rev. D}\ }\textbf {\bibinfo {volume} {80}},\ \bibinfo
  {pages} {064012} (\bibinfo {year} {2009})}\BibitemShut {NoStop}%
\bibitem [{\citenamefont {Wang}(2002)}]{wang_2002_sfv}%
  \BibitemOpen
  \bibfield  {author} {\bibinfo {author} {\bibfnamefont {Z.}~\bibnamefont
  {Wang}},\ }\href {\doibase 10.1006/jcph.2002.7041} {\bibfield  {journal}
  {\bibinfo  {journal} {Journal of Computational Physics}\ }\textbf {\bibinfo
  {volume} {178}},\ \bibinfo {pages} {210} (\bibinfo {year}
  {2002})}\BibitemShut {NoStop}%
\bibitem [{\citenamefont {Touil}\ \emph {et~al.}(2007)\citenamefont {Touil},
  \citenamefont {Hussaini},\ and\ \citenamefont {Sussman}}]{touil_2007_tdh}%
  \BibitemOpen
  \bibfield  {author} {\bibinfo {author} {\bibfnamefont {H.}~\bibnamefont
  {Touil}}, \bibinfo {author} {\bibfnamefont {M.}~\bibnamefont {Hussaini}}, \
  and\ \bibinfo {author} {\bibfnamefont {M.}~\bibnamefont {Sussman}},\ }\href
  {\doibase 10.1016/j.jcp.2007.02.016} {\bibfield  {journal} {\bibinfo
  {journal} {J. Comput. Phys.}\ }\textbf {\bibinfo {volume} {225}},\ \bibinfo
  {pages} {1810} (\bibinfo {year} {2007})}\BibitemShut {NoStop}%
\bibitem [{\citenamefont {Berger}\ \emph {et~al.}(2005)\citenamefont {Berger},
  \citenamefont {Aftosmis},\ and\ \citenamefont {Murman}}]{berger_2005_asl}%
  \BibitemOpen
  \bibfield  {author} {\bibinfo {author} {\bibfnamefont {M.}~\bibnamefont
  {Berger}}, \bibinfo {author} {\bibfnamefont {M.~J.}\ \bibnamefont
  {Aftosmis}}, \ and\ \bibinfo {author} {\bibfnamefont {S.~M.}\ \bibnamefont
  {Murman}},\ }\href@noop {} {\emph {\bibinfo {title} {Analysis of Slope
  Limiters on Irregular Grids}}},\ \bibinfo {type} {NAS Technical Report}\
  \bibinfo {number} {2005-0490}\ (\bibinfo  {institution} {American Institute
  of Aeronautics and Astronautics},\ \bibinfo {year} {2005})\BibitemShut
  {NoStop}%
\bibitem [{\citenamefont {Mart{\'\i}}\ and\ \citenamefont
  {M{\"u}ller}(2003)}]{Marti03}%
  \BibitemOpen
  \bibfield  {author} {\bibinfo {author} {\bibfnamefont {J.~M.}\ \bibnamefont
  {Mart{\'\i}}}\ and\ \bibinfo {author} {\bibfnamefont {E.}~\bibnamefont
  {M{\"u}ller}},\ }\href {http://www.livingreviews.org/lrr-2003-7} {\bibfield
  {journal} {\bibinfo  {journal} {Living Rev. Relativ.}\ }\textbf {\bibinfo
  {volume} {6}},\ \bibinfo {pages} {7} (\bibinfo {year} {2003})}\BibitemShut
  {NoStop}%
\bibitem [{\citenamefont {{Del Zanna}}\ \emph {et~al.}(2007)\citenamefont {{Del
  Zanna}}, \citenamefont {{Zanotti}}, \citenamefont {{Bucciantini}},\ and\
  \citenamefont {{Londrillo}}}]{DelZanna2007}%
  \BibitemOpen
  \bibfield  {author} {\bibinfo {author} {\bibfnamefont {L.}~\bibnamefont {{Del
  Zanna}}}, \bibinfo {author} {\bibfnamefont {O.}~\bibnamefont {{Zanotti}}},
  \bibinfo {author} {\bibfnamefont {N.}~\bibnamefont {{Bucciantini}}}, \ and\
  \bibinfo {author} {\bibfnamefont {P.}~\bibnamefont {{Londrillo}}},\ }\href
  {\doibase 10.1051/0004-6361:20077093} {\bibfield  {journal} {\bibinfo
  {journal} {Astron. Astrophys.}\ }\textbf {\bibinfo {volume} {473}},\ \bibinfo
  {pages} {11} (\bibinfo {year} {2007})}\BibitemShut {NoStop}%
\bibitem [{\citenamefont {Suresh}\ and\ \citenamefont
  {Huynh}(1997)}]{suresh_1997_amp}%
  \BibitemOpen
  \bibfield  {author} {\bibinfo {author} {\bibfnamefont {A.}~\bibnamefont
  {Suresh}}\ and\ \bibinfo {author} {\bibfnamefont {H.~T.}\ \bibnamefont
  {Huynh}},\ }\href {\doibase DOI: 10.1006/jcph.1997.5745} {\bibfield
  {journal} {\bibinfo  {journal} {Journal of Computational Physics}\ }\textbf
  {\bibinfo {volume} {136}},\ \bibinfo {pages} {83} (\bibinfo {year}
  {1997})}\BibitemShut {NoStop}%
\bibitem [{\citenamefont {Zhou}\ \emph {et~al.}(2001)\citenamefont {Zhou},
  \citenamefont {Li},\ and\ \citenamefont {Shu}}]{zhou_2001_ncw}%
  \BibitemOpen
  \bibfield  {author} {\bibinfo {author} {\bibfnamefont {T.}~\bibnamefont
  {Zhou}}, \bibinfo {author} {\bibfnamefont {Y.}~\bibnamefont {Li}}, \ and\
  \bibinfo {author} {\bibfnamefont {C.-W.}\ \bibnamefont {Shu}},\ }\href
  {\doibase 10.1023/A:1012282706985} {\bibfield  {journal} {\bibinfo  {journal}
  {Journal of Scientific Computing}\ }\textbf {\bibinfo {volume} {16}},\
  \bibinfo {pages} {145} (\bibinfo {year} {2001})}\BibitemShut {NoStop}%
\bibitem [{\citenamefont {Noh}(1987)}]{noh_1987_ecs}%
  \BibitemOpen
  \bibfield  {author} {\bibinfo {author} {\bibfnamefont {W.}~\bibnamefont
  {Noh}},\ }\href {\doibase 10.1016/0021-9991(87)90074-X} {\bibfield  {journal}
  {\bibinfo  {journal} {Journal of Computational Physics}\ }\textbf {\bibinfo
  {volume} {72}},\ \bibinfo {pages} {78} (\bibinfo {year} {1987})}\BibitemShut
  {NoStop}%
\bibitem [{\citenamefont {Rider}(2000)}]{rider_2000_rwh}%
  \BibitemOpen
  \bibfield  {author} {\bibinfo {author} {\bibfnamefont {W.}~\bibnamefont
  {Rider}},\ }\href {\doibase 10.1006/jcph.2000.6544} {\bibfield  {journal}
  {\bibinfo  {journal} {Journal of Computational Physics}\ }\textbf {\bibinfo
  {volume} {162}},\ \bibinfo {pages} {395} (\bibinfo {year}
  {2000})}\BibitemShut {NoStop}%
\bibitem [{\citenamefont {Michel}(1972)}]{Michel72}%
  \BibitemOpen
  \bibfield  {author} {\bibinfo {author} {\bibfnamefont {F.~C.}\ \bibnamefont
  {Michel}},\ }\href@noop {} {\bibfield  {journal} {\bibinfo  {journal}
  {Astrophys. Spa. Sci.}\ }\textbf {\bibinfo {volume} {15}},\ \bibinfo {pages}
  {153} (\bibinfo {year} {1972})}\BibitemShut {NoStop}%
\bibitem [{\citenamefont {Tooper}(1965)}]{tooper_1965_afs}%
  \BibitemOpen
  \bibfield  {author} {\bibinfo {author} {\bibfnamefont {R.~F.}\ \bibnamefont
  {Tooper}},\ }\href@noop {} {\bibfield  {journal} {\bibinfo  {journal}
  {Astrophysical Journal}\ }\textbf {\bibinfo {volume} {142}},\ \bibinfo
  {pages} {1541} (\bibinfo {year} {1965})}\BibitemShut {NoStop}%
\bibitem [{\citenamefont {Yoshida}\ and\ \citenamefont
  {Eriguchi}(2001)}]{Yoshida01}%
  \BibitemOpen
  \bibfield  {author} {\bibinfo {author} {\bibfnamefont {S.}~\bibnamefont
  {Yoshida}}\ and\ \bibinfo {author} {\bibfnamefont {Y.}~\bibnamefont
  {Eriguchi}},\ }\href@noop {} {\bibfield  {journal} {\bibinfo  {journal} {Mon.
  Not. R. Astron. Soc.}\ }\textbf {\bibinfo {volume} {322}},\ \bibinfo {pages}
  {389} (\bibinfo {year} {2001})}\BibitemShut {NoStop}%
\bibitem [{\citenamefont {Agrez}(2007)}]{agrez_2007_dfe}%
  \BibitemOpen
  \bibfield  {author} {\bibinfo {author} {\bibfnamefont {D.}~\bibnamefont
  {Agrez}},\ }\href {\doibase 10.1109/TIM.2007.908240} {\bibfield  {journal}
  {\bibinfo  {journal} {IEEE Transactions on Instrumentation and Measurement}\
  }\textbf {\bibinfo {volume} {56}},\ \bibinfo {pages} {2111} (\bibinfo {year}
  {2007})}\BibitemShut {NoStop}%
\bibitem [{\citenamefont {Cerda-Duran}(2009)}]{cerda-duran_2009_nvh}%
  \BibitemOpen
  \bibfield  {author} {\bibinfo {author} {\bibfnamefont {P.}~\bibnamefont
  {Cerda-Duran}},\ }\href
  {http://www.citebase.org/abstract?id=oai:arXiv.org:0912.1774} {\enquote
  {\bibinfo {title} {Numerical viscosity in hydrodynamics simulations in
  general relativity},}\ } (\bibinfo {year} {2009}),\ \Eprint
  {http://arxiv.org/abs/0912.1774v2} {0912.1774v2} \BibitemShut {NoStop}%
\bibitem [{\citenamefont {Giacomazzo}\ and\ \citenamefont
  {Rezzolla}(2007)}]{Giacomazzo:2007ti}%
  \BibitemOpen
  \bibfield  {author} {\bibinfo {author} {\bibfnamefont {B.}~\bibnamefont
  {Giacomazzo}}\ and\ \bibinfo {author} {\bibfnamefont {L.}~\bibnamefont
  {Rezzolla}},\ }\href@noop {} {\bibfield  {journal} {\bibinfo  {journal}
  {Class. Quantum Grav.}\ }\textbf {\bibinfo {volume} {24}},\ \bibinfo {pages}
  {S235} (\bibinfo {year} {2007})}\BibitemShut {NoStop}%
\bibitem [{\citenamefont {Pollney}\ \emph {et~al.}(2007)\citenamefont
  {Pollney}, \citenamefont {Reisswig}, \citenamefont {Rezzolla}, \citenamefont
  {Szil{\'a}gyi}, \citenamefont {Ansorg}, \citenamefont {Deris}, \citenamefont
  {Diener}, \citenamefont {Dorband}, \citenamefont {Koppitz}, \citenamefont
  {Nagar},\ and\ \citenamefont {Schnetter}}]{Pollney:2007ss}%
  \BibitemOpen
  \bibfield  {author} {\bibinfo {author} {\bibfnamefont {D.}~\bibnamefont
  {Pollney}}, \bibinfo {author} {\bibfnamefont {C.}~\bibnamefont {Reisswig}},
  \bibinfo {author} {\bibfnamefont {L.}~\bibnamefont {Rezzolla}}, \bibinfo
  {author} {\bibfnamefont {B.}~\bibnamefont {Szil{\'a}gyi}}, \bibinfo {author}
  {\bibfnamefont {M.}~\bibnamefont {Ansorg}}, \bibinfo {author} {\bibfnamefont
  {B.}~\bibnamefont {Deris}}, \bibinfo {author} {\bibfnamefont
  {P.}~\bibnamefont {Diener}}, \bibinfo {author} {\bibfnamefont {E.~N.}\
  \bibnamefont {Dorband}}, \bibinfo {author} {\bibfnamefont {M.}~\bibnamefont
  {Koppitz}}, \bibinfo {author} {\bibfnamefont {A.}~\bibnamefont {Nagar}}, \
  and\ \bibinfo {author} {\bibfnamefont {E.}~\bibnamefont {Schnetter}},\
  }\href@noop {} {\bibfield  {journal} {\bibinfo  {journal} {Phys. Rev. D}\
  }\textbf {\bibinfo {volume} {76}},\ \bibinfo {pages} {124002} (\bibinfo
  {year} {2007})}\BibitemShut {NoStop}%
\bibitem [{\citenamefont {Schnetter}\ \emph {et~al.}(2004)\citenamefont
  {Schnetter}, \citenamefont {Hawley},\ and\ \citenamefont
  {Hawke}}]{Schnetter-etal-03b}%
  \BibitemOpen
  \bibfield  {author} {\bibinfo {author} {\bibfnamefont {E.}~\bibnamefont
  {Schnetter}}, \bibinfo {author} {\bibfnamefont {S.~H.}\ \bibnamefont
  {Hawley}}, \ and\ \bibinfo {author} {\bibfnamefont {I.}~\bibnamefont
  {Hawke}},\ }\href@noop {} {\bibfield  {journal} {\bibinfo  {journal} {Class.
  Quantum Grav.}\ }\textbf {\bibinfo {volume} {21}},\ \bibinfo {pages} {1465}
  (\bibinfo {year} {2004})}\BibitemShut {NoStop}%
\bibitem [{\citenamefont {Dimmelmeier}\ \emph {et~al.}(2002)\citenamefont
  {Dimmelmeier}, \citenamefont {Font},\ and\ \citenamefont
  {M{\"u}ller}}]{Dimmelmeier02a}%
  \BibitemOpen
  \bibfield  {author} {\bibinfo {author} {\bibfnamefont {H.}~\bibnamefont
  {Dimmelmeier}}, \bibinfo {author} {\bibfnamefont {J.~A.}\ \bibnamefont
  {Font}}, \ and\ \bibinfo {author} {\bibfnamefont {E.}~\bibnamefont
  {M{\"u}ller}},\ }\href@noop {} {\bibfield  {journal} {\bibinfo  {journal}
  {Astron. Astrophys.}\ }\textbf {\bibinfo {volume} {388}},\ \bibinfo {pages}
  {917} (\bibinfo {year} {2002})}\BibitemShut {NoStop}%
\bibitem [{\citenamefont {Baiotti}\ \emph {et~al.}(2003)\citenamefont
  {Baiotti}, \citenamefont {Hawke}, \citenamefont {Montero},\ and\
  \citenamefont {Rezzolla}}]{Baiotti03a}%
  \BibitemOpen
  \bibfield  {author} {\bibinfo {author} {\bibfnamefont {L.}~\bibnamefont
  {Baiotti}}, \bibinfo {author} {\bibfnamefont {I.}~\bibnamefont {Hawke}},
  \bibinfo {author} {\bibfnamefont {P.}~\bibnamefont {Montero}}, \ and\
  \bibinfo {author} {\bibfnamefont {L.}~\bibnamefont {Rezzolla}},\ }in\
  \href@noop {} {\emph {\bibinfo {booktitle} {Computational Astrophysics in
  Italy: Methods and Tools}}},\ Vol.~\bibinfo {volume} {1},\ \bibinfo {editor}
  {edited by\ \bibinfo {editor} {\bibfnamefont {R.}~\bibnamefont
  {Capuzzo-Dolcetta}}}\ (\bibinfo  {publisher} {MSAIt},\ \bibinfo {address}
  {Trieste},\ \bibinfo {year} {2003})\ p.\ \bibinfo {pages} {210}\BibitemShut
  {NoStop}%
\bibitem [{\citenamefont {{Cordero-Carri{\'o}n}}\ \emph
  {et~al.}(2009)\citenamefont {{Cordero-Carri{\'o}n}}, \citenamefont
  {{Cerd{\'a}-Dur{\'a}n}}, \citenamefont {{Dimmelmeier}}, \citenamefont
  {{Jaramillo}}, \citenamefont {{Novak}},\ and\ \citenamefont
  {{Gourgoulhon}}}]{Cordero2009}%
  \BibitemOpen
  \bibfield  {author} {\bibinfo {author} {\bibfnamefont {I.}~\bibnamefont
  {{Cordero-Carri{\'o}n}}}, \bibinfo {author} {\bibfnamefont {P.}~\bibnamefont
  {{Cerd{\'a}-Dur{\'a}n}}}, \bibinfo {author} {\bibfnamefont {H.}~\bibnamefont
  {{Dimmelmeier}}}, \bibinfo {author} {\bibfnamefont {J.~L.}\ \bibnamefont
  {{Jaramillo}}}, \bibinfo {author} {\bibfnamefont {J.}~\bibnamefont
  {{Novak}}}, \ and\ \bibinfo {author} {\bibfnamefont {E.}~\bibnamefont
  {{Gourgoulhon}}},\ }\href {\doibase 10.1103/PhysRevD.79.024017} {\bibfield
  {journal} {\bibinfo  {journal} {Phys. Rev. D}\ }\textbf {\bibinfo {volume}
  {79}},\ \bibinfo {pages} {024017} (\bibinfo {year} {2009})}\BibitemShut
  {NoStop}%
\bibitem [{\citenamefont {Gourgoulhon}(1992)}]{gourgholon_1992_nra}%
  \BibitemOpen
  \bibfield  {author} {\bibinfo {author} {\bibfnamefont {E.}~\bibnamefont
  {Gourgoulhon}},\ }\href {\doibase 10.1088/0264-9381/9/S/005} {\bibfield
  {journal} {\bibinfo  {journal} {Classical and Quantum Gravity}\ }\textbf
  {\bibinfo {volume} {9}},\ \bibinfo {pages} {S117} (\bibinfo {year}
  {1992})}\BibitemShut {NoStop}%
\bibitem [{\citenamefont {Novak}(2001)}]{novak_2001_vic}%
  \BibitemOpen
  \bibfield  {author} {\bibinfo {author} {\bibfnamefont {J.}~\bibnamefont
  {Novak}},\ }\href
  {http://www.citebase.org/abstract?id=oai:arXiv.org:gr-qc/0107045} {\bibfield
  {journal} {\bibinfo  {journal} {Astron. Astrophys.}\ }\textbf {\bibinfo
  {volume} {376}},\ \bibinfo {pages} {606} (\bibinfo {year}
  {2001})}\BibitemShut {NoStop}%
\bibitem [{\citenamefont {{Liebling}}\ \emph {et~al.}(2010)\citenamefont
  {{Liebling}}, \citenamefont {{Lehner}}, \citenamefont {{Neilsen}},\ and\
  \citenamefont {{Palenzuela}}}]{liebling_2010_emr}%
  \BibitemOpen
  \bibfield  {author} {\bibinfo {author} {\bibfnamefont {S.~L.}\ \bibnamefont
  {{Liebling}}}, \bibinfo {author} {\bibfnamefont {L.}~\bibnamefont
  {{Lehner}}}, \bibinfo {author} {\bibfnamefont {D.}~\bibnamefont {{Neilsen}}},
  \ and\ \bibinfo {author} {\bibfnamefont {C.}~\bibnamefont {{Palenzuela}}},\
  }\href {\doibase 10.1103/PhysRevD.81.124023} {\bibfield  {journal} {\bibinfo
  {journal} {Phys. Rev. D}\ }\textbf {\bibinfo {volume} {81}},\ \bibinfo
  {pages} {124023} (\bibinfo {year} {2010})}\BibitemShut {NoStop}%
\bibitem [{\citenamefont {{Radice}}\ \emph {et~al.}(2010)\citenamefont
  {{Radice}}, \citenamefont {{Rezzolla}},\ and\ \citenamefont
  {{Kellerman}}}]{Radice:10}%
  \BibitemOpen
  \bibfield  {author} {\bibinfo {author} {\bibfnamefont {D.}~\bibnamefont
  {{Radice}}}, \bibinfo {author} {\bibfnamefont {L.}~\bibnamefont
  {{Rezzolla}}}, \ and\ \bibinfo {author} {\bibfnamefont {T.}~\bibnamefont
  {{Kellerman}}},\ }\href {\doibase 10.1088/0264-9381/27/23/235015} {\bibfield
  {journal} {\bibinfo  {journal} {Classical and Quantum Gravity}\ }\textbf
  {\bibinfo {volume} {27}},\ \bibinfo {pages} {235015} (\bibinfo {year}
  {2010})}\BibitemShut {NoStop}%
\bibitem [{\citenamefont {Duez}\ \emph {et~al.}(2006)\citenamefont {Duez},
  \citenamefont {Liu}, \citenamefont {Shapiro}, \citenamefont {Shibata},\ and\
  \citenamefont {Stephens}}]{Duez:2005cj}%
  \BibitemOpen
  \bibfield  {author} {\bibinfo {author} {\bibfnamefont {M.~D.}\ \bibnamefont
  {Duez}}, \bibinfo {author} {\bibfnamefont {Y.~T.}\ \bibnamefont {Liu}},
  \bibinfo {author} {\bibfnamefont {S.~L.}\ \bibnamefont {Shapiro}}, \bibinfo
  {author} {\bibfnamefont {M.}~\bibnamefont {Shibata}}, \ and\ \bibinfo
  {author} {\bibfnamefont {B.~C.}\ \bibnamefont {Stephens}},\ }\href {\doibase
  10.1103/PhysRevLett.96.031101} {\bibfield  {journal} {\bibinfo  {journal}
  {Phys. Rev. Lett.}\ }\textbf {\bibinfo {volume} {96}},\ \bibinfo {pages}
  {031101} (\bibinfo {year} {2006})}\BibitemShut {NoStop}%
\bibitem [{\citenamefont {Baiotti}\ and\ \citenamefont
  {Rezzolla}(2006)}]{Baiotti06}%
  \BibitemOpen
  \bibfield  {author} {\bibinfo {author} {\bibfnamefont {L.}~\bibnamefont
  {Baiotti}}\ and\ \bibinfo {author} {\bibfnamefont {L.}~\bibnamefont
  {Rezzolla}},\ }\href@noop {} {\bibfield  {journal} {\bibinfo  {journal}
  {Phys. Rev. Lett.}\ }\textbf {\bibinfo {volume} {97}},\ \bibinfo {pages}
  {141101} (\bibinfo {year} {2006})}\BibitemShut {NoStop}%
\bibitem [{\citenamefont {Baiotti}\ \emph {et~al.}(2007)\citenamefont
  {Baiotti}, \citenamefont {Hawke},\ and\ \citenamefont
  {Rezzolla}}]{Baiotti07}%
  \BibitemOpen
  \bibfield  {author} {\bibinfo {author} {\bibfnamefont {L.}~\bibnamefont
  {Baiotti}}, \bibinfo {author} {\bibfnamefont {I.}~\bibnamefont {Hawke}}, \
  and\ \bibinfo {author} {\bibfnamefont {L.}~\bibnamefont {Rezzolla}},\
  }\href@noop {} {\bibfield  {journal} {\bibinfo  {journal} {Class. Quantum
  Grav.}\ }\textbf {\bibinfo {volume} {24}},\ \bibinfo {pages} {S187} (\bibinfo
  {year} {2007})}\BibitemShut {NoStop}%
\bibitem [{\citenamefont {{Noble}}\ and\ \citenamefont
  {{Choptuik}}(2008)}]{Noble08a}%
  \BibitemOpen
  \bibfield  {author} {\bibinfo {author} {\bibfnamefont {S.~C.}\ \bibnamefont
  {{Noble}}}\ and\ \bibinfo {author} {\bibfnamefont {M.~W.}\ \bibnamefont
  {{Choptuik}}},\ }\href {\doibase 10.1103/PhysRevD.78.064059} {\bibfield
  {journal} {\bibinfo  {journal} {Phys. Rev. D}\ }\textbf {\bibinfo {volume}
  {78}},\ \bibinfo {pages} {064059} (\bibinfo {year} {2008})}\BibitemShut
  {NoStop}%
\bibitem [{\citenamefont {{Cordero-Carri{\'o}n}}\ \emph
  {et~al.}(2010)\citenamefont {{Cordero-Carri{\'o}n}}, \citenamefont
  {{Cerd{\'a}-Dur{\'a}n}},\ and\ \citenamefont {{Mar{\'{\i}}a
  Ib{\'a}{\~n}ez}}}]{Cordero10}%
  \BibitemOpen
  \bibfield  {author} {\bibinfo {author} {\bibfnamefont {I.}~\bibnamefont
  {{Cordero-Carri{\'o}n}}}, \bibinfo {author} {\bibfnamefont {P.}~\bibnamefont
  {{Cerd{\'a}-Dur{\'a}n}}}, \ and\ \bibinfo {author} {\bibfnamefont
  {J.}~\bibnamefont {{Mar{\'{\i}}a Ib{\'a}{\~n}ez}}},\ }\href {\doibase
  10.1088/1742-6596/228/1/012055} {\bibfield  {journal} {\bibinfo  {journal}
  {Journal of Physics Conference Series}\ }\textbf {\bibinfo {volume} {228}},\
  \bibinfo {pages} {012055} (\bibinfo {year} {2010})}\BibitemShut {NoStop}%
\bibitem [{\citenamefont {{O'Connor}}\ and\ \citenamefont
  {{Ott}}(2010)}]{OConnor10}%
  \BibitemOpen
  \bibfield  {author} {\bibinfo {author} {\bibfnamefont {E.}~\bibnamefont
  {{O'Connor}}}\ and\ \bibinfo {author} {\bibfnamefont {C.~D.}\ \bibnamefont
  {{Ott}}},\ }\href {\doibase 10.1088/0264-9381/27/11/114103} {\bibfield
  {journal} {\bibinfo  {journal} {Classical and Quantum Gravity}\ }\textbf
  {\bibinfo {volume} {27}},\ \bibinfo {pages} {114103} (\bibinfo {year}
  {2010})}\BibitemShut {NoStop}%
\bibitem [{\citenamefont {{Thierfelder}}\ \emph {et~al.}(2010)\citenamefont
  {{Thierfelder}}, \citenamefont {{Bernuzzi}}, \citenamefont {{Hilditch}},
  \citenamefont {{Bruegmann}},\ and\ \citenamefont
  {{Rezzolla}}}]{Thierfelder10}%
  \BibitemOpen
  \bibfield  {author} {\bibinfo {author} {\bibfnamefont {M.}~\bibnamefont
  {{Thierfelder}}}, \bibinfo {author} {\bibfnamefont {S.}~\bibnamefont
  {{Bernuzzi}}}, \bibinfo {author} {\bibfnamefont {D.}~\bibnamefont
  {{Hilditch}}}, \bibinfo {author} {\bibfnamefont {B.}~\bibnamefont
  {{Bruegmann}}}, \ and\ \bibinfo {author} {\bibfnamefont {L.}~\bibnamefont
  {{Rezzolla}}},\ }\href@noop {} {\bibfield  {journal} {\bibinfo  {journal}
  {Phys. Rev. D}\ }\textbf {\bibinfo {volume} {83}},\ \bibinfo {pages} {064022}
  (\bibinfo {year} {2010})}\BibitemShut {NoStop}%
\bibitem [{\citenamefont {Shu}(2001)}]{Shu01}%
  \BibitemOpen
  \bibfield  {author} {\bibinfo {author} {\bibfnamefont {C.~W.}\ \bibnamefont
  {Shu}},\ }\href@noop {} {\emph {\bibinfo {title} {{H}igh order finite
  difference and finite volume WENO schemes and discontinuous Galerkin methods
  for CFD}}},\ \bibinfo {type} {Tech. Rep.}\ \bibinfo {number} {ICASE Report
  2001-11; NASA CR-2001-210865}\ (\bibinfo  {institution} {NASA Langley
  Research Center},\ \bibinfo {year} {2001})\BibitemShut {NoStop}%
\end{thebibliography}
\end{document}